\documentclass[12pt,english]{article}
\usepackage[T1]{fontenc}
\usepackage[latin9]{inputenc}
\usepackage{geometry}
\usepackage{babel}
\usepackage{array}
\usepackage{booktabs}
\usepackage{mathtools}
\usepackage{multirow}
\usepackage{amsmath}
\usepackage{amsthm}
\usepackage{amssymb}
\usepackage{graphicx}
\usepackage{setspace}
\usepackage[authoryear]{natbib}
\usepackage[unicode=true,pdfusetitle,
 bookmarks=true,bookmarksnumbered=false,bookmarksopen=false,
 breaklinks=false,pdfborder={0 0 1},backref=false,colorlinks=false]
 {hyperref}
\hypersetup{
 urlbordercolor={0 1 0}, linkbordercolor={0 1 0}, citebordercolor={0 1 0}}

\makeatletter

\providecommand{\tabularnewline}{\\}

\usepackage{algorithmicx,algorithm} 
\usepackage[noend]{algpseudocode} 
\usepackage[dvipsnames,svgnames,x11names,hyperref]{xcolor}
\usepackage{threeparttable}

\newcommand{\indep}{\perp \!\!\! \perp}
\usepackage{appendix}
\usepackage{chngpage}
\usepackage{lipsum}

\definecolor{harvardcrimson}{rgb}{0.79, 0.0, 0.09}
\definecolor{princetonorange}{rgb}{1.0, 0.56, 0.0}
\definecolor{brickred}{rgb}{0.8, 0.25, 0.33}
\definecolor{brightmaroon}{rgb}{0.76, 0.13, 0.28}
\definecolor{brownweb}{rgb}{0.65, 0.16, 0.16}
\definecolor{burgundy}{rgb}{0.5, 0.0, 0.13}
\definecolor{burntumber}{rgb}{0.54, 0.2, 0.14}
\definecolor{carmine}{rgb}{0.59, 0.0, 0.09}
\definecolor{carnelian}{rgb}{0.7, 0.11, 0.11}
\definecolor{airforceblue}{rgb}{0.36, 0.54, 0.66}
\definecolor{ceil}{rgb}{0.57, 0.63, 0.81}
\definecolor{cordovan}{rgb}{0.54, 0.25, 0.27}
\definecolor{cornellred}{rgb}{0.7, 0.11, 0.11}
\definecolor{crimsonglory}{rgb}{0.75, 0.0, 0.2}
\definecolor{darkpastelred}{rgb}{0.76, 0.23, 0.13}
\definecolor{oucrimsonred}{rgb}{0.6, 0.0, 0.0}
\definecolor{darkpastelred}{rgb}{0.76, 0.23, 0.13}
\definecolor{darkterracotta}{rgb}{0.8, 0.31, 0.36}
\definecolor{deepchestnut}{rgb}{0.73, 0.31, 0.28}
\definecolor{indianred}{rgb}{0.8, 0.36, 0.36}
\definecolor{maroon}{rgb}{0.5, 0.0, 0.0}
\definecolor{mediumcarmine}{rgb}{0.69, 0.25, 0.21}
\definecolor{tuftsblue}{rgb}{0.28, 0.57, 0.81}
\definecolor{yaleblue}{rgb}{0.06, 0.3, 0.57}
\definecolor{darkpastelblue}{rgb}{0.47, 0.62, 0.8}
\definecolor{bluegray}{rgb}{0.4, 0.6, 0.8}
\definecolor{airforceblue}{rgb}{0.36, 0.54, 0.66}

\@ifundefined{showcaptionsetup}{}{%
 \PassOptionsToPackage{caption=false}{subfig}}
\usepackage{subfig}
\makeatother

\begin{document}
\pagenumbering{gobble}
\title{Ensemble Learning with Statistical and Structural Models\thanks{We thank Panle Jia Barwick, Whitney Newey, and seminar audiences for
many helpful discussions and suggestions. Mao acknowledges financial
support by the national natural science foundation of China.}}
\author{Jiaming Mao\thanks{Corresponding author. Xiamen University, Email: \texttt{\protect\href{mailto:jmao@xmu.edu.cn}{jmao@xmu.edu.cn}}}
\and Jingzhi Xu\thanks{Xiamen University, Email: \texttt{\protect\href{mailto:jingzhixu@stu.xmu.edu.cn}{jingzhixu@stu.xmu.edu.cn}}}}
\maketitle
\begin{abstract}
\begin{onehalfspace}
Statistical and structural modeling represent two distinct approaches
to data analysis. In this paper, we propose a set of novel methods
for combining statistical and structural models for improved prediction
and causal inference. Our first proposed estimator has the doubly
robustness property in that it only requires the correct specification
of either the statistical or the structural model. Our second proposed
estimator is a weighted ensemble that has the ability to outperform
both models when they are both misspecified. Experiments demonstrate
the potential of our estimators in various settings, including fist-price
auctions, dynamic models of entry and exit, and demand estimation
with instrumental variables. 
\end{onehalfspace}
\end{abstract}
\newpage{}

\pagenumbering{arabic}

\section{Introduction}

In economics as well as many other scientific disciplines, statistical
and structural modeling represent two distinct approaches to data
analysis \citep{heckman_causal_2000}. The structural approach draws
a direct link between data and theory. It estimates structural models,
or \emph{scientific models} \citep{shalizi_advanced_2013}, that specify
the causal mechanisms generating the observed data. A complete structural
model in economics describes economic and social phenomena as the
outcomes of individual behavior in specific economic and social environments
\citep{heckman_econometric_2007,reiss_structural_2007}. Once estimated,
these models can be used for making predictions, evaluating causal
effects, and conducting normative welfare analyses \citep{low_use_2017}.

In contrast to the structural approach, the statistical approach to
data analysis relies on the use of statistical models for prediction
and causal inference. While recent advances in machine learning have
focused on predictive tasks \citep{athey_beyond_2017}, a large literature
in causal inference across multiple disciplines\footnote{e.g. the social sciences, the biomedical sciences, statistics, and
computer science.} has proposed statistical methods for estimating causal effects from
experimental and observational data \citep{imbens_causal_2015}\footnote{In the statistical approach to causal inference, causal knowledge
is used not to specify a complete structural model, but to inform
research designs that can identify the causal effects of interest
by exploiting exogenous variations in the data.}. In economics, this statistical approach to causal inference is informally
referred to as the \emph{reduced-form} approach \citep{chetty_sufficient_2009}\footnote{As \citet{chetty_sufficient_2009} pointed out, the term ``reduced-form''
is largely a misnomer, whose meaning in the econometrics literature
today has departed from its historical root. Historically, a reduced-form
model is an alternative representation of a structural model. Given
a structural model $\mathcal{M}(x,y,\epsilon)=0$, where $x$ is exogenous,
$y$ is endogenous, and $\epsilon$ is unobserved, if we write $y$
as a function of $x$ and $\epsilon$, $y=f(x,\epsilon)$, then $f$
is the\emph{ reduced-form} of $\mathcal{M}$ \citep{reiss_structural_2007}.
Today, however, applied economists typically refer to nonstructural,
statistical treatment effect models as ``reduced-form'' models.
Perhaps reflecting the informal nature of the terminology today, \citet{rust_limits_2014}
gave the following definitions of the two approaches: ``At the risk
of oversimplifying, empirical work that takes theory \textquotedblleft seriously\textquotedblright{}
is referred to as \emph{structural econometrics} whereas empirical
work that avoids a tight integration of theory and empirical work
is referred to as \emph{reduced form econometrics}.''}. Methods such as controlling for observed confounding and instrumental
variables regression are widely used in applied economic analyses
\citep{athey_state_2017}.

Which approach should be preferred -- the statistical or the structural
-- has been the subject of a long-standing debate within the economics
profession \citep{angrist_credibility_2010,deaton_instruments_2010,keane_structural_2010,keane_structural_2010-1,nevo_taking_2010,wolpin_limits_2013}.
For predictive tasks, statistical and machine learning models often
fit the observed data well and have advantages in \emph{in-domain}
prediction, where the training and the test data have the same distribution\footnote{Using the terminology of transfer learning, a \emph{domain} is a joint
distribution governing the input and output variables \citep{muandet_domain_2013}.
A key limitation with most statistical and machine learning models
is that they require the distributions governing the training data
(the \emph{source} domain) and the test data (the \emph{target} domain)
to be the same in order to guarantee performance \citep{ben-david_theory_2010}. }. On the other hand, a main advantage of structural estimation lies
in its ability to make \emph{out-of-domain} predictions\footnote{In this paper, we distinguish between the notions of \emph{out-of-domain}
and \emph{out-of-sample}. Out-of-sample data are test data drawn from
the same distribution as the training data.}. As long as the same causal mechanism governs data generation, a
correctly specified structural model provides a way to extrapolate
from the training data to the test data even if the distributions
have changed\footnote{Traditionally, economists emphasize the ability of structural models
to make \emph{counterfactual} predictions. We note that counterfactual
predictions can be viewed as a special type of out-of-domain predictions.}. Similarly, in causal inference, reduced-form methods that exploit
credible sources of identifying information deliver estimates of causal
effects with high \emph{internal validity}\footnote{\citet{angrist_credibility_2010} offered an account of what they
call ``the credibility revolution'' -- the increasing popularity
of quasi-experimental methods that seek natural experiments as sources
of identifying information. Our definition of reduced-form methods
include both quasi-experimental and more traditional, non-quasi-experimental
statistical methods that use expert knowledge to locate exogenous
sources of variation.}, while structural estimates may have more claims to \emph{external
validity}. 

The relative strengths of the two approaches point to a complementarity
that provides the motivation for this paper. Of course, the reason
that any approach may outperform the other in certain aspects of data
analysis is fundamentally due to model misspecification\footnote{By model misspecification, we refer to both incorrect functional form
and distributional assumptions and, in the case of causal inference,
incorrect causal assumptions.} -- if any model captures the true distributions governing the source
and the target domains, then no improvement is possible. Indeed, one
can argue that researchers on both sides of the methodological debate
are motivated by a shared concern over model misspecification. Proponents
for the statistical approach are concerned about misspecifications
due to the often strong and unrealistic assumptions -- both causal
and parametric -- made in structural models, while those advocating
for the structural approach are concerned about misspecifications
due to not incorporating theoretical insight -- functional forms
such as constant elasticity of substitution (CES) aggregation and
the gravity equation of trade, for example, often encode important
prior economic knowledge that sophisticated statistical and machine
learning methods would not be able to capture based on training data
alone\footnote{\citet{rust_limits_2014}: \textquotedblleft Notice the huge difference
in world views. The primary concern of Leamer, Manski, Pischke, and
Angrist is that we rely too much on assumptions that could be wrong,
and which could result in incorrect empirical conclusions and policy
decisions. Wolpin argues that assumptions and models could be right,
or at least they may provide reasonable first approximations to reality.\textquotedblright{}}.

In this paper, we propose a set of methods for combining the statistical
and structural approaches for improved prediction and causal inference.
Our first proposed estimator, which we call the \emph{doubly robust
statistical-structural} (\emph{DRSS}) estimator, provides a consistent
in-domain estimate as long as either the structural or the (reduced-form)
statistical model is correctly specified. Our second proposed estimator,
which we call the \emph{ensemble statistical-structural} (\emph{ESS})
estimator, is a weighted ensemble that has the ability to outperform
both the structural and the (reduced-form) statistical model, both
\emph{in-domain} and \emph{out-of-domain}, when both are misspecified. 

Our methods build on several intuitions. First, statistically speaking,
a structural model is a \emph{generative }model \citep{jebara_machine_2012}.
Given a structural model that specifies the data-generating mechanism
of $\left(x_{1},\ldots,x_{p}\right)\in\mathcal{O}$, we can generate
predictions of \emph{discriminative} relationships $\mathbb{E}\left[\left.x_{j}\right|x_{i}\right]$
or $\mathbb{E}\left[x_{j}^{x_{i}=a}\right]$ for any $\left(x_{i},x_{j}\right)\subset\left(x_{1},\ldots,x_{p}\right)$,
where $x_{j}^{x_{i}=a}$ denotes the \emph{potential outcome} of $x_{j}$
under the intervention of $x_{i}=a$\footnote{In this paper, we mainly adopt the notations of the Rubin causal model
\citep{rubin_estimating_1974} in discussing causal inference. Equivalently,
using the notation of \citep{pearl_causality_2009}, $\mathbb{E}\left[x_{j}^{x_{i}=a}\right]$
can be expressed as $\mathbb{E}\left[\left.x_{j}\right|\text{do}\left(x_{i}=a\right)\right]$. }. These structurally derived relationships can then be considered
as \emph{competitors} to (reduced-form) statistical models that explicitly
model these relationships. This allows us to leverage the large statistical
literature on dealing with competing models. One popular method used
in causal inference is the doubly robust estimator that combines an
outcome regression model with a treatment assignment model in the
estimation of causal effects \citep{bang_doubly_2005}. The doubly
robust estimator is consistent if either of the two models is correctly
specified, thus providing an insurance against model misspecification.
\citet{lewbel_general_2019} generalized the classic doubly robust
method to allow the combination of any parametric models. Their method
provides a basis for our DRSS estimator.

Second, the complementary properties of statistical and structural
models suggest that a model combination approach may yield superior
results \citep{kellogg_combining_2020}. In the Bayesian paradigm,
model averaging has long been proposed as an alternative to model
selection \citep{hoeting_bayesian_1999}. Given a set of candidate
models, bayesian model averaging produces a weighted average, with
each model weighted by its posterior probability. Doing so accounts
for the model uncertainty that is ignored by the standard practice
of selecting a single model. More recently, in the machine learning
literature, ensemble methods such as stacking, bagging, and boosting
are proposed that seek to combine models to improve prediction so
that the ensemble performs better than any of its individual members
\citep{dietterich_ensemble_2000}. These methods work by not only
incorporating model uncertainty but expanding the space of representable
functions \citep{minka_bayesian_2000}\footnote{When the models being combined are complex and high-dimensional for
which global optima are hard to obtain, the ensemble approach also
produces gains by averaging local optima produced by local search
\citep{dietterich_ensemble_2000}.}. As \citet{breiman_stacked_1996} pointed out, ensemble methods benefit
the most from the use of diverse and dissimilar models, which is exactly
the case when we combine statistical and structural models. 

In this paper, we provide two ensemble estimators. The first, which
we call \emph{ESS-LN}, is a linear ensemble based on the method of
\emph{stacking} \citep{wolpert_stacked_1992}, or \emph{jackknife
averaging} \citep{hansen_jackknife_2012}, which produces an optimal
linear combination of a set of models by minimizing a cross-validated
loss criterion such as expected mean squared error. We show how to
use the method both for prediction and causal inference. Our second
ensemble estimator, \emph{ESS-NP}, goes beyond linear combinations
and builds a nonparametric ensemble of statistical and structural
models. For conditional mean estimation, it employs the random forest
algorithm introduced by \citet{breiman_random_2001}, which allows
for the modeling of nonlinear relationships and complex interactions
by building a large number of regression trees that adaptively partition
the input space and combining them through bootstrap aggregation.
The method can be viewed as an adaptive locally weighted estimator
\citep{athey_generalized_2019}, allowing us to assign different weights
to different regions of the input space depending on which model --
the statistical or the structural -- performs better in that region.
The resulting ensemble has the ability to combine the strengths of
statistical and structural models while defending against their weaknesses.

\paragraph*{Example}

\begin{figure}
\subfloat[\label{fig:consumption_a}]{\includegraphics[width=0.5\columnwidth]{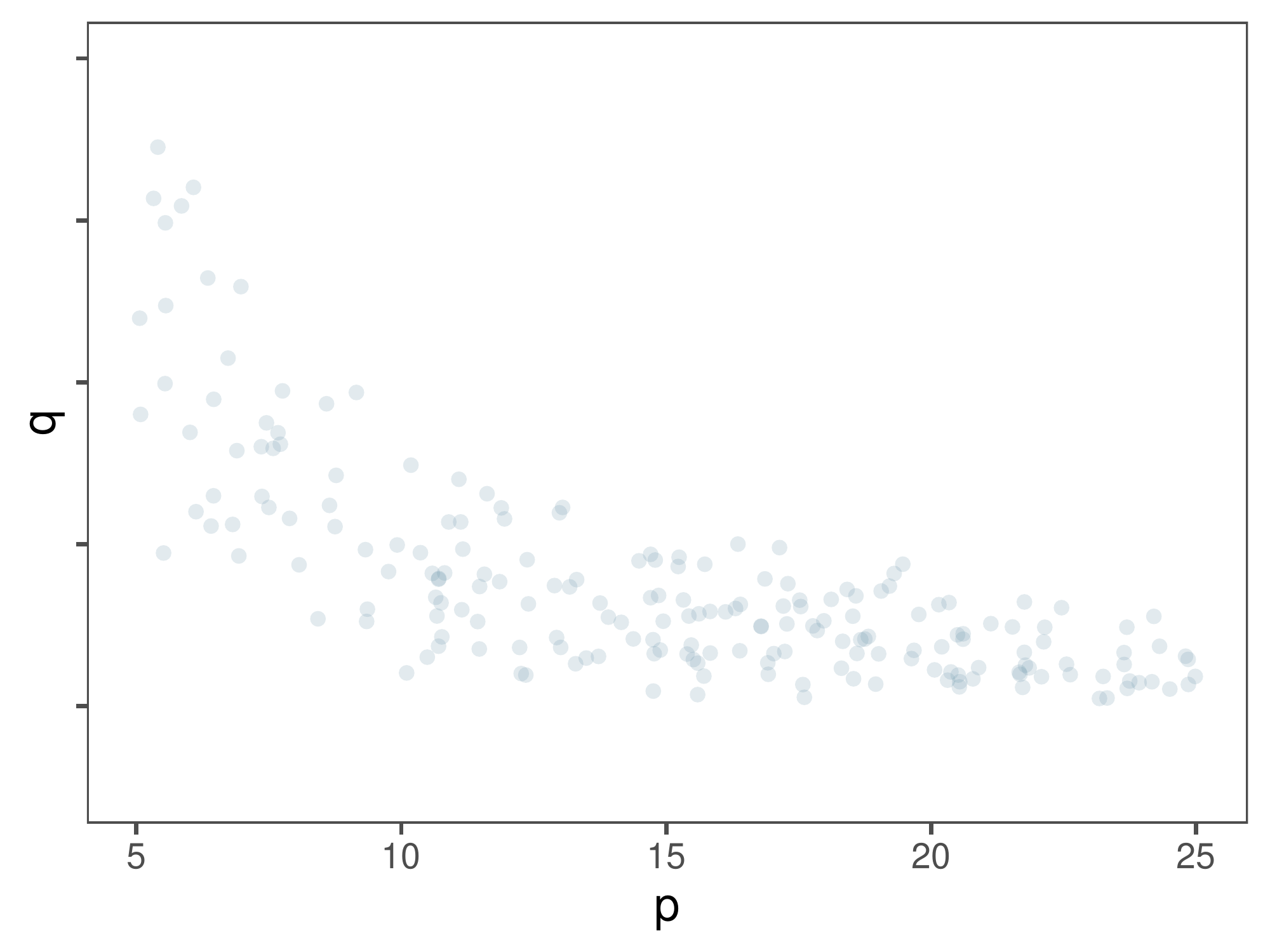}

}\subfloat[\label{fig:consumption_b}]{\includegraphics[width=0.5\columnwidth]{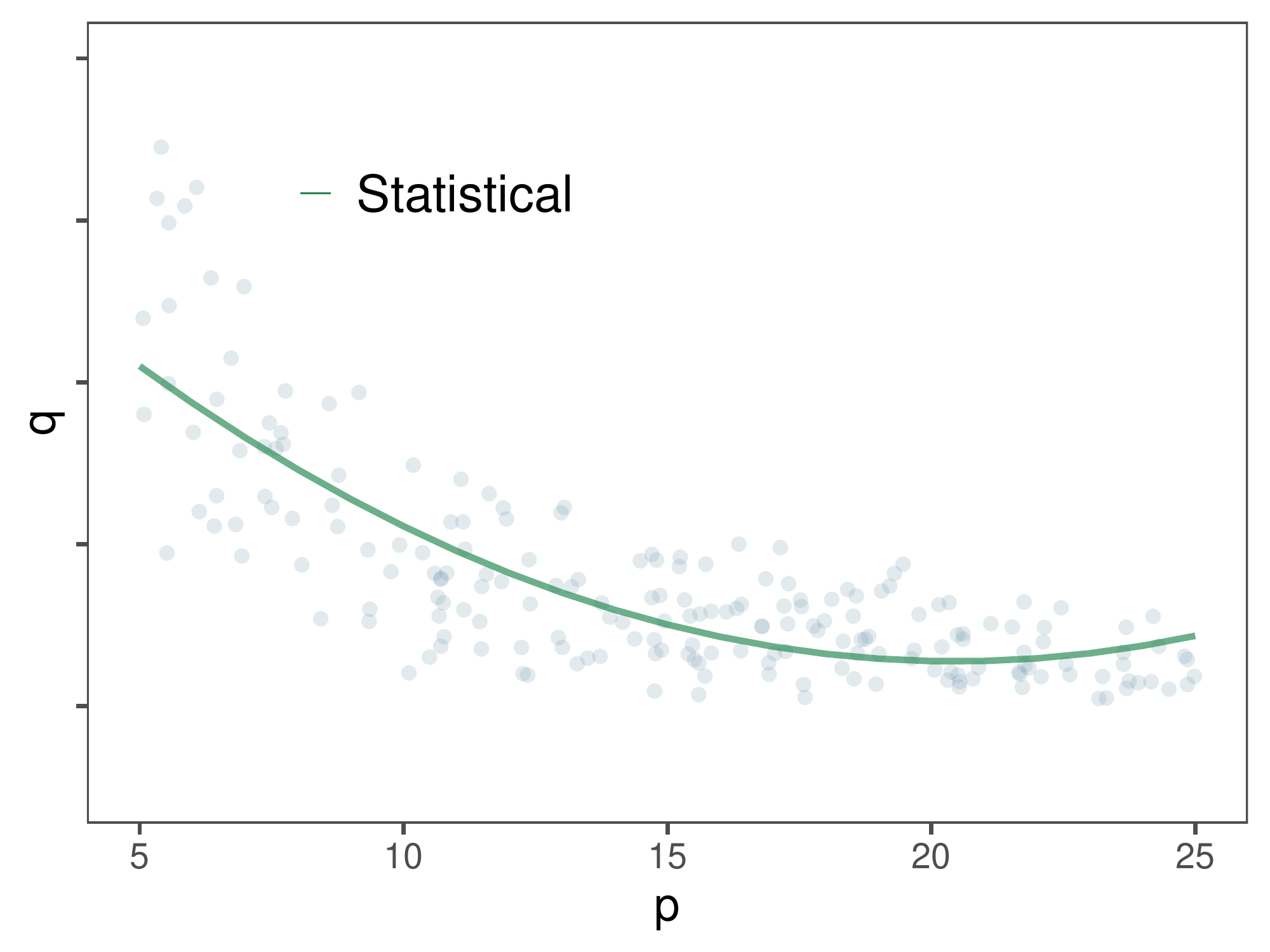}

}

\subfloat[\label{fig:consumption_c}]{\includegraphics[width=0.5\columnwidth]{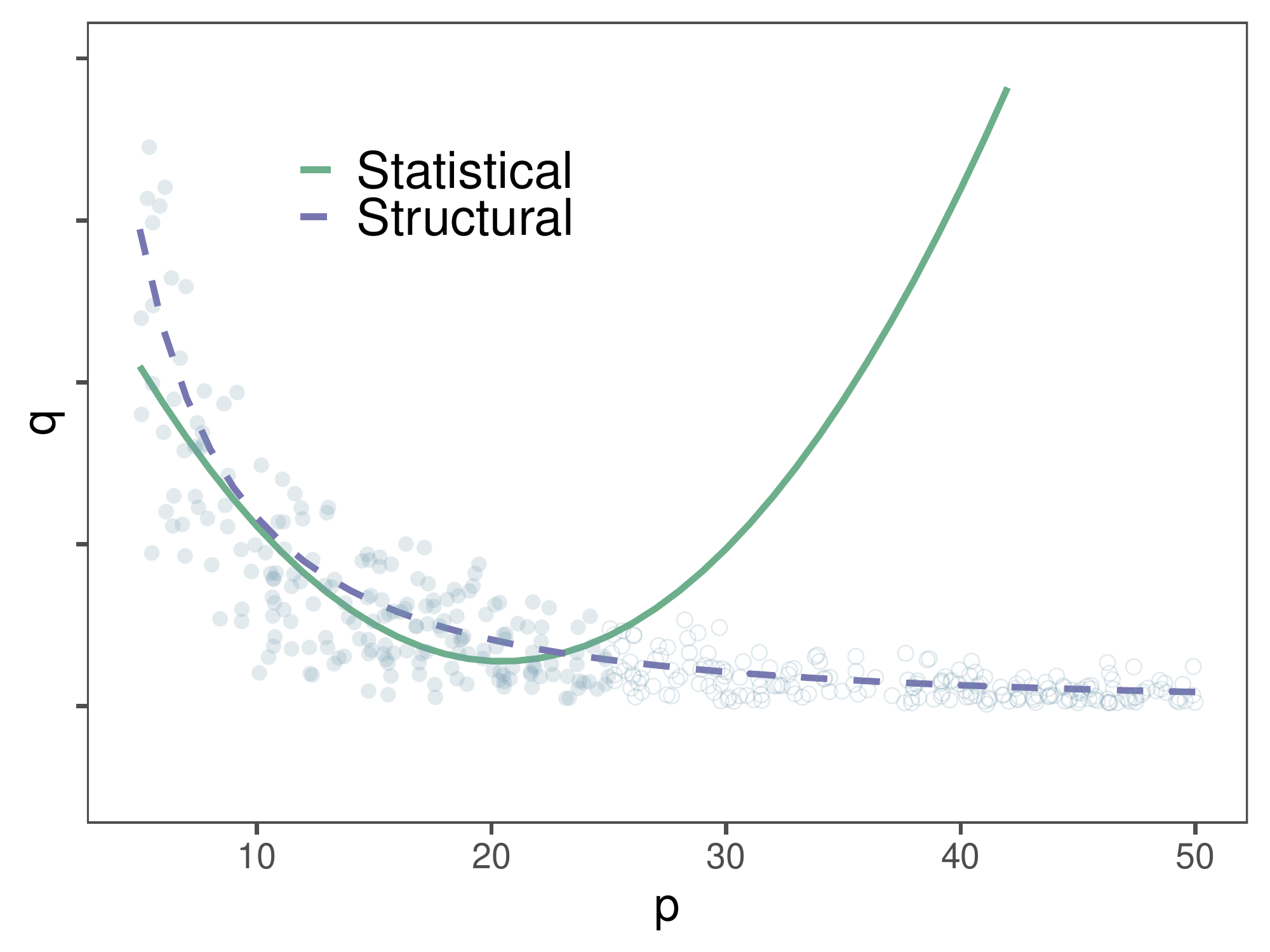}

}\subfloat[\label{fig:consumption_d}]{\includegraphics[width=0.5\columnwidth]{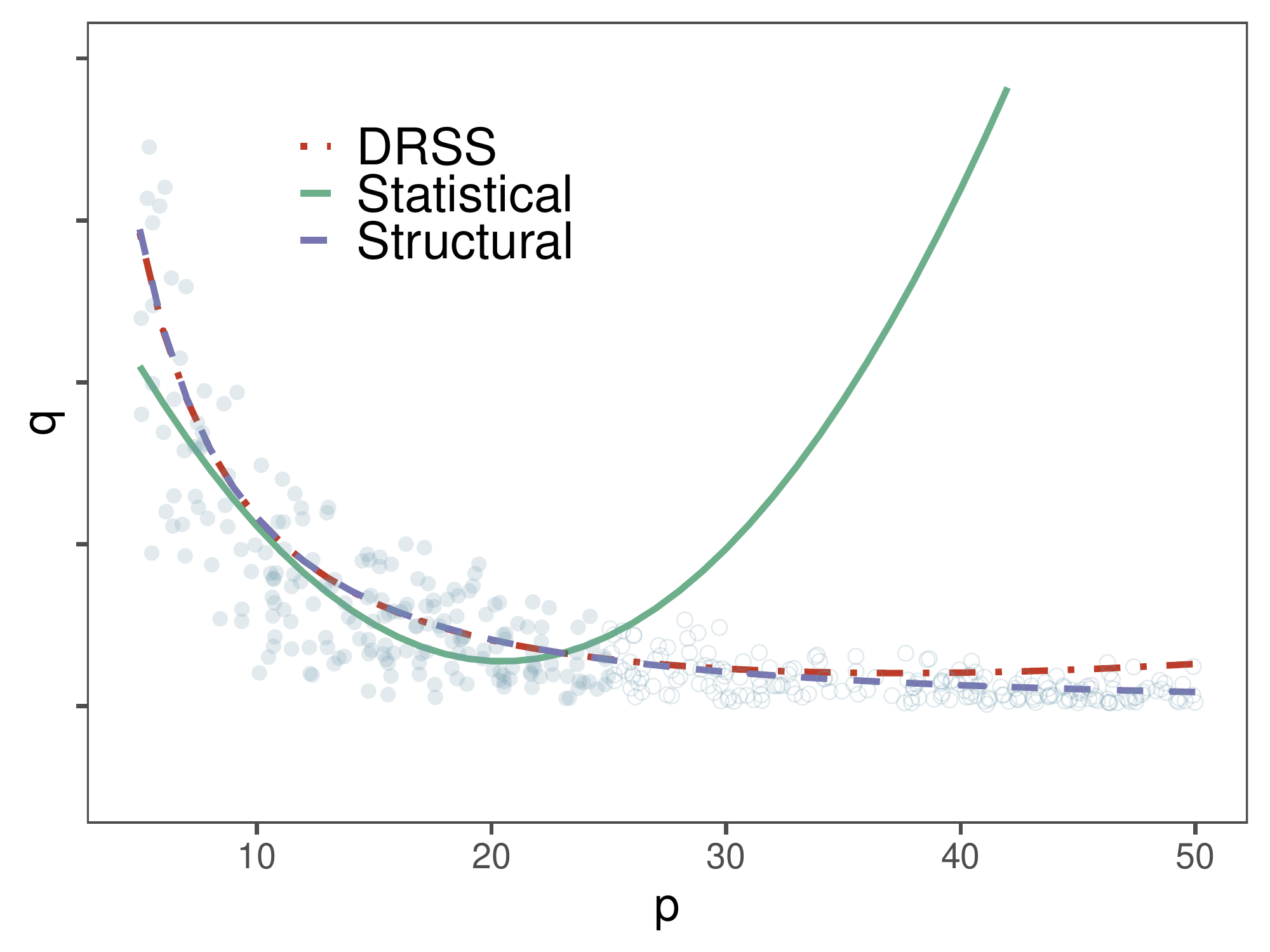}

}

\subfloat[\label{fig:consumption_e}]{\includegraphics[width=0.5\columnwidth]{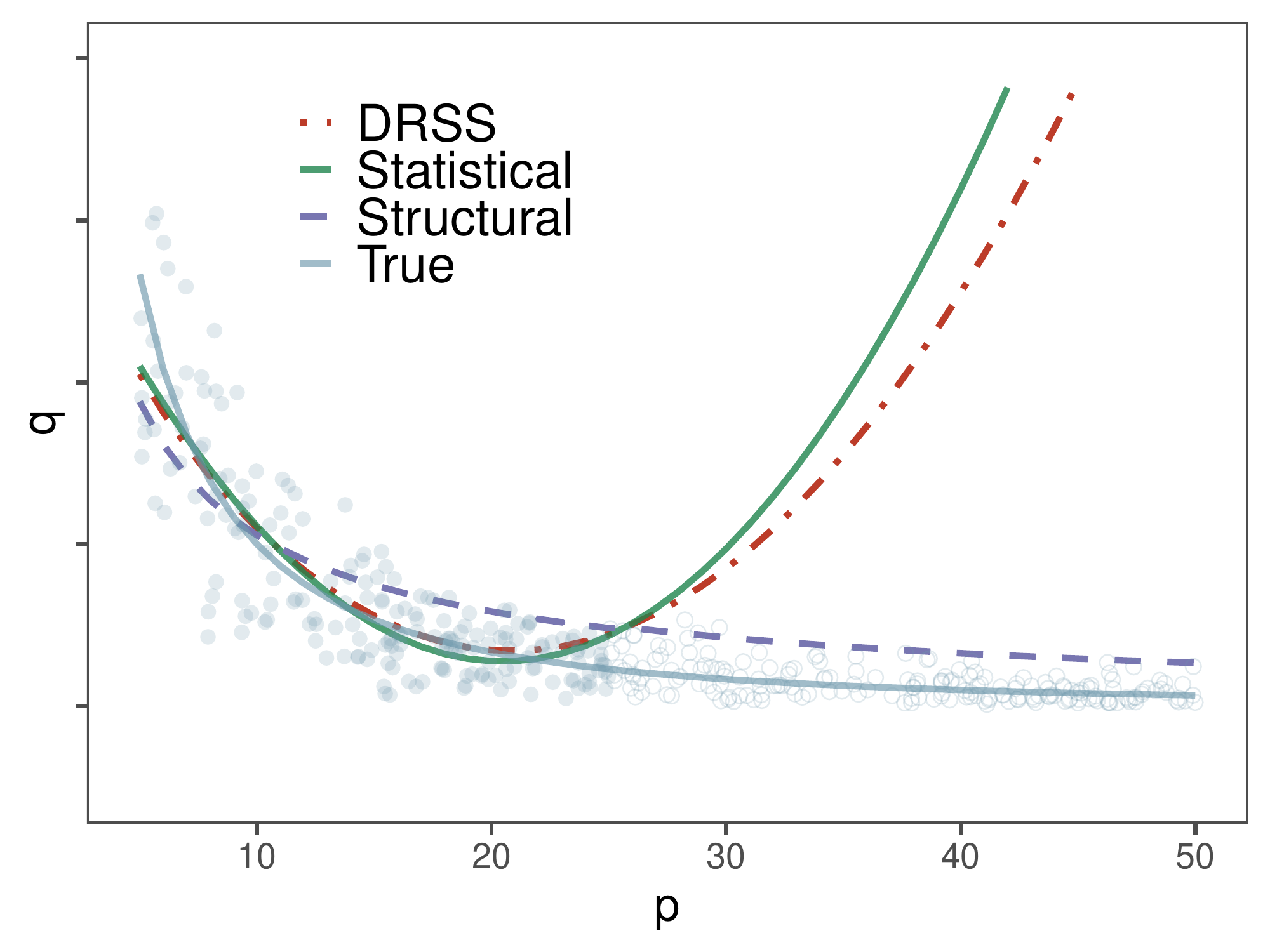}

}\subfloat[\label{fig:consumption_f}]{\includegraphics[width=0.5\columnwidth]{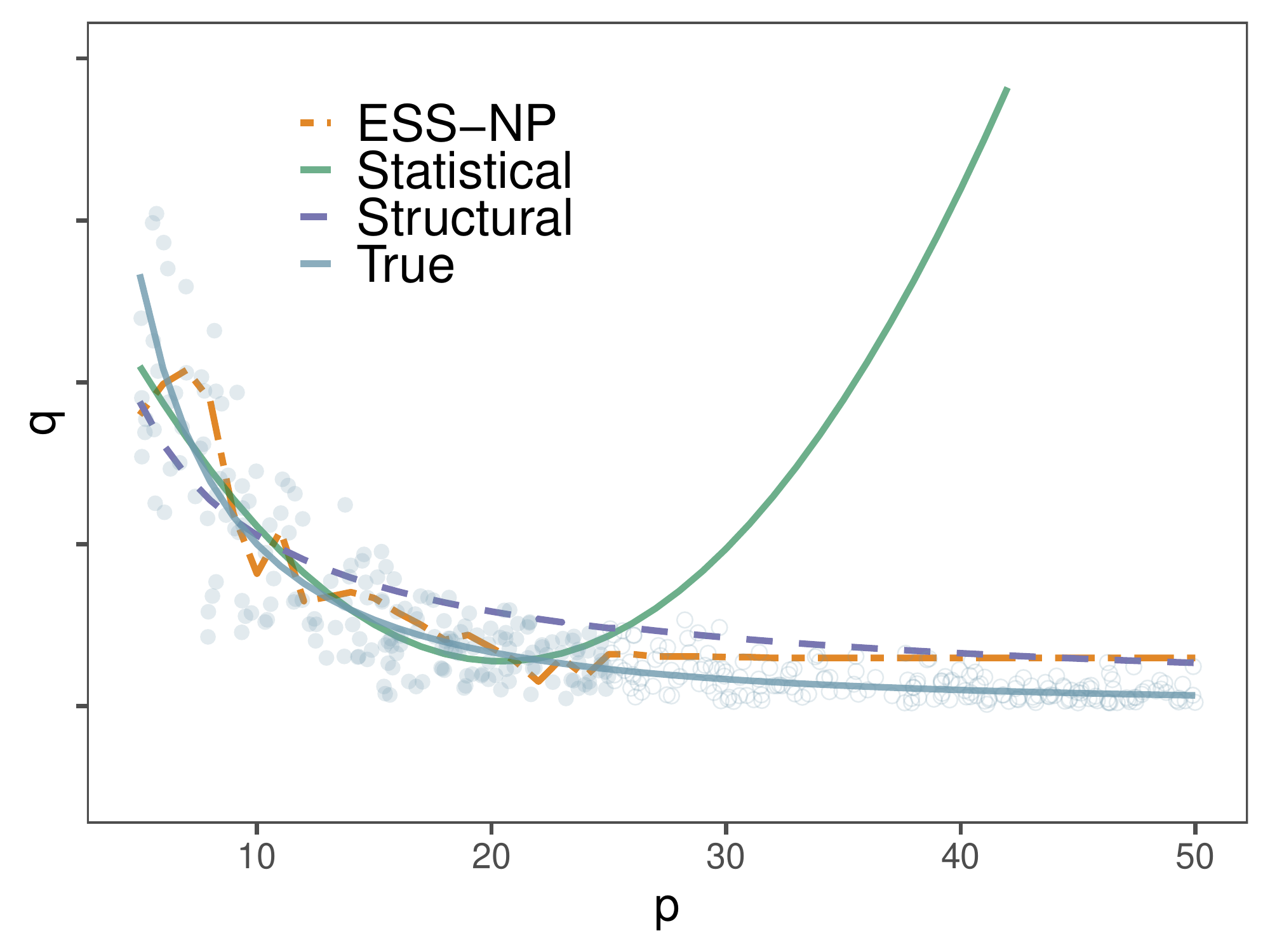}

}

\caption{{\small{}Demand Estimation. Filled circles represent training data.
Unfilled circles represent out-of-domain test data.}}
\end{figure}

To illustrate our methods, consider the setting of a simple demand
estimation problem. We observe the prices and quantities sold of a
good $x$, as plotted in Figure \ref{fig:consumption_a}. Suppose
the data are generated by the consumption decisions of $n$ consumers
who purchased $x$ at different prices. Each consumer had fixed income
$I$ and decided how much to purchase by solving the problem:
\begin{equation}
\max_{q,q^{o}}u_{i}\left(q_{i},q_{i}^{o}\right)\quad\text{subject to }p_{i}q_{i}+p_{i}^{o}q_{i}^{o}\le I\label{eq:motivate}
\end{equation}
, where $\left(p_{i},q_{i},p_{i}^{o},q_{i}^{o}\right)$ denote respectively
the price and quantity of good $x$ and of an outside good $o$. The
consumer utility function is given by the following CES function:
\begin{equation}
u_{i}\left(q_{i},q_{i}^{o}\right)=\left[\alpha_{i}q_{i}^{\rho}+\left(1-\alpha_{i}\right)\left(q_{i}^{o}\right)^{\rho}\right]^{\frac{1}{\rho}}
\end{equation}
, where $\rho=\frac{1}{2}$, suggesting an elasticity of substitution
of $2$\footnote{$\left\{ \alpha_{i}\right\} $ are generated as follows: 
\[
\alpha_{i}=\frac{\exp\left(\xi_{i}\right)}{1+\exp\left(\xi_{i}\right)},\quad\xi_{i}\sim\mathcal{N}\left(0,0.5\right)
\]
}.

We can fit the following statistical model to the data:
\begin{equation}
q_{i}=\beta_{0}+\beta_{1}p_{i}+\beta_{2}p_{i}^{2}+\epsilon_{i}\label{eq:appetizier02}
\end{equation}

The result is plotted in Figure \ref{fig:consumption_b}. Under the
causal assumption that prices are exogenous to the consumers, \eqref{eq:appetizier02}
represents a \emph{reduced-form} estimate of the individual demand
curve. The model appears to fit the data quite well. However, once
we extrapolate beyond the observed ranges of prices, its predictions
become very bad (Figure \ref{fig:consumption_c}). On the other hand,
structurally estimating the parameters of model \eqref{eq:motivate}
would yield a demand curve that has both \emph{internal} and \emph{external
validity} (Figure \ref{fig:consumption_c}). This is not surprising
as \eqref{eq:motivate} describes the true data-generating mechanism.
In practice, given two competing models, the (reduced-form) statistical
model \eqref{eq:appetizier02} and the structural model \eqref{eq:motivate},
we may not know which one is correctly specified. The DRSS resolves
this issue by combining the two models and providing a consistent
estimate as long as one of them is correctly specified. Figure \ref{fig:consumption_d}
plots the DRSS fit. In this case, the DRSS estimator is able to ``pick
up'' the right model and hews closely to the true structural fit.

In reality, of course, most often all our models are misspecified.
In Figure \ref{fig:consumption_e}, we plot the results of estimating
model \eqref{eq:motivate} but assuming $\rho=-\frac{1}{2}$\footnote{That is, instead of estimating both $\left(\alpha_{i},\rho\right)$
from the data, we estimate $\alpha_{i}$ only while treating $\rho=-0.5$
as an assumption of the model. The assumption, of course, is incorrect
in this case.}. The resulting structural fit now deviates pronouncedly from the
true model, highlighting the fact that the validity of the structural
approach hinges crucially on the model being correct. The DRSS estimator
that combines this misspecified structural model with the (reduced-form)
statistical model \eqref{eq:appetizier02} now puts most of its weight
on the latter and is no longer consistent (Figure \ref{fig:consumption_e}).
Note, however, compared to \eqref{eq:appetizier02}, the misspecified
structural model has worse fit \emph{in-domain}, but still performs
significantly better \emph{out-of-domain}. This provides the motivation
for our ensemble approach. Intuitively, although we misspecify the
utility function, the theory of consumer utility maximization subject
to budget constraints still provides important prior information on
the likely shape of the demand curve -- such as its downward-slopingness
-- that can be used to regulate the behavior of statistical models.
In Figure \ref{fig:consumption_f}, we show the results of our ESS-NP
estimator based on a random forest ensemble of the misspecified structural
model and the (reduced-form) statistical model. The ESS-NP fit is
closer to the true model and performs well both \emph{in-domain} and
\emph{out-of-domain}. Thus in this example, the ensemble approach\footnote{The ESS-LN method produces similar results as the ESS-NP in this example.}
is able to deliver optimal performance when both the structural and
the (reduced-form) statistical models are incorrect.

In section \ref{sec:Applications}, we demonstrate the effectiveness
of our methods using a set of simulation experiments under a variety
of more realistic settings in applied economic analyses, including
first-price auctions and dynamic models of entry and exit. We also
revisit this demand estimation problem and show how to apply our methods
to estimating the demand curve with the help of instrumental variables
when prices are endogenous. For each experiment, we report the performance
of the DRSS and ESS estimators when either or both of a structural
model and a (reduced-form) statistical model is misspecified. 

\paragraph*{Related Literature}

This paper is related to several strands of literature. The doubly
robust estimator was proposed by \citet{robins_estimation_1994,robins_semiparametric_1995,scharfstein_adjusting_1999}
as a means of estimating the average treatment effect by combining
an outcome regression model with a treatment assignment model so that
the estimator remains consistent as long as one of the models is correctly
specified. In general, an estimator is said to have the doubly robustness
property if it is consistent for the target parameter when any one
of two nuisance parameters is consistently estimated \citep{benkeser_doubly_2017}.
Subsequent developments in doubly robust estimation include \citet{bang_doubly_2005,tan_bounded_2010,okui_doubly_2012,farrell_robust_2015,vermeulen_bias-reduced_2015,benkeser_doubly_2017,arkhangelsky_double-robust_2019}.
\citet{chernozhukov_double_2016,chernozhukov_doubledebiasedneyman_2017}
showed that the doubly robust estimator can be viewed as being based
on Neyman-orthogonal moment conditions that are first-order robust
to errors in nuisance parameter estimation. More recently, \citet{lewbel_general_2019}
proposed the \emph{general doubly robust} (\emph{GDR}) method that
provides a general technique for constructing a doubly robust combination
out of any parametric models, which forms the basis of our DRSS estimator. 

Our paper is also related to the literature on model averaging and
ensemble methods. Model averaging provides a natural response to model
uncertainty in the Bayesian framework and has long been considered
an alternative to model selection. See \citet{hoeting_bayesian_1999}
for a comprehensive review of bayesian model averaging methods. In
machine learning, \citet{wolpert_stacked_1992} proposed the method
of stacking, or stacked generalization\footnote{Also see \citet{breiman_stacked_1996}. When weights are restricted
under a simplex constraint, stacking can be considered a frequentist
model averaging technique. \citet{van_der_laan_super_2007} and \citet{hansen_jackknife_2012}
provided theory on its asymptotic optimality. These authors also gave
different names to the method: \emph{super learning} \citep{van_der_laan_super_2007}
and \emph{jackknife model averaging} \citep{hansen_jackknife_2012}.}. \citep{breiman_bagging_1996} proposed bagging, or bootstrap aggregation.
\citet{freund_experiments_1996} introduced boosting. These ensemble
methods are constructed with the explicit goal of maximizing predictive
accuracy and achieve their effectiveness by incorporating model uncertainty,
averaging local optima, and enriching the model space \citep{dietterich_ensemble_2000}.
More recently, there has also been a growing body of research in the
statistics and econometrics literature on asymptotically optimal frequentist
model averaging. See \citet{claeskens_focused_2003,hjort_frequentist_2003,hansen_least_2007,hansen_jackknife_2012,kitagawa_model_2016,zhang_optimal_2016,ando_weight-relaxed_2017}.
\citet{moral-benito_model_2015,steel_model_2019} provided overviews
of the use of model averaging in economics. 

Both the DRSS and the ESS estimators can be used to improve out-of-domain
statistical predictions relative to a pure statistical approach. Our
paper thus makes a contribution to the literature on transfer learning,
which studies the problem of applying a model trained on a source
domain to a target\emph{ }domain where the data-generating distribution
may have changed\footnote{The problem of transfer learning is closely related to the problem
of \emph{sampling bias} or the\emph{ sample selection} problem --
a general problem that arises when we try to make inference, whether
statistical or causal, about a population using data collected from
another population.}. See \citet{pan_survey_2010} for a survey on transfer learning and
\citet{ben-david_theory_2010} for theory on learning from different
domains. A majority of research on transfer learning so far has focused
on \emph{domain adaptation}, where the marginal distributions of the
input variables vary across domains and are observed, but the conditional
outcome distribution is assumed to be the same. Methods that have
been proposed aim to reduce the difference in input distributions
either by sample-reweighting \citep{zadrozny_learning_2004,huang_correcting_2007,jiang_instance_2007,sugiyama_direct_2008}
or by finding a domain-invariant transformation \citep{pan_domain_2010,gopalan_domain_2011}\footnote{This includes the more recent deep domain adaptation literature that
employs deep neural networks for domain adaptation. See \citet{glorot_domain_2011,chopra_dlid_2013,ganin_unsupervised_2014,tzeng_deep_2014,long_learning_2015}.
\citet{wang_deep_2018} provides an overview of this literature in
the context of computer vision.}. Our methods, however, can be viewed as tackling the more difficult
problem of \emph{domain generalization}, where the target domain is
unknown at the time of training and where both the marginal and the
conditional distributions are allowed to vary. Intuitively, we achieve
this by incorporating theory into statistical modeling\footnote{Transfer learning has also been referred to \emph{knowledge transfer}
\citep{pan_survey_2010}. We note, however, that true knowledge transfer
must involve causal knowledge as encapsulated in theory.}. The effectiveness of our approach hinges on the stability of the
underlying causal mechanism and on the availability of a structural
model that is informative, if not correctly specified\footnote{\citet{rojas-carulla_invariant_2018,kuang_stable_2020} also proposed
methods for domain generalization by assuming stability in causal
relationships. Both studies rely on the assumption that a subset of
the input variables $v\subseteq x$ have a causal relation with the
outcome $y$ and the conditional probability $p\left(y|v\right)$
is invariant across domains. However, it is \emph{not} true that having
a causal relationship implies $p\left(y|v\right)$ is domain-invariant.
Let $w=x\backslash v$. The assumption only holds under very limited
and untestable conditions, namely that $y\perp w|v$ and that the
causal effect of $v$ on $y$ is homogeneous.}. 

A main contribution of this paper is to the literature on combining
structural and reduced-form estimation. Many authors in economics
have called for combining these two approaches to harness their respective
strengths\footnote{\citet{chetty_sufficient_2009}: ``The structural and statistic methods
can be combined to address the short-comings of each strategy ...
By combining the two methods in this manner, researchers can pick
a point in the interior of the continuum between reduced-form and
structural estimation, without being pinned to one endpoint or the
other.''}$^{,}$\footnote{Mirroring the debate in economics on structural vs. reduced-form estimation,
there has long been a debate in the machine learning literature on
generative vs. discriminative models as well as efforts to combine
them. See \citet{ng_discriminative_2002,bishop_generative_2007}. }. Early efforts include \citep{chetty_sufficient_2009,heckman_building_2010}.
Their solution is to use structural models to derive sufficient statistics
for the intended analysis and then use reduced-form methods to estimate
them. In comparison, we offer a set of general algorithms rather than
relying on ad hoc derivations\footnote{However, our method cannot be used to conduct welfare analysis, which
is the focus of \citet{chetty_sufficient_2009}.}. More recently, \citet{fessler_how_2019,mao_structural_2020} proposed
shrinkage methods that combine statistical and structural models by
shrinking the former toward the latter. Their methods can be viewed
as complementary to ours. Indeed, there is a connection between shrinkage
and model averaging \citep{hansen_least_2007}. By combining models
of different complexities, a model averaging procedure effectively
shrinks the more complex models toward the less complex ones. 

Compared to \citet{fessler_how_2019,mao_structural_2020}, our approach
arguably also has several advantages. First, their methods are \emph{asymmetric}
with respect to the complexities of statistical and structural models.
Specifically, they require the specification of complex statistical
models to be regularized with structural models. In contrast, our
approach is \emph{symmetric}, allowing researchers to combine structural
models with simple linear reduced-form models frequently used in applied
research. Second, when the structural models are complex and high-dimensional,
our ensemble methods can provide effective regularization. This can
be most easily seen in the case of the stacking estimator ESS-LN.
When the structural model is more complex than the statistical model,
the ESS-LN effectively regularizes the former with the latter by averaging
the two. This is relevant since many structural models used in empirical
applications today are highly complicated and prone to overfitting
as researchers strive for ever more ``realistic'' models\footnote{Importantly, the best model to describe a given data set may not be
the model that truthfully describes the data-generating mechanism.
This is because the true model may well be too complex for the amount
of the data we have, in which case the model will be poorly fit on
the limited sample and generate unreliable predictions. We therefore
echo \citet{hansen_method_2015}: ``it remains an important challenge
for econometricians to devise methods for infusing empirical credibility
into \textquoteleft highly stylized\textquoteright{} models of dynamical
economic systems. Dismissing this problem through advocating only
the analysis of more complicated \textquoteleft empirically realistic\textquoteright{}
models will likely leave econometrics and statistics on the periphery
of important applied research.''}. 

The rest of this paper is organized as follows. Section \ref{sec:Methodology}
lays out the details of our algorithm. In section \ref{sec:Applications}
we apply our method to three sets of simulation experiments in the
settings of first-price auctions, dynamic models of entry and exit,
and demand estimation with instrumental variables and report their
results. Section \ref{sec:Conclusion} concludes. 

\section{Methodology\label{sec:Methodology}}

\subsection{Doubly Robust Statistical-Structural Estimation\label{subsec:DRSS}}

The DRSS builds on the GDR method of \citet{lewbel_general_2019}.
In this section, we discuss the estimator first in the context of
statistical prediction and then in causal inference. In both contexts,
we first assume that we have access to a \emph{representative} data
set, i.e. the target domain on which we wish to make inference is
the same as the source domain from which the data are drawn. We then
consider the case that our data is \emph{non-representative} and discuss
its implications on the \emph{external validity} or \emph{out-of-domain}
performance of our algorithms.

\paragraph*{Statistical Prediction}

Given variables $\left(x,y\right)\in\mathcal{X\times\mathbb{R}}$,
assume first that our goal is to learn the conditional expectation
function $\mu\left(x\right)=\mathbb{E}\left[y|x\right]$. We have
at our disposal two parametric models for $\mu\left(x\right)$: $h\left(x;\theta_{h}\right)$
and $g\left(x;\theta_{g}\right)$, where $\theta_{h}\in\mathbb{R}^{p_{h}},\theta_{g}\in\mathbb{R}^{p_{g}}$.
One of these models is correctly specified, but we do not know which
one. Let $f\in\left\{ h,g\right\} $ index the correct model. Suppose
the true parameter $\theta_{f}^{0}$ is identified by a set of $\ell_{f}\times1,\ \ell_{f}>p_{f}$
moment conditions $\mathbb{E}\left[\psi_{f}\left(x,y;\theta_{f}^{0}\right)\right]=0$.
Given a sample of $n$ \emph{i.i.d.} observations, we can then construct
the following (adjusted) moment distance functions:
\begin{equation}
Q_{m}\left(\theta_{m}\right)=\kappa_{m}^{-1}\overline{\psi}_{m}\left(\theta_{m}\right)'\Omega_{m}\overline{\psi}_{m}\left(\theta_{m}\right),\;m\in\left\{ h,g\right\} \label{eq:DR01}
\end{equation}
, where $\overline{\psi}_{m}\left(\theta_{m}\right)\doteq\frac{1}{n}\sum_{i=1}^{n}\psi_{m}\left(x_{i},y_{i};\theta_{m}\right)$,
$\Omega_{m}$ is a $\ell_{m}\times\ell_{m}$ positive definite weight
matrix\footnote{\citet{lewbel_general_2019} recommend the use of $\Omega=\widehat{\mathbb{E}}\left[\psi\left(\theta_{0}\right)\psi\left(\theta_{0}\right)'\right]^{-1}$,
the (estimated) efficient GMM weight of \citet{hansen_large_1982}.
However, it may not be the optimal weight for the GDR or for our DRSS.
We leave the characterization of the optimal weight matrix to future
work. \label{fn:OptimalW}}, and $\kappa_{m}=\ell_{m}-p_{m}$ is the degrees of freedom of the
$\chi^{2}$ statistic that the unadjusted $\text{\ensuremath{Q_{m}}}$
equals if $m$ is the true model.

Let $\widehat{\theta}_{m}=\underset{\theta_{m}}{\arg\min}\ Q_{m}\left(\theta_{m}\right),\ m\in\left\{ h,g\right\} $.
A doubly robust estimator for $\mu\left(x\right)$ can be constructed
as follows:
\begin{equation}
\widehat{\mu}\left(x\right)=w_{h}h\left(x;\widehat{\theta}_{h}\right)+w_{g}g\left(x;\widehat{\theta}_{g}\right)\label{eq:DR02}
\end{equation}
, where
\begin{equation}
w_{h}=\frac{Q_{g}\left(\widehat{\theta}_{g}\right)}{Q_{h}\left(\widehat{\theta}_{h}\right)+Q_{g}\left(\widehat{\theta}_{g}\right)},\enskip w_{g}=1-w_{h}\label{eq:DR03}
\end{equation}

Under regularity conditions, as long as one of the two models, $h$
or $g$, is correctly specified, it can be shown that $\widehat{\mu}\left(x\right)\rightarrow^{p}\mu\left(x\right)$.
The proof is based on Theorem 1 of \citet{lewbel_general_2019} (see
Appendix \hyperlink{APP}{A.1}). The intuition is simple: if one of
the models, say $h$, is correctly specified but $g$ is not, then
$Q_{h}\left(\widehat{\theta}_{h}\right)\rightarrow^{p}0$ while $Q_{g}\left(\widehat{\theta}_{g}\right)$
will have a nonzero limit. Thus in the limit, $w_{h}$ will be $1$
and $\widehat{\mu}\left(x\right)$ becomes $h\left(x;\widehat{\theta}_{h}\right)$
-- the consistently estimated correct model for $\mu\left(x\right)$. 

Adapting the doubly robust estimator \eqref{eq:DR02} to combining
statistical and structural models is straightforward: let $\mathcal{M}\left(x,y;\theta_{\mathcal{M}}\right)$
be a structural model that specifies the data-generating mechanism
of $\left(x,y\right)$. From this \emph{generative} structural model,
we can derive its prediction of the \emph{discriminative} function
$\mu\left(x\right)$. Let $g\left(x;\theta_{\mathcal{M}}\right)=\mathbb{E}^{\mathcal{M}}\left[y|x\right]$
be the \emph{implied} conditional mean of $y$ according to $\mathcal{M}$.
We can then combine $g\left(x;\theta_{\mathcal{M}}\right)$ with any
statistical model $h\left(x;\theta_{h}\right)$ according to \eqref{eq:DR02}.
The resulting estimator is the DRSS estimator for $\mu\left(x\right)$. 

In practice, there are two ways to construct $\psi_{g}\left(x,y;\theta_{\mathcal{M}}\right)$
for the structurally derived discriminative model $g\left(x;\theta_{\mathcal{M}}\right)$.
If $\mathcal{M}$ is the true model and $\theta_{\mathcal{M}}^{0}$
is the true parameter value, $\psi_{g}$ needs to satisfy $\mathbb{E}\left[\psi_{g}\left(x,y;\theta_{\mathcal{M}}^{0}\right)\right]=0$.
Therefore, we can either directly specify a set of moment conditions
that identify $\mathcal{M}$ or let $\psi_{g}\left(x,y;\theta_{\mathcal{M}}\right)=\phi\left(x\right)\left(y-g\left(x;\theta_{\mathcal{M}}\right)\right)$
for any function $\phi\left(.\right)$. We can then construct $Q_{g}\left(\theta_{\mathcal{M}}\right)$
based on $\text{\ensuremath{\psi_{g}\left(x,y;\theta_{\mathcal{M}}\right)}}$
and compute $\left(w_{h},w_{g}\right)$ based on $\text{\ensuremath{\left(Q_{h}\left(\widehat{\theta}_{h}\right),Q_{g}\left(\widehat{\theta}_{\mathcal{M}}\right)\right)}}$,
where $\left(\widehat{\theta}_{h},\widehat{\theta}_{\mathcal{M}}\right)$
are obtained from separate first stage estimation of the statistical
model $h$ and the structural model $\mathcal{M}$. 

\paragraph{Sample Splitting}

The DRSS method as outlined above is a two-stage procedure, where
$\left(\widehat{\theta}_{h},\widehat{\theta}_{\mathcal{M}}\right)$
are obtained in a first stage and the estimator is constructed according
to \eqref{eq:DR02} in a second stage. If both stages are conducted
on the same sample of data, however, finite sample bias from the first
stage will be carried over to the second stage, especially when complex
statistical or structural models, prone to overfitting, are estimated
in the first stage. To avoid bias from overfitting and ensure good
statistical behavior, we can use separate data sets for the two stages
of the procedure. This can be accomplished by, for example, splitting
the observed data randomly into two parts. This is known as \emph{sample-splitting}
\citep{angrist_split-sample_1995}\footnote{The idea of sample-splitting is of course closely related to the idea
of using separate training and validation data sets for fitting model-
and hyper-parameters in machine learning. Indeed, the weights $\left(w_{h},w_{g}\right)$
can be viewed as the \emph{hyperparameters} of the DRSS model.}. This way, from the perspective of the second stage, $\left(\widehat{\theta}_{h},\widehat{\theta}_{\mathcal{M}}\right)$
are exogenously given, so that when we evaluate the moment distance
functions $Q_{h}$ and $Q_{g}$ -- critical for computing the DRSS
weights -- we do not suffer an optimistic bias due to $\left(\widehat{\theta}_{h},\widehat{\theta}_{\mathcal{M}}\right)$
being obtained from the same data. 

There is an efficiency cost involved in sample-splitting, as half
of the data are wasted in each stage. The results can also be highly
variable due to the whims of a single random split. To improve efficiency,
we can perform sample-splitting multiple times and average their results.
This is the idea behind \emph{cross-validation} and \emph{cross-fitting}
\citet{chernozhukov_double_2016,chernozhukov_doubledebiasedneyman_2017}
and can be described as follows for our DRSS estimator: randomly partition
the data into $K$ equal-sized parts. For $k=1,\cdots,K$, let $\mathcal{D}_{k}$
denote the data of the $k$th partition and let $\mathcal{D}_{-k}$
denote the data not in $\mathcal{D}_{k}$. We use $\mathcal{D}_{-k}$
for the first stage estimation of $\theta_{h}$ and $\theta_{\mathcal{M}}$.
This gives us $\left(\widehat{\theta}_{h}^{\left(-k\right)},\widehat{\theta}_{\mathcal{M}}^{\left(-k\right)}\right)$.
We then use $\mathcal{D}_{k}$ to evaluate $Q_{h}$ and $Q_{g}$ at
$\left(\widehat{\theta}_{h}^{\left(-k\right)},\widehat{\theta}_{\mathcal{M}}^{\left(-k\right)}\right)$.
This gives us $\text{\ensuremath{\left(Q_{h}^{\left(k\right)}\left(\widehat{\theta}_{h}^{\left(-k\right)}\right),Q_{g}^{\left(k\right)}\left(\widehat{\theta}_{\mathcal{M}}^{\left(-k\right)}\right)\right)}}$.
Finally, for cross-validation, $w$ is determined as 
\begin{equation}
w_{h}=\frac{\overline{Q}_{g}}{\overline{Q}_{h}+\overline{Q}_{g}},\enskip w_{g}=1-w_{h}\label{eq:DRSSCV}
\end{equation}
, where $\overline{Q}_{m}\doteq\frac{1}{K}\sum_{k=1}^{K}Q_{m}^{\left(k\right)}\left(\widehat{\theta}_{m}^{\left(-k\right)}\right),\ m\in\left\{ h,g/\mathcal{M}\right\} $
are cross-validated moment distances. For cross-fitting, let $w_{h}^{\left(k\right)}$
be constructed from $\text{\ensuremath{\left(Q_{h}^{\left(k\right)}\left(\widehat{\theta}_{h}^{\left(-k\right)}\right),Q_{g}^{\left(k\right)}\left(\widehat{\theta}_{\mathcal{M}}^{\left(-k\right)}\right)\right)}}$
according to \eqref{eq:DR03}. Then the cross-fitted weight is\footnote{Both methods are consistent. See \citet{li_asymptotic_1987,chernozhukov_double_2016}.
Although to our knowledge, their asymptotic efficiency and finite
sample performance have not been compared in existing studies.} 
\begin{equation}
w_{h}=\frac{1}{K}\sum_{k=1}^{K}w_{h}^{\left(k\right)},\enskip w_{g}=1-w_{h}\label{eq:DRSSCF}
\end{equation}

\paragraph*{Causal Inference}

We now discuss the problem of causal effect estimation under unconfoundedness.
Let the observed variables be $\left(y,d,v\right)\in\mathbb{R}\times\mathbb{R}\times\mathcal{V}$,
where $y$ is the outcome variable, $d$ is the treatment variable,
and $v$ is a set of control variables. We are interested in the causal
effect of $d$ on $y$. Specifically, let our target be the average
treatment effect (ATE) denoted by $\tau$. We allow $\tau$ to be
fully nonlinear and heterogeneous, i.e. $\tau=\tau\left(d,v\right)$.
Then
\begin{equation}
\tau\left(d,v\right)=\frac{\partial}{\partial d}\mathbb{E}\left[\left.y^{d}\right|v\right]\label{eq:ATE}
\end{equation}
, where $y^{d}$ is the potential outcome of $y$ under treatment
$d$. 

Under the unconfoundedness assumption of \citet{rosenbaum_central_1983}\footnote{Suppose the treatment variable $d$ takes on a discrete set of values,
$d\in\left\{ 1,\ldots,D\right\} $, then the \emph{unconfoundedness}
-- or \emph{conditional exchangeability} -- assumption can be stated
as \[\left.d\indep\left(y^{d=1},\ldots,y^{d=D}\right)\right|v\]This
assumption is satisfied if $d$ is not associated with any other causes
of $y$ conditional on $v$, in which case we say $d$ is \emph{exogenous}
to $y$ conditional on $v$. A more precise statement on the sufficient
conditions for satisfying this assumption, made in the language of
causal graphical models based on directed acyclic graphs (DAGs), is
that $v$ satisfies the \emph{back-door criterion} \citep{pearl_causality_2009}.}, $\mathbb{E}\left[\left.y^{d}\right|v\right]=\mathbb{E}\left[\left.y\right|d,v\right]$.
Let $x=\left(d,v\right)$. The task of estimating $\tau\left(d,v\right)$
is thus equivalent to the task of estimating $\mathbb{E}\left[y|x\right]$.
Suppose now that we have a reduced-form model $h\left(x;\theta_{h}\right)$
for $\mathbb{E}\left[y|x\right]$ and a structural model $\mathcal{M}\left(x,y;\theta_{\mathcal{M}}\right)$,
both supporting the unconfoundedness condition\footnote{i.e. (1) the design of $h$ is based on the unconfoundedness condition;
(2) in the causal structure assumed by $\mathcal{M}$, $v$ satisfies
the back-door criterion.}, then we can use the DRSS to produce an estimate of $\mathbb{E}\left[y|x\right]$
by combining these two models, from which we can derive $\widehat{\tau}\left(d,v\right)$\footnote{Technically, $\tau\left(d,v\right)$ is the conditional ATE. With
a slight abuse of notation, the population ATE $\tau\left(d\right)=\mathbb{E}_{v}\left[\tau\left(d,v\right)\right]$.}.

When the unconfoundedness condition does not hold so that $d$ is
\emph{endogenous} conditional on $v$, one of the most widely used
strategies in reduced-form inference is to rely on the use of instrumental
variables, which are auxiliary sources of randomness that can be used
to identify causal effects. Let $h\left(x;\theta_{h}\right),\ x=\left(d,v\right)$
be a reduced-form model for $\mathbb{E}\left[\left.y^{d}\right|v\right]$.
We can write $y=h\left(x;\theta_{h}\right)+\epsilon$, where $\epsilon$
is \emph{defined} as $y-h\left(x;\theta_{h}\right)$ and may be correlated
with $d$\footnote{\emph{By definition}, when $\mathbb{E}\left[d\epsilon\right]\ne0$,
the received treatment $d$ is related to unobserved factors that
affect potential outcomes $y^{d}$, thus violating the unconfoundedness
condition.}. If we have access to a variable $z$ that is correlated with $d$
(conditional on $v$) and satisfies $\mathbb{E}\left[z\epsilon\right]=0$,
then $z$ can serve as an instrument for $d$\footnote{On a causal graph, this translates into the requirement that $z$
is correlated with $d$ and that every open path connecting $z$ and
$y$ has an arrow pointing into $d$.}. In general, given $\theta_{h}\in\mathbb{R}^{p_{h}}$, let $\psi_{h}\left(x,y,z;\theta_{h}\right)=\phi\left(z\right)\left(y-h\left(x;\theta_{h}\right)\right)$
be a set of $\ell_{h}>p_{h}$ functions, where $\phi\left(z\right)$
is any function of $z$. If $h$ is the true model and $\theta_{h}^{0}$
is the true parameter, then $\theta_{h}^{0}$ can be identified via
the following moment conditions:
\begin{equation}
\mathbb{E}\left[\psi_{h}\left(x,y,z;\theta_{h}\right)\right]=0\label{eq:ivmoment}
\end{equation}

Now let $\mathcal{M}\left(x,y,z;\theta_{\mathcal{M}}\right)$ be a
structural model for the data-generating mechanism of the observed
variables\footnote{$\mathcal{M}$ does not have to contain $z$. See e.g. section \eqref{subsec:Demand-Estimation}
for an example. If $\mathcal{M}$ \emph{does }contain $z$, $z$ needs
to satisfy the IV requirement in the causal structure of $\mathcal{M}$,
i.e. $z$ is correlated with $d$ and that every open path connecting
$z$ and $y$ has an arrow pointing into $d$. If $\mathcal{M}$ is
a model for $\left(x,y\right)$ only, in the case that it is the true
model, the DRSS estimator for $\mathbb{E}\left[\left.y^{d}\right|v\right]$
will be based both on the causal assumptions in $\mathcal{M}$ and
on the additional assumption that $z$ is a variable satisfying the
IV requirement.}. Let $g\left(x;\theta_{\mathcal{M}}\right)=\mathbb{E}^{\mathcal{M}}\left[\left.y^{d}\right|v\right]$
be the model derived conditional expectation of the potential outcome
under treatment $d$. Let $\psi_{g}\left(x,y,z;\theta_{\mathcal{M}}\right)$
be either a set of moment functions for $\mathcal{M}$ or let $\psi_{g}\left(x,y,z;\theta_{\mathcal{M}}\right)=\phi\left(z\right)\left(y-g\left(x;\theta_{\mathcal{M}}\right)\right)$.
We can then construct $Q_{h}\left(\theta_{h}\right)$ and $Q_{g}\left(\theta_{\mathcal{M}}\right)$
based on $\psi_{h}\left(x,y,z;\theta_{h}\right)$ and $\text{\ensuremath{\psi_{g}\left(x,y,z;\theta_{\mathcal{M}}\right)}}$,
and combine $h\left(x;\theta_{h}\right)$ and $g\left(x;\theta_{\mathcal{M}}\right)$
according to \eqref{eq:DR02} to produce a DRSS estimate of $\mathbb{E}\left[\left.y^{d}\right|v\right]$\footnote{The difference is that in \eqref{eq:DR02}, by combining $h$ and
$g$, we get $\widehat{\mathbb{E}}\left[\left.y\right|d,v\right]$.
Here we get $\widehat{\mathbb{E}}\left[\left.y^{d}\right|v\right]$.}, from which we can obtain $\widehat{\tau}\left(d,v\right)$.

\paragraph*{Discussion}

The goal of doubly robust estimation is to ensure consistency when
one of two candidate models is correctly specified but we do not know
which one. When both models are misspecified, however, doubly robust
estimators can perform poorly \citep{kang_demystifying_2007}. This
is not surprising as these estimators are not constructed to optimize
performance based on a loss criterion such as expected mean squared
error. In fact, the DRSS estimator can be viewed as a weighted average
of its candidate models (see \eqref{eq:DR02}) and bears a close resemblance
to bayesian model averaging, which is known to be flawed in $\mathcal{M}$-open
settings in which none of the candidate models is true \citep{clyde_bayesian_2013,yao_using_2018}\footnote{More precisely, bayesian model averaging is appropriate for $\mathcal{M}$-closed
settings rather than $\mathcal{M}$-complete or $\mathcal{M}$-open
settings. Following the definitions of \citet{bernardo_bayesian_2009},
given a list of candidate models, the $\mathcal{M}$-closed setting
is the one in which the true model is in the list. In the $\mathcal{M}$-complete
setting, the true model can be specified but for tractability of computations
or other reasons is not included in the model list. The $\mathcal{M}$-open
setting refers to the situation in which we know the true model is
not in the list and have no idea what it looks like.}.

In our presentation so far, we have also assumed that we have access
to a representative sample drawn from the population of interest,
i.e. the source domain is the same as the target domain. In practice,
however, this is often not the case. In particular, we are often interested
in making inference on populations that are much larger than the population
from which we draw our sample, i.e. we care about the \emph{external
validity} or \emph{out-of-domain} performance of our estimators. The
DRSS however assures only \emph{in-domain} consistency if one of its
candidate models is correctly specified. In general, no similar guarantees
on out-of-domain consistency can be obtained without further assumptions\footnote{This can be readily seen by considering two models that produce the
same fit in-domain but behave completely differently out-of-domain.
Without further assumptions, there is no way to tell them apart using
observed data.}. 

If our goal is not to achieve consistency on a target population,
but rather to improve predictive accuracy as much as possible, then
note that simply averaging a statistical model that fits well in-domain
with an approximately correct structural model could improve the in-domain
fit of the latter and the out-of-domain fit of the former. This observation
applies to the DRSS as well, as it is also a weighted average method.
The weights of the DRSS, however, are not constructed to optimize
a performance criterion. This brings us to the ensemble estimators
that we introduce in the next section, which are explicitly constructed
to do so. As we will see, even though the criteria are evaluated on
observed data, the ensemble estimators often produce superior in-domain
\emph{and} out-of-domain results relative to both of its candidate
models and the DRSS approach, especially when both individual models
are misspecified.

\subsection{Ensemble Statistical-Structural Estimation\label{subsec:ESS}}

\subsubsection{ESS-LN}

Given variables $\left(x,y\right)\in\mathcal{X\times\mathbb{R}}$,
again assume that our goal is to learn the conditional expectation
function $\mu\left(x\right)=\mathbb{E}\left[y|x\right]$ and we have
at our disposal two parametric models $h\left(x;\theta_{h}\right)$
and $g\left(x;\theta_{g}\right)$. Let $\widehat{h}\left(x\right)\doteq h\left(x;\widehat{\theta}_{h}\right)$
and $\widehat{g}\left(x\right)\doteq g\left(x;\widehat{\theta}_{g}\right)$
be their fitted values on the observed sample. The linear ensemble,
\emph{ESS-LN}, combines the two linearly to form an estimate of $\mu\left(x\right)$:
\begin{equation}
\mu\left(x\right)=w_{0}+w_{1}\widehat{h}\left(x\right)+w_{2}\widehat{g}\left(x\right)\label{eq:ESSLN01}
\end{equation}

To choose the optimal weights $w=\left(w_{0},w_{1},w_{2}\right)$,
we can simply run a least squares regression of $y$ on $\widehat{h}\left(x\right)$
and $\widehat{g}\left(x\right)$. At the population level, combining
models this way never make things worse \citep{hastie_elements_2009}.
On finite sample, however, we need to take into consideration differences
in model complexity and avoid carrying over any biases in the first
stage estimation of $\left(\widehat{\theta}_{h},\widehat{\theta}_{g}\right)$
into the choice of $w$. To this end, one can use the method of \emph{stacking}
\citep{wolpert_stacked_1992} and obtain $w$ via leave-one-out cross
validation: 
\begin{equation}
\widehat{w}=\underset{w}{\arg\min}\left\{ \sum_{i=1}^{n}\left(y_{i}-w_{0}-w_{1}\widehat{h}^{-i}\left(x_{i}\right)-w_{2}\widehat{g}^{-i}\left(x_{i}\right)\right)^{2}\right\} \label{eq:ESSLN02}
\end{equation}
, where $\widehat{h}^{-i}\left(x_{i}\right)$ and $\widehat{g}^{-i}\left(x_{i}\right)$
are respectively the predictions at $x_{i}$ using $h$ and $g$ that
are estimated on the training data with the $i$th observation removed.
The cross-validated error gives a better approximation of the expected
error, allowing an optimal combination. In practice, one can also
account for model complexity via the use of sample-splitting or cross-fitting,
or use $K-$fold instead of leave-one-out cross validation.

To adapt the stacking method to combining statistical and structural
models, as in the construction of the DRSS estimator, we let $g\left(x;\theta_{\mathcal{M}}\right)=\mathbb{E}^{\mathcal{M}}\left[y|x\right]$
be the implied conditional mean of $y$ according to the structural
model $\mathcal{M}\left(x,y;\theta_{\mathcal{M}}\right)$. We then
combine $g\left(x;\theta_{\mathcal{M}}\right)$ with statistical model
$h\left(x;\theta_{h}\right)$ according to \eqref{eq:ESSLN01}. With
regard to the choice of $w$, in \citep{wolpert_stacked_1992}, no
restrictions are placed and $\widehat{w}$ is given by least squares
regression of $y_{i}$ on $\widehat{h}^{-i}\left(x_{i}\right)$ and
$\widehat{g}^{-i}\left(x_{i}\right)$\footnote{The stacking method as proposed by \citep{wolpert_stacked_1992} is
therefore a general model combination or ensemble method rather than
a model averaging method.}. \citet{hansen_jackknife_2012} proved the asymptotic optimality
of stacking for linear models under a model averaging constraint that
$w_{0}=0,w_{1},w_{2}\ge0,w_{1}+w_{2}=1$. \citet{ando_weight-relaxed_2017}
proved asymptotic optimality for generalized linear models with weight
restrictions relaxed to $w_{0}=0,w_{1},w_{2}\in\left[0,1\right]$.
In this paper, we follow the original stacking method and do not place
restrictions on $w$\footnote{In particular, both \citet{hansen_jackknife_2012} and \citet{ando_weight-relaxed_2017}
assumed individual (generalized) linear models with intercept terms,
so that their prediction errors have mean $0$. In our case, we do
not require misspecified structural models to generate predictions
of $y$ that have mean $0$ error. We thus need an additional intercept
term $w_{0}$.}. 

We now discuss the use of ESS-LN for causal effect estimation. As
discussed in section \ref{subsec:DRSS}, given treatment variable
$d$, outcome variable $y$, and control variables $v$, the task
of estimating the conditional ATE under unconfoundedness is equivalent
to the task of estimating the conditional expectation $\mathbb{E}\left[y|d,v\right]$\footnote{Technically, the conditional ATE $\tau\left(d,v\right)=\left.\partial\mathbb{E}\left[y|d,v\right]\right/\partial d$
under unconfoundedness.}$^{,}$\footnote{When the unconfoundedness condition does not hold, a number of reduced-form
strategies are often employed to identify causal effects. In addition
to the use of instrumental variables, which we detail below, these
methods include difference-in-differences (DID) and regression discontinuity
(RD). Statistically, both DID and RD can be cast as a conditional
mean estimation problem given specific designs and thus can be combined
with their structurally-derived counterpart using the ensemble method
we have described.}. Procedurally, the causal inference problem is thus the same as the
statistical prediction problem in this case\footnote{We note that in current practice, the goal of causal inference is
typically to produce an unbiased estimate of the treatment effect,
while in predictive modeling, the goal is to often to minimize an
expected $\mathcal{L}_{2}$ loss. However, whether causal effect estimation
should aim for unbiasedness or precision remains an unsettled question.}$^{,}$\footnote{Importantly, in the case of ensemble estimators, even if the ensemble
model estimates causal effects based on the unconfoundedness assumption,
the structural model in the ensemble does not have to support the
assumption. Whatever the causal assumptions are made by the structural
model, we use its derived functional form for $\mathbb{E}\left[y|d,v\right]$
as an input into the ensemble. Thus, the final ensemble estimate is
still based on the unconfoundedness assumption. If this assumption
holds true but is unsupported by a member model in the ensemble, then
that model is simply misspecified. \label{fn:ESSMImportant}}. 

In general, however, without assuming unconfoundedness, our goal is
to produce an estimate of $\mathbb{E}\left[\left.y^{d}\right|v\right]$
based on a reduced-form model $\widehat{h}\left(x\right)=h\left(x;\widehat{\theta}_{h}\right)$
and a structurally-derived model $\widehat{g}\left(x\right)=g\left(x;\widehat{\theta}_{\mathcal{M}}\right)$:
\begin{equation}
\mathbb{E}\left[\left.y^{d}\right|v\right]=w_{0}+w_{1}\widehat{h}\left(x\right)+w_{2}\widehat{g}\left(x\right),\ x=\left(d,v\right)\label{eq:ESSLN03}
\end{equation}
, from which we can obtain $\widehat{\tau}\left(d,v\right)=\left.\partial\widehat{\mathbb{E}}\left[\left.y^{d}\right|v\right]\right/\partial d$.

When $d$ is endogenous -- when there is \emph{unmeasured} confounding,
if we observe a variable $z$ that can serve as a valid instrument
for $d$, then we can specify the following $\ell\times1,\ \ell\ge3$
moment conditions:
\begin{equation}
\mathbb{E}\left[\phi\left(z\right)\left(y-w_{0}^{0}+w_{1}^{0}\widehat{h}\left(x\right)+w_{2}^{0}\widehat{g}\left(x\right)\right)\right]=0\label{eq:ivESS}
\end{equation}
, where $\phi\left(z\right)$ is any function of $z$ and $w^{0}=\left(w_{0}^{0},w_{0}^{1},w_{0}^{2}\right)$
are the true values of $w$\footnote{Assuming that \eqref{eq:ESSLN03} is the true model.}$^{,}$\footnote{The structural model $\mathcal{M}$ from which $\widehat{g}\left(x\right)$
is derived does not have to contain $z$, and if it does, $z$ does
not need to satisfy the IV requirement in the causal structure assumed
by $\mathcal{M}$. See footnote \ref{fn:ESSMImportant}.}. 

Let $\psi\left(x,y,z;w\right)\doteq\phi\left(z\right)\left(y-w_{0}+w_{1}\widehat{h}\left(x\right)+w_{2}\widehat{g}\left(x\right)\right)$.
Let $\overline{\psi}\left(w\right)\doteq\frac{1}{n}\sum_{i=1}^{n}\psi\left(x_{i},y_{i},z_{i};w\right)$.
Let $Q\left(w\right)\doteq\overline{\psi}\left(w\right)'\Omega\overline{\psi}\left(w\right)$,
where $\Omega$ is a $\ell\times\ell$ positive definite weight matrix\footnote{e.g. the efficient GMM weight of \citet{hansen_large_1982}.}.
The optimal $w$ can then be obtained by minimizing the GMM objective
function:
\begin{equation}
\widehat{w}=\underset{w}{\arg\min\ }Q\left(w\right)\label{eq:GMMESS}
\end{equation}

In practice, as in the case of conditional mean modeling, given finite
sample, we want to account for model complexity and avoid carrying
any bias in the first stage estimation of $\widehat{h}$ and $\widehat{g}$
into the determination of $w$. This can be accomplished by using
the strategies of either sample-splitting, cross-validation, or cross-fitting. 

\subsubsection{ESS-NP}

The ESS-LN is a linear ensemble. Our ESS-NP estimator goes one step
further and allows any nonlinear combinations of individual models.
In conditional mean estimation, let 
\begin{equation}
\mu\left(x\right)=f\left(\widehat{h}\left(x\right),\widehat{g}\left(x\right);w\right)\label{eq:ESSRF01}
\end{equation}
, where $f\left(.,.\right)$ is any function. Statistically, this
amounts to regressing the outcome $y$ nonparametrically on the predictions
obtained from individual models $h$ and $g$. 

While a large class of nonparametric models can be used for $f$,
in this paper we adopt the random forest model of \citet{breiman_random_2001}.
The random forest is based on decision tree models. A decision tree
is constructed by repeatedly splitting or partitioning the predictor
space into different regions in order to maximize fit. In each region,
a constant model is fit so that the predicted value is simply the
mean of the observed outcomes in that region. Thus, in its simplest
form, with a predetermined number of splits (such as in the case of
a \emph{stump}), a decision tree is a piecewise-constant model. When
splits are adaptively chosen to minimize prediction error, the decision
tree becomes a nonparametric model whose complexity grows with data
and is related to kernels and nearest-neighbor methods in that its
predictions are based on the values of neighborhood observations,
except that it chooses the neighborhoods (regions) in a data-driven
way \citep{athey_generalized_2019}. 

In contrast to conventional trees, in the ESS-NP, the predictor space
is formed by $\widehat{h}\left(x\right)$ and $\widehat{g}\left(x\right)$
-- the predictions obtained from statistical model $h$ and structurally-derived
model $g$. A tree constructed out of $\widehat{h}\left(x\right)$
and $\widehat{g}\left(x\right)$ carves up the space formed by $\widehat{h}\left(x\right)$
and $\widehat{g}\left(x\right)$, which in turn, implies a partition
of the underlying input space $x$. The ESS-NP can therefore be viewed
as allowing us to adaptively assign different weights to different
regions of the input space depending on which model -- the statistical
or the structural -- performs better.

While decision trees are powerful tools for capturing nonlinear relations
and complex interactions, they tend to suffer from high variance and
instability. Random forests improve upon decision trees by building
and combining a large number of trees through bootstrap aggregation,
thereby reducing variance and increasing predictive accuracy\footnote{The random forest is an ensemble of individual trees. In our ESS-NP
estimator, each tree is in turn an ensemble of $h$ and $g$. The
ESS-NP is therefore an ``ensemble of ensembles''.}. Additional randomness can be introduced to further de-correlate
individuals trees via random split selection that restricts the variables
available for consideration in each split\footnote{See \citet{loh_fifty_2014,biau_random_2016} for overviews of decision
trees and forest-based methods. Consistency results on random forests
are obtained in \citet{biau_analysis_2012,scornet_consistency_2015,scornet_asymptotics_2016}.}. In the ESS-NP estimator \eqref{eq:ESSRF01}, $f$ is therefore based
on the random forest model.

The conditional mean ESS-NP estimator can be used for prediction and
causal effect estimation under unconfoundedness\footnote{The estimator can also be used to combine structural models with reduced-form
models based on statistical designs such as DID and RD when there
is unmeasured confounding.}. When there is unmeasured confounding, as in the case of ESS-LN,
it is conceptually possible to adapt the ESS-NP to perform instrumental
variables estimation based on the following conditional moment restrictions:
\begin{equation}
\mathbb{E}\left[\left.\left(y-f\left(\widehat{h}\left(x\right),\widehat{g}\left(x\right);w\right)\right)\right|z\right]=0\label{eq:ivESSNP}
\end{equation}
, where $f\left(.,.\right)$ is again any function. The type of nonparametric
IV regression defined by \eqref{eq:ivESSNP}, however, is known to
suffer from poor statistical performance due to the ill-posed inverse
problem \citep{newey_nonparametric_2013}. Applying the random forest
method to this task is also not straight-forward\footnote{Methods for estimating heterogeneous causal effects with semiparametric
IV regression based on random forests have recently been proposed
in \citet{athey_generalized_2019}.}. Therefore, in this paper, we do not propose an ESS-NP method for
IV estimation.

\section{Experiments\label{sec:Applications}}

\renewcommand\thesubsection{\Alph{subsection}}

In this section, we demonstrate the effectiveness of our methods and
compare their finite-sample performances using three sets of simulated
experiments. Taken together, these exercises cover prediction and
causal inference problems, static and dynamic settings, and individual
behavior that deviates in various ways from perfect rationality. 

\subsection{First-Price Auction}

In our first experiment, we consider first-price sealed-bid auctions.
Auctions are one of the most important market allocation mechanisms.
Empirical analysis of auction data has been transformed in recent
years by structural estimation of auction models based on games of
incomplete information\footnote{See \citet{paarsch_introduction_2006,athey_nonparametric_2007,hickman_structural_2012,perrigne_econometrics_2019}
for surveys on econometric analysis of auction data}. Structural analysis of auction data views the observed bids as equilibrium
outcomes and attempts to recover the distribution of bidders' private
values by estimating relationships derived directly from equilibrium
bid functions. This approach, while offering a tight integration of
theory and observations, relies on a set of strong assumptions on
the information structure and rationality of bidders \citep{bajari_are_2005}. 

In this exercise, we conduct three experiments by simulating auction
data with varying number of participants under three scenarios. The
first scenario features rational bidders with independent private
values drawn from a uniform distribution. The second scenario features
rational bidders whose values are drawn from a beta distribution.
The third scenario features boundedly-rational bidders whose bids
deviate from optimal bidding strategies. In each experiment, we're
interested in the effect of the number of bidders $n$ on the winning
bid $b^{*}$, $\mathbb{E}\left[\left.b^{*}\right|n\right]$. We estimate
this target function using (a) a statistical model, (b) a structural
model, (c) the DRSS estimator, (d) the ESS estimators (ESS-LN, ESS-NP),
and compare their performances. For all experiments, we use a structural
model that assumes rational bidders with uniformly distributed values.
The model is thus correctly specified for experiment 1, but is misspecified
in experiment 2 and 3. Table \ref{tab:AuctionSetup} summarizes this
setup. Below we detail the data-generating models of the three experiments.

\begin{table}
\centering

\begin{threeparttable}

\caption{First-price Auction - Setup\tnote{a}\label{tab:AuctionSetup}}

\medskip{}

\begin{tabular}{clcc}
\toprule 
\toprule Experiment & \multicolumn{1}{c}{True Mechanism} & Structural Model & Statistical Model\tabularnewline
\midrule 
1 & $v_{i}\overset{\text{i.i.d.}}{\sim}U(0,1)$, $b_{i}=b\left(v_{i}\right)$ & \multirow{3}{*}{$v_{i}\overset{\text{i.i.d.}}{\sim}U(0,1)$, $b_{i}=b\left(v_{i}\right)$} & \multirow{3}{*}{see \eqref{eq:AuctionStatModel}}\tabularnewline
2 & $v_{i}\overset{\text{i.i.d.}}{\sim}\text{Beta}(2,5)$, $b_{i}=b\left(v_{i}\right)$ &  & \tabularnewline
3 & $v_{i}\overset{\text{i.i.d.}}{\sim}U(0,1)$, $b_{i}=\eta_{i}\cdot b\left(v_{i}\right)$ &  & \tabularnewline
\bottomrule
\end{tabular}

\medskip{}

\begin{tablenotes} 
\small
\item [a] $b\left(v_{i}\right)$ is the equilibrium bid function \eqref{eq:auctionBidding}. $\eta_{i}\overset{\text{i.i.d.}}{\sim}\text{TN}\left(0,0.25,0,\infty\right)$.
\end{tablenotes}

\end{threeparttable}
\end{table}

\paragraph{Setup}

Consider a first-price sealed-bid auction with $n$ risk-neutral bidders
with independent private value $v_{i}\sim^{i.i.d.}F(v)$. Each bidder
submits a bid $b_{i}$ to maximize her expected return 
\begin{equation}
\pi_{i}=\left(v_{i}-b_{i}\right)\times\Pr\left(b_{i}>\max\left\{ b_{-i}\right\} \right)\label{eq:auctionProfit}
\end{equation}
, where $b_{-i}$ denotes the other submitted bids. In Bayesian-Nash
equilibrium, each bidder's bidding strategy is given by 
\begin{equation}
b\left(v\right)=v-\frac{1}{F\left(v\right)^{n-1}}\int_{0}^{v_{i}}F(x)^{n-1}dx\label{eq:auctionBidding}
\end{equation}

For experiment 1 and 3, we let $F$ be $U\left(0,1\right)$. In this
case the equilibrium bid function simplifies to:
\begin{equation}
b\left(v\right)=\frac{n-1}{n}v\label{eq:simplebid}
\end{equation}

For experiment 2, we let $F$ be $\text{Beta}\left(2,5\right)$. In
each experiment, we simulate repeated auctions with varying number
of bidders\footnote{Assuming the same object is being repeatedly auctioned.}.
For experiment 1 and 2, the observed bids $b_{i}$ are the equilibrium
outcomes, i.e. $b_{i}=b\left(v_{i}\right)$. For experiment 3, we
let $b_{i}=\eta_{i}\cdot b\left(v_{i}\right)$, where $\eta_{i}$
follows a normal distribution left-truncated at $0$, $\eta_{i}\overset{\text{i.i.d.}}{\sim}\text{TN}\left(0,0.25,0,\infty\right)$.
Bidders in experiment 3 thus ``overbid'' relative to the Bayesian-Nash
equilibrium. 

\paragraph*{Simulation}

For each experiment, we simulate $M=500$ auctions with number of
bidders $n_{m}$ varying between $5$ and $25$. The observed data
thus consist of $\text{\ensuremath{\mathcal{D}}}=\left\{ \left\{ b_{i}^{m}\right\} _{i=1}^{n_{m}}\right\} _{m=1}^{M}$.
In this exercise, our goal is to learn $\mathbb{E}\left[\left.b^{*}\right|n\right]$,
the relationship between the number of bidders and the winning bid.
To assess the performance of various estimators, we use the true data-generating
models to compute $\mathbb{E}\left[\left.b^{*}\right|n\right]$ for
$n\in\left[5,50\right]$, so that we can compare the predictions of
each method with the true values both in-domain and out-of-domain. 

\paragraph*{Statistical Model}

To estimate $\mathbb{E}\left[\left.b^{*}\right|n\right]$ using a
statistical model\footnote{Since $n$ is exogenous, $\mathbb{E}\left[\left.b^{*}\right|n\right]$
is also a causal relationship and \eqref{eq:AuctionStatModel} can
also be thought of a reduced-form model of the effect of the number
of bidders on the winning bid.}, the data we need are $\left\{ \left(n_{m},b_{m}^{*}\right)\right\} _{m=1}^{M}$,
where $b_{m}^{*}$ is the winning bid of auction $m$. We adopt the
following second degree polynomial as the model for $\mathbb{E}\left[\left.b^{*}\right|n\right]$:
\begin{equation}
b_{m}^{*}=\beta_{0}+\beta_{1}n_{m}+\beta_{1}n_{m}^{2}+e_{m}\label{eq:AuctionStatModel}
\end{equation}

\paragraph*{Structural Model}

Our structural model assumes that bidders are rational, risk-neutral,
and have independent private values drawn from a $U\left(0,1\right)$
distribution. Under these assumptions, the bidders' private values
can be easily identified from the observed bids in each auction by
$v_{i}=\frac{n}{n-1}b_{i}$\footnote{In general, if we do not impose the assumption that $v_{i}\overset{\text{i.i.d.}}{\sim}U(0,1)$
and assume instead that $v_{i}\overset{\text{i.i.d.}}{\sim}F\left(v\right)$,
with $F$ unknown, then we can identify and estimate $v_{i}$ using
the following strategy based on \citet{guerre_optimal_2000}: let
$G\left(b\right)$ and $g\left(b\right)$ be the distribution and
density of the bids. \eqref{eq:auctionBidding} implies 
\[
v_{i}=b_{i}+\frac{1}{n-1}\frac{G\left(b_{i}\right)}{g\left(b_{i}\right)}
\]
Thus, by nonparametrically estimating $G\left(b\right)$ and $g\left(b\right)$
from the observed bids, we can obtain an estimate of $v_{i}$.}. The structural model makes it even easier to make predictions on
the winning bid. The model implies that:
\begin{equation}
\mathbb{E}\left[\left.b^{*}\right|n\right]=\frac{n}{n+1}\label{eq:auctionSTPred}
\end{equation}
No estimation is necessary. 

\paragraph*{Results}

\begin{figure}
\subfloat[{\small{}Experiment 1: Training Sample}\label{fig:auction1_a}]{\includegraphics[width=0.5\columnwidth]{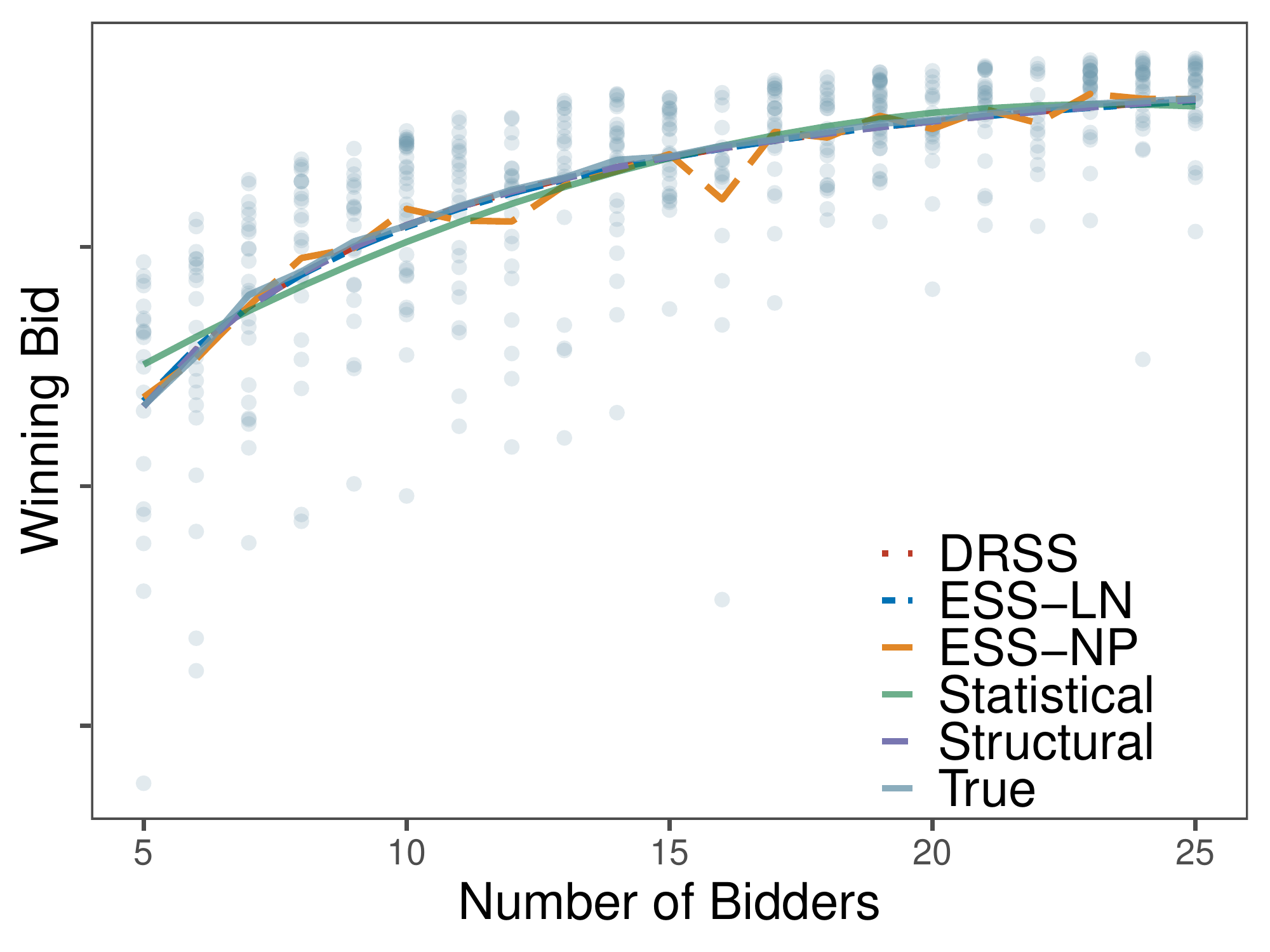}

}\subfloat[{\small{}Experiment 1: Test Sample}\label{fig:auction1_b}]{\includegraphics[width=0.5\columnwidth]{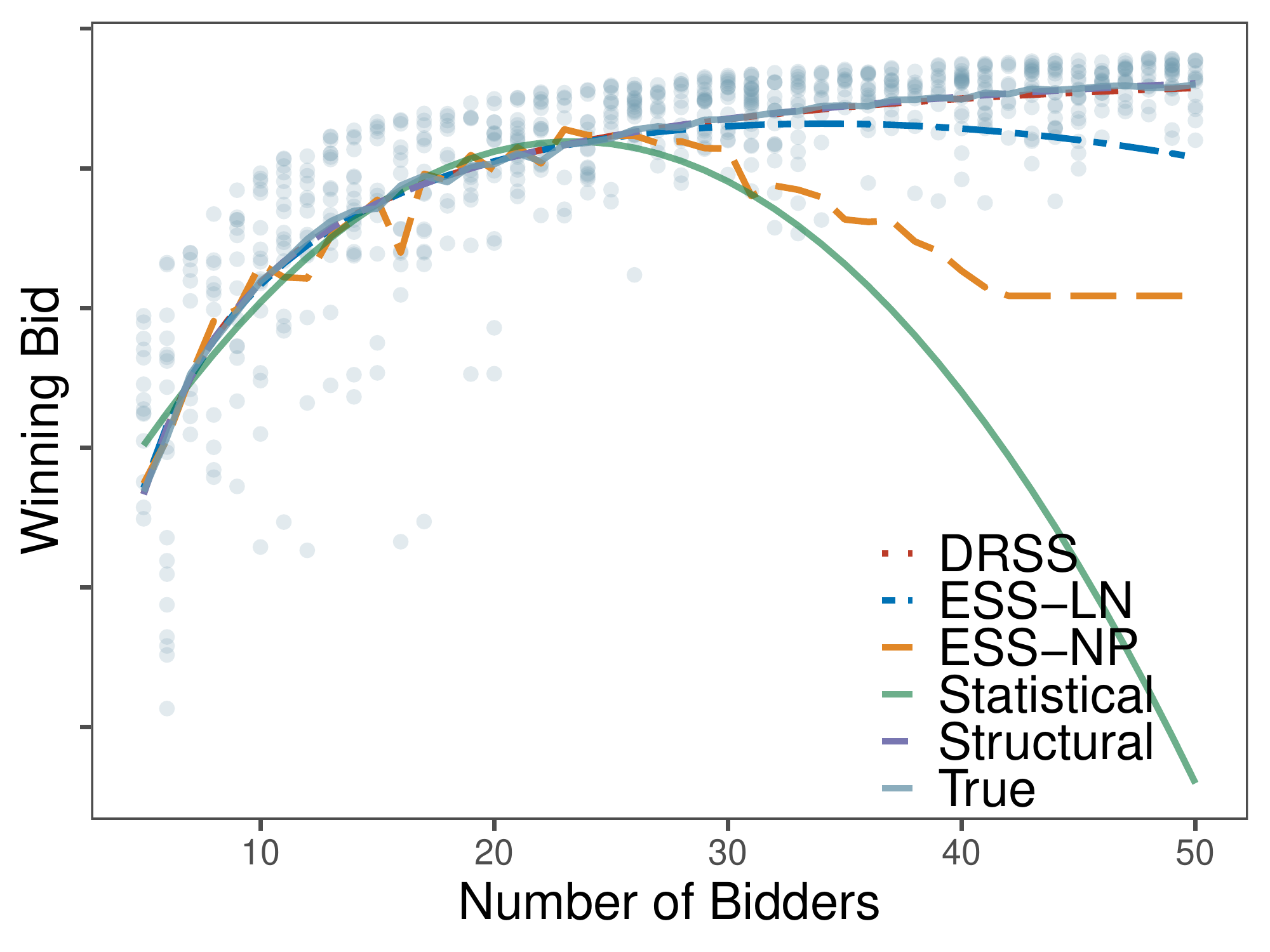}

}

\subfloat[{\small{}Experiment 2: Training Sample}\label{fig:auction2_a}]{\includegraphics[width=0.5\columnwidth]{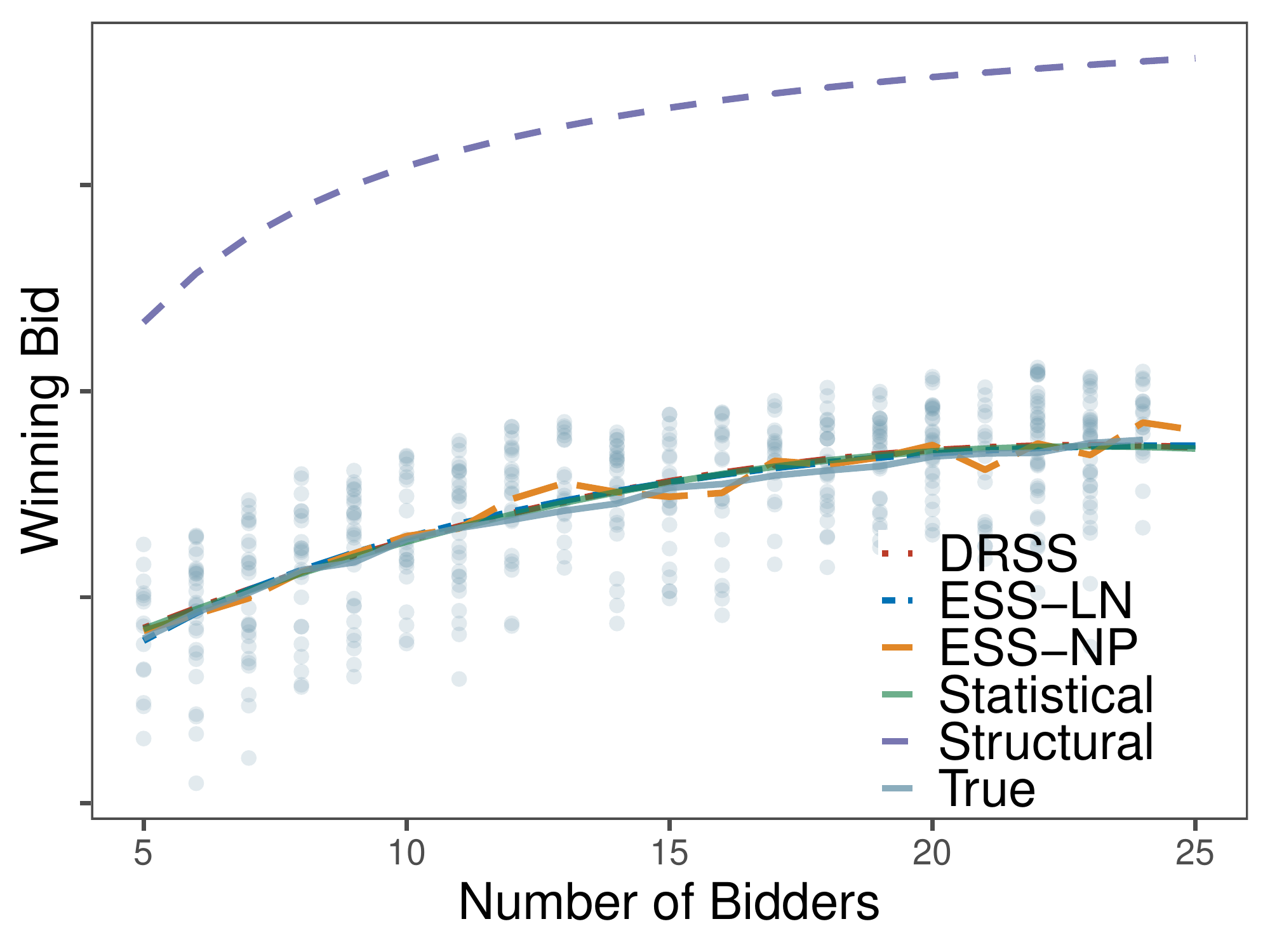}

}\subfloat[{\small{}Experiment 2: Test Sample}\label{fig:auction2_b}]{\includegraphics[width=0.5\columnwidth]{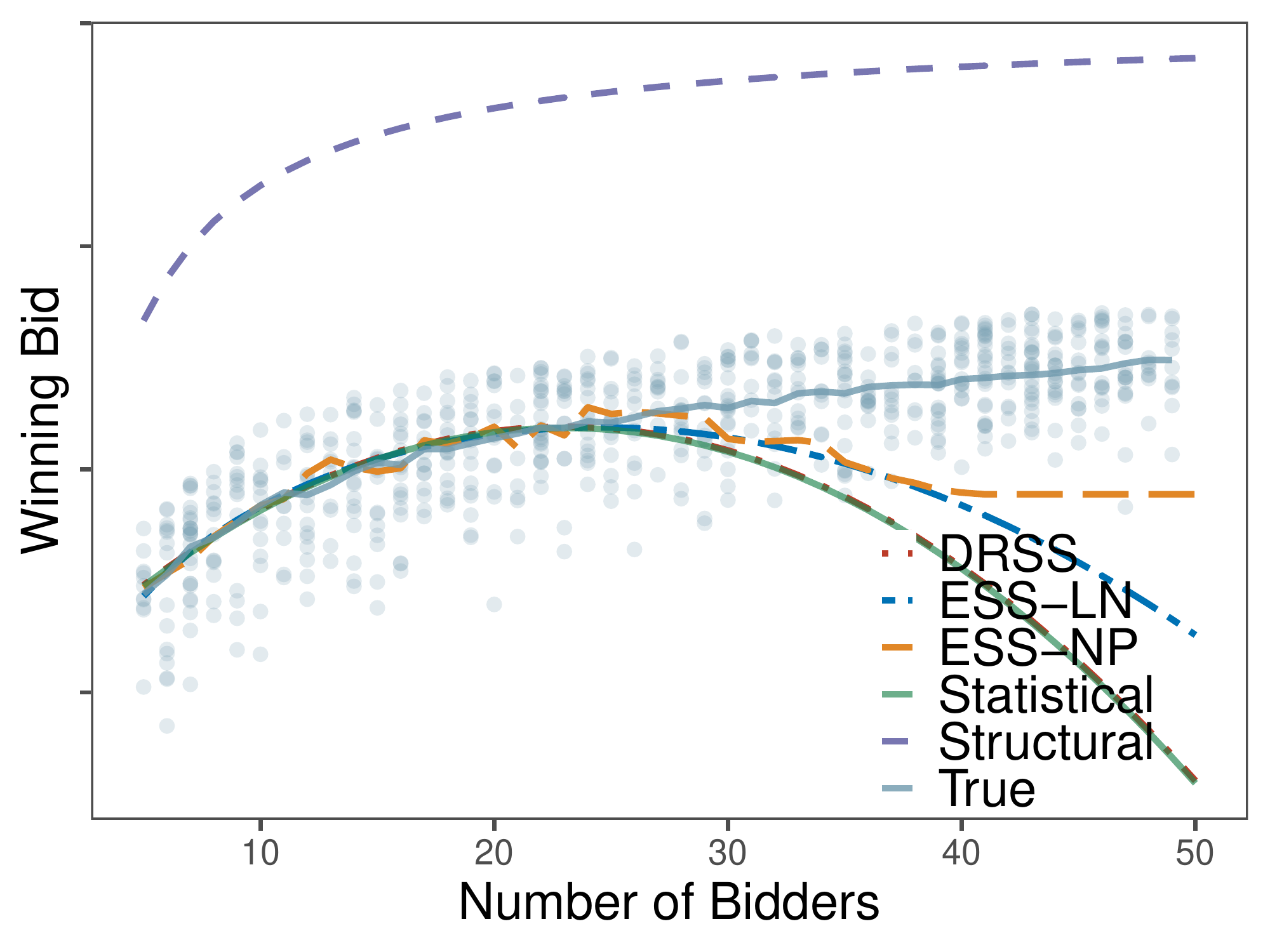}

}

\subfloat[{\small{}Experiment 3: Training Sample}\label{fig:auction3_a}]{\includegraphics[width=0.5\columnwidth]{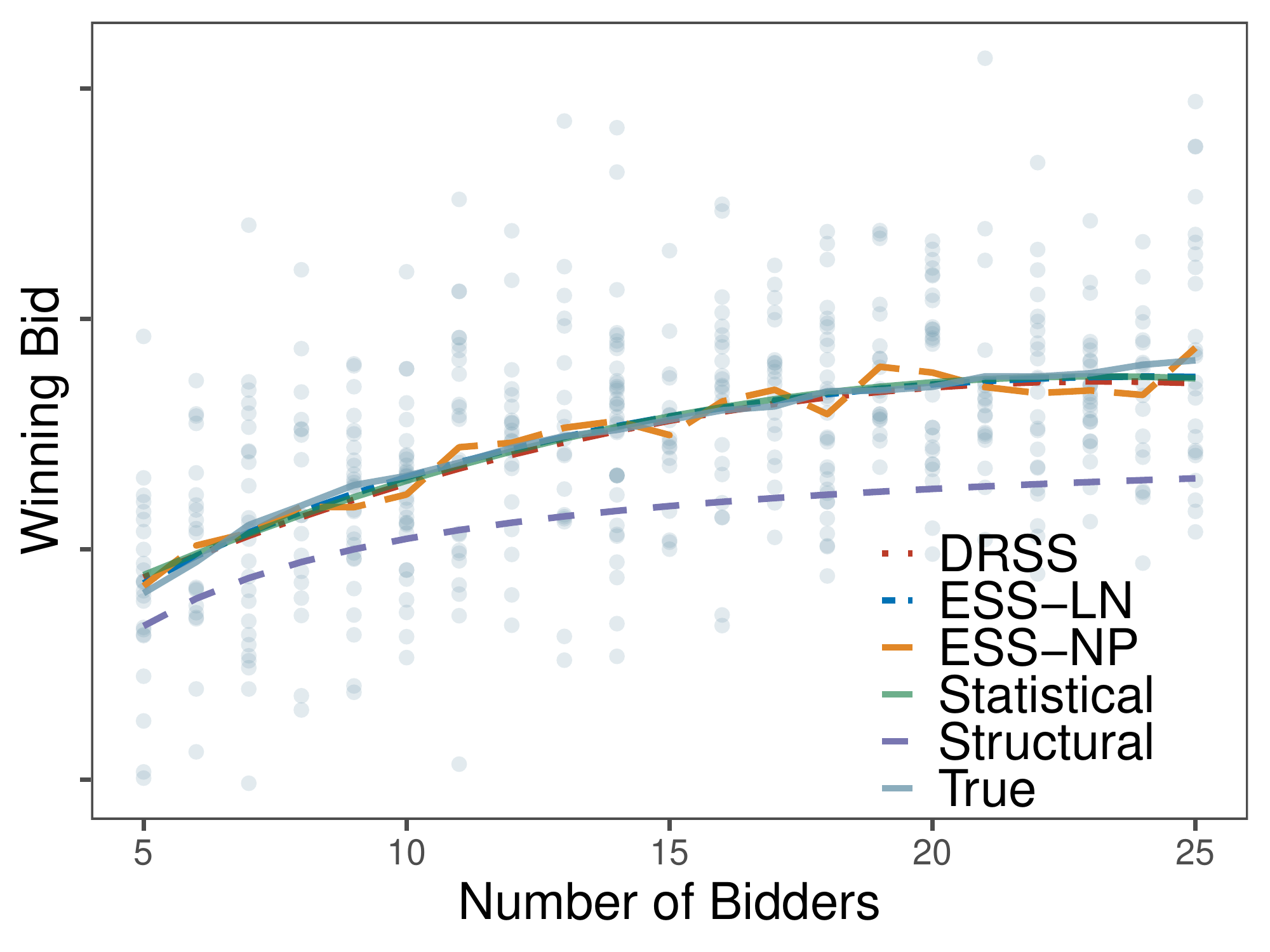}

}\subfloat[{\small{}Experiment 3: Test Sample}\label{fig:auction3_b}]{\includegraphics[width=0.5\columnwidth]{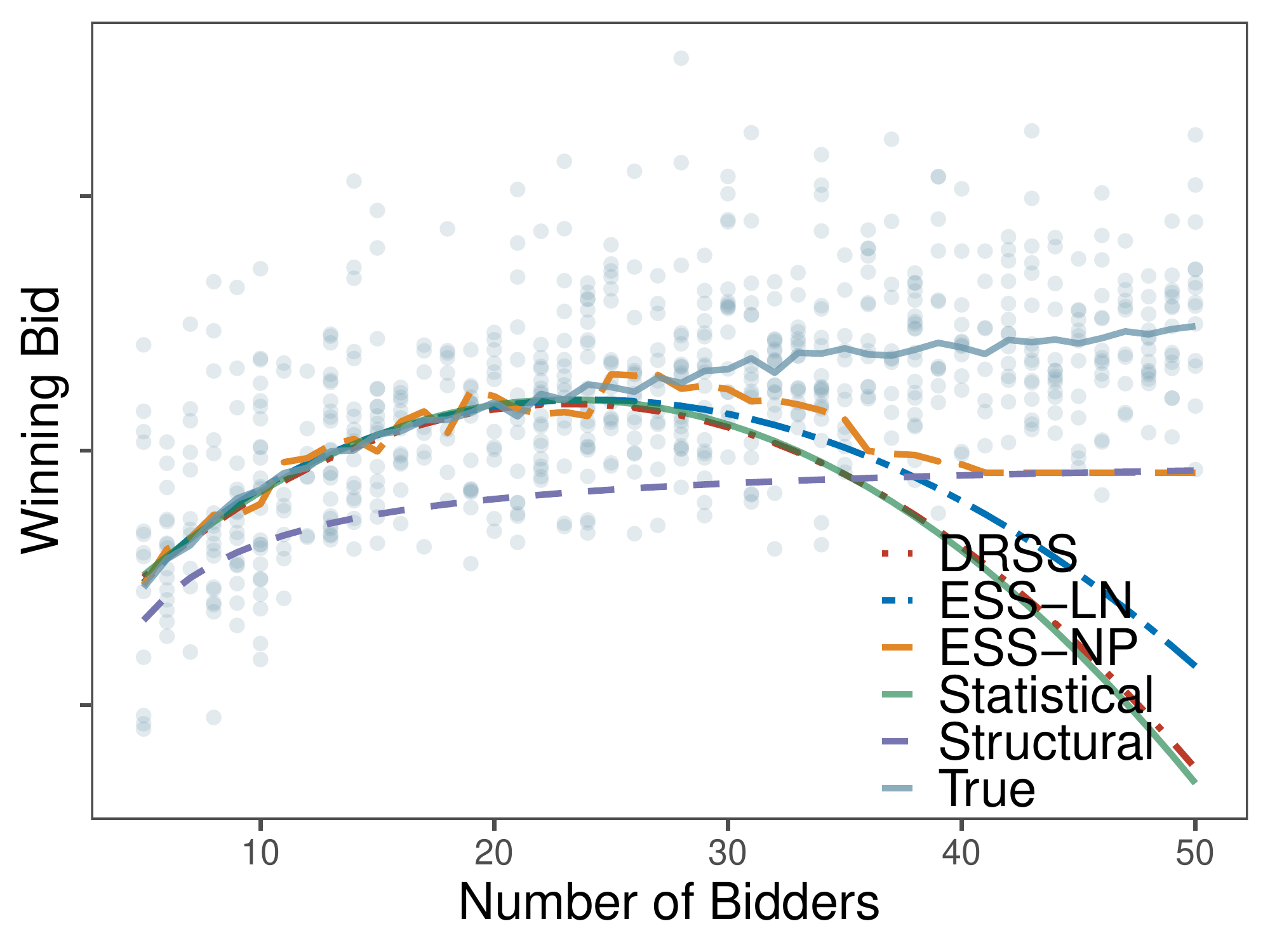}

}

\caption{{\small{}First-price Auction - The relationship between the number
of bidders and the winning bid.}}
\end{figure}

\begin{table}
\centering

\begin{threeparttable}

\caption{First-price Auction - Results\tnote{a}\label{tab:AuctionTable}}
\medskip{}

\begin{tabular}{lrrrrrrr}
\toprule 
\toprule & \multicolumn{3}{c}{In-Domain} &  & \multicolumn{3}{c}{Out-of-Domain}\tabularnewline
\cmidrule{2-4} \cmidrule{3-4} \cmidrule{4-4} \cmidrule{6-8} \cmidrule{7-8} \cmidrule{8-8} 
 & MSE & Bias & Var & \  & MSE & Bias & Var\tabularnewline
\midrule 
 &  &  &  &  &  &  & \tabularnewline
\multicolumn{8}{l}{\textit{Experiment 1}}\tabularnewline
 &  &  &  &  &  &  & \tabularnewline
Structural & 0.00 & 0.00 & 0.00 &  & 0.00 & 0.00 & 0.00\tabularnewline
Statistical & 1.27 & 86.36 & 0.29 &  & 871.37 & 2320.12 & 31.56\tabularnewline
DRSS & 0.17 & 20.72 & 0.11 &  & 126.83 & 566.38 & 77.66\tabularnewline
ESS-LN & 0.38 & 41.70 & 0.36 &  & 123.18 & 730.98 & 115.23\tabularnewline
ESS-NP & 1.89 & 104.09 & 1.88 &  & 123.69 & 965.51 & 2.08\tabularnewline
 &  &  &  &  &  &  & \tabularnewline
\midrule
 &  &  &  &  &  &  & \tabularnewline
\multicolumn{8}{l}{\textit{Experiment 2}}\tabularnewline
 &  &  &  &  &  &  & \tabularnewline
Structural & 1311.47 & 3617.50 & 0.00 &  & 1252.17 & 3537.75 & 0.00\tabularnewline
Statistical & 0.47 & 53.58 & 0.21 &  & 326.01 & 1392.57 & 28.52\tabularnewline
DRSS & 0.47 & 53.41 & 0.21 &  & 324.93 & 1389.32 & 28.48\tabularnewline
ESS-LN & 0.37 & 46.91 & 0.24 &  & 138.98 & 908.81 & 20.92\tabularnewline
ESS-NP & 1.37 & 91.36 & 1.32 &  & 98.16 & 836.85 & 3.86\tabularnewline
 &  &  &  &  &  &  & \tabularnewline
\midrule
 &  &  &  &  &  &  & \tabularnewline
\multicolumn{8}{l}{\textit{Experiment 3}}\tabularnewline
 &  &  &  &  &  &  & \tabularnewline
Structural & 214.41 & 1394.45 & 0.00 &  & 602.32 & 2443.56 & 0.00\tabularnewline
Statistical & 3.70 & 144.71 & 1.66 &  & 1245.99 & 2630.90 & 156.97\tabularnewline
DRSS & 3.63 & 143.92 & 1.69 &  & 1227.77 & 2624.21 & 151.48\tabularnewline
ESS-LN & 2.88 & 130.51 & 1.98 &  & 460.74 & 1480.48 & 132.90\tabularnewline
ESS-NP & 13.95 & 290.35 & 13.16 &  & 323.67 & 1483.44 & 24.96\tabularnewline
 &  &  &  &  &  &  & \tabularnewline
\bottomrule
\end{tabular}

\medskip{}

\begin{tablenotes} 
\small
\item [a] Results are based on 100 simulation trials. All numbers are on the scale of $10^{-4}$. Since the structural model predicts $\mathbb{E}\left[\left.b^{*}\right|n\right]=\left.\left(n-1\right)\right/\left(n+1\right)$, its predictions have zero variance and are the true values in experiment 1.
\end{tablenotes}

\end{threeparttable}
\end{table}

Figure \ref{fig:auction1_a} and \ref{fig:auction1_b} show the results
of the first experiment. In Figure \ref{fig:auction1_a}, we plot
the number of participants $n$ against the winning bid $b^{*}$,
the true relationship $\mathbb{E}\left[\left.b^{*}\right|n\right]$,
and the predictions obtained from five models: statistical, structural,
DRSS, ESS-LN, and ESS-NP. Since the structural model is the true model
in this experiment, it predicts the true expected winning bids. The
other four models, however, all fit relatively well. Figure \ref{fig:auction1_b}
plots the results of extrapolating the model predictions from $n\in\left[5,25\right]$
to $n\in\left[2,50\right]$. While the structural predictions still
hold true, the statistical fit becomes very bad, as can be expected.
Because the structural model is correctly specified while the statistical
model is not, the DRSS puts most of the weight on the structural model
and closely approximates its performance. The two ensemble estimators,
ESS-LN and ESS-NP, are also able to significantly outperform the statistical
model out-of-domain. In the first panel of Table \ref{tab:AuctionTable},
we report the bias, variance, and mean squared error of all the estimators
for $100$ simulation runs\footnote{Given an estimator $f$, let $f^{\left(r\right)}\left(n\right)$ denote
the estimator's prediction of the winning bid in simulation $r$,
then
\begin{align*}
\text{bias\ensuremath{\left(f\right)}} & =\mathbb{E}_{n}\left[\mathbb{E}_{r}\left[\left|f^{\left(r\right)}\left(n\right)-\mathbb{E}\left[\left.b^{*}\right|n\right]\right|\right]\right]\\
\text{var}\left(f\right) & =\mathbb{E}_{n}\left[\mathbb{E}_{r}\left[\left(f^{\left(r\right)}\left(n\right)-\mathbb{E}_{r}\left[f^{\left(r\right)}\left(n\right)\right]\right)^{2}\right]\right]\\
\text{mse}\left(f\right) & =\mathbb{E}_{n}\left[\mathbb{E}_{r}\left[\left(f^{\left(r\right)}\left(n\right)-\mathbb{E}\left[\left.b^{*}\right|n\right]\right)^{2}\right]\right]
\end{align*}

Reported are their empirical estimates.}. \emph{In domain}, compared to the true structural model, the DRSS
provides the best fit, followed by the ESS-LN. Both the statistical
and the ESS-NP models fit well as well. \emph{Out of domain}, the
statistical model has by far the worst performance. The three proposed
estimators all have similar MSE and achieve significant gains in performance
over the statistical model. Out of the three, the DRSS has the smallest
bias. Thus, the DRSS estimator appears to work the best in this experiment.
This is not surprising as one of its candidate models is correctly
specified, satisfying the condition for DRSS consistency. 

Figure \ref{fig:auction2_a} $-$ \ref{fig:auction3_b} show the results
of experiment 2 and 3. The results tell as similar story. In both
experiments, the structural model is misspecified. In experiment 2,
it misspecifies the private value distribution. In experiment 3, it
assumes that bidders are rational and the observed bids are Bayesian-Nash
equilibrium outcomes when they are not. As a consequence, in both
cases, the structural fit deviates from the true model significantly.
The statistical model, like in experiment 1, is able to fit well in-domain
but poorly out-of-domain. Since both of its candidate models are misspecified
in these experiments, the DRSS does not perform well. As the statistical
model has better in-domain fit relative to the misspecified structural,
the DRSS puts the majority of its weight on the statistical model.
In comparison, the two ensemble estimators are able to both fit well
in-domain and extrapolate better than the statistical, the structural,
and the DRSS models. In the second and third panels of Table \ref{tab:AuctionTable},
we observe the performance of these estimators over $100$ simulation
runs. In both experiments, the ESS-LN produces the best in-domain
fit, while the ESS-NP produces the best out-of-domain fit. Intuitively,
the ensemble methods are able to achieve these performance gains due
to a complementarity that exists between the statistical and the structural
models in these two experiments: the statistical model fits well in-domain,
while the structural model, though misspecified, provides useful guidance
on the functional form of $\mathbb{E}\left[\left.b^{*}\right|n\right]$
when we extrapolate beyond the observed domain, as evidenced in Figure
\ref{fig:auction2_b}, \ref{fig:auction3_b}.

\subsection{Dynamic Entry and Exit\label{subsec:DDCM}}

Our second application concerns the modeling and estimation of firm
entry and exit dynamics. Structural analysis of dynamic firm behavior
based on dynamic discrete choice (DDC) and dynamic game models has
been an important part of empirical industrial organization\footnote{See \citet{aguirregabiria_dynamic_2010,bajari_game_2013} for surveys
on structural estimation of dynamic discrete choice and dynamic game
models.}. These dynamic structural models capture the path dependence and
forward-looking behavior of agents, but pays the price of imposing
strong behavioral and parametric assumptions for tractability and
computational convenience. 

In this exercise, we focus our attention on the rational expectations
assumption that has been a key building block of dynamic structural
models in macro- and microeconomic analyses. The assumption and its
variants state that agents have expectations that do not systematically
differ from the realized outcomes\footnote{More precisely, rational expectations are mathematical expectations
based on information and probabilities that are model-consistent \citep{muth_rational_1961}.}. Despite having long been criticized as unrealistic, the rational
expectations paradigm has remained dominant due to a lack of tractable
alternatives and the fact that economists still know preciously little
about belief formation.

We conduct three experiments in the context of the dynamic entry and
exit of firms in competitive markets in non-stationary environments.
Our data-generating models are DDC models of entry and exit with entry
costs and exogenously evolving economic conditions. In our first experiment,
agents have rational expectations about future economic conditions.
In the second experiment, agents have a simple form of adaptive expectations
that assume the future is always like the past. The third experiment
features myopic agents who optimize only their current period returns.
In all experiments, we are interested in predicting the number of
firms that are operating in the market each period. To this end, we
estimate (a) a statistical model, (b) a structural model, and combine
them using (c) the DRSS estimator, and (d) the ESS estimators (ESS-LN,
ESS-NP). The structural model we estimate assumes rational expectations
and is thus correctly specified only in experiment 1. Table \ref{tab:DDCSetup}
summarizes this setup.

\begin{table}
\caption{Dynamic Entry and Exit - Setup\label{tab:DDCSetup}}

\medskip{}

\centering{}%
\begin{tabular}{clcc}
\toprule 
\toprule Experiment & \multicolumn{1}{c}{True Mechanism} & Structural Model & Statistical Model\tabularnewline
\midrule 
1 & Rational Expectations & \multirow{3}{*}{Rational Expectations} & \multirow{3}{*}{see \eqref{eq:DDCStatModel}}\tabularnewline
2 & Adaptive Expectations &  & \tabularnewline
3 & Myopic &  & \tabularnewline
\bottomrule
\end{tabular}
\end{table}

\paragraph{Setup}

Consider a market with $N$ firms. In each period, the market structure
consists of $n_{t}$ incumbent firms and $N-n_{t}$ potential entrants.
The profit to operating in the market at time $t$ is $R_{t}$, which
we assume to be exogenous and time-varying. At the beginning of each
period, both incumbents and potential entrants observe the current
period payoff $R_{t}$ and each draws an idiosyncratic utility shock
$\epsilon_{it}$. Incumbent firms then decide whether to remain or
exit the market by weighing the expected present values of each option,
while potential incumbents decide whether or not to enter the market,
which will incur a one-time entry cost $c$. Specifically, let the
entry status of a firm be represented by $\left(0,1\right)$. The
time-$t$ flow utility of a firm, who is in state $j\in\left\{ 0,1\right\} $
in time $t-1$ and state $k\in\left\{ 0,1\right\} $ in time $t$,
is given by
\begin{equation}
u_{it}^{jk}=\pi_{t}^{jk}+\epsilon_{it}^{k}\label{eq:DDCUtility}
\end{equation}
, where 
\begin{equation}
\pi_{t}^{jk}=\left(\mu+\alpha\cdot R_{t}-c\cdot\mathcal{I}\left(j=0\right)\right)\cdot\mathcal{I}\left(k=1\right)\label{eq:DDCPayoff}
\end{equation}
is the deterministic payoff function and $\epsilon_{it}=\left(\epsilon_{it}^{0},\epsilon_{it}^{1}\right)$
are idiosyncratic shocks, which we assume are \emph{i.i.d.} type-I
extreme value distributed. The parameter $\alpha$ measures the importance
of operating profits to entry-exit decisions relative to the idiosyncratic
utility shocks. 

The \emph{ex-ante} value function of a firm at the beginning of a
period is given by
\begin{align}
V_{t}^{j}\left(\epsilon_{it}\right) & =\max_{k\in\left\{ 0,1\right\} }\left\{ \pi_{t}^{jk}+\epsilon_{it}^{k}+\beta\cdot\mathbb{E}_{t}\left[\overline{V}_{t+1}^{k}\right]\right\} \label{eq:DDCMValueFcn}\\
 & =\max_{k\in\left\{ 0,1\right\} }\left\{ \mathcal{V}_{t}^{jk}+\epsilon_{it}^{k}\right\} 
\end{align}
, where $j$ is the firm's state in $t-1$, $\beta$ is the discount
factor, $\overline{V}_{t}^{j}\coloneqq\mathbb{E}_{\epsilon}\left[V_{t}^{j}\left(\epsilon_{it}\right)\right]$
is the expected value integrated over idiosyncratic shocks, and $\mathcal{V}_{t}^{jk}\coloneqq\pi_{t}^{jk}+\beta\cdot\mathbb{E}_{t}\left[\overline{V}_{t+1}^{k}\right]$
is the choice-specific conditional value function. 

At the beginning of each period, after idiosyncratic shocks are realized,
each firm thus chooses its action, $a_{it}\in\left\{ 0,1\right\} $,
by solving the following problem:
\begin{equation}
a_{it}=\underset{k\in\left\{ 0,1\right\} }{\arg\max}\left\{ \mathcal{V}_{t}^{jk}+\epsilon_{it}^{k}\right\} \label{eq:DDCAction}
\end{equation}
, which gives rise to the conditional choice probability (CCP) function:
\begin{equation}
p_{t}\left(k|j\right)\coloneqq\Pr\left(\left.a_{it}=k\right|a_{i,t-1}=j\right)=\frac{e^{\mathcal{V}_{t}^{jk}}}{\sum_{\ell=0}^{1}e^{\mathcal{V}_{t}^{j\ell}}}\label{eq:DDCCCP}
\end{equation}
, which follows from the extreme value distribution assumption. 

Since the value function involves the continuation values $\mathbb{E}_{t}\left[\overline{V}_{t+1}^{k}\right]$,
which requires expectations of the future profits $\left(R_{t+1},R_{t+2},\ldots\right)$,
its solution requires us to specify how such expectations are formed.
In experiment 1, we assume firms have \emph{perfect foresight} on
$R_{t}$. This is a stronger form of rational expectations that assumes
individuals \emph{knows} the future realized values. Firms can then
compute $\overline{V}_{t}^{j}=\mathbb{E}_{\epsilon}\left[V_{t}^{j}\left(\epsilon_{it}\right)\right],\ j\in\left\{ 0,1\right\} $
in a model-consistent way, i.e. based on the distributional assumption
of $\epsilon_{it}$. In experiment 2, we assume firms have a form
of \emph{adaptive expectations}, according to which beliefs about
the future are formed based on past values. Here for simplicity, we
assume that firms expect future profits to be always the same as in
current period, i.e. $R_{t}=R_{t+1}=R_{t+2}=\cdots$. Finally, in
experiment 3, we allow firms to be \emph{myopic}, so that they do
not care about the future and only maximize current payoffs.

\paragraph*{Simulation}

\begin{figure}
\begin{centering}
\includegraphics[width=0.8\columnwidth]{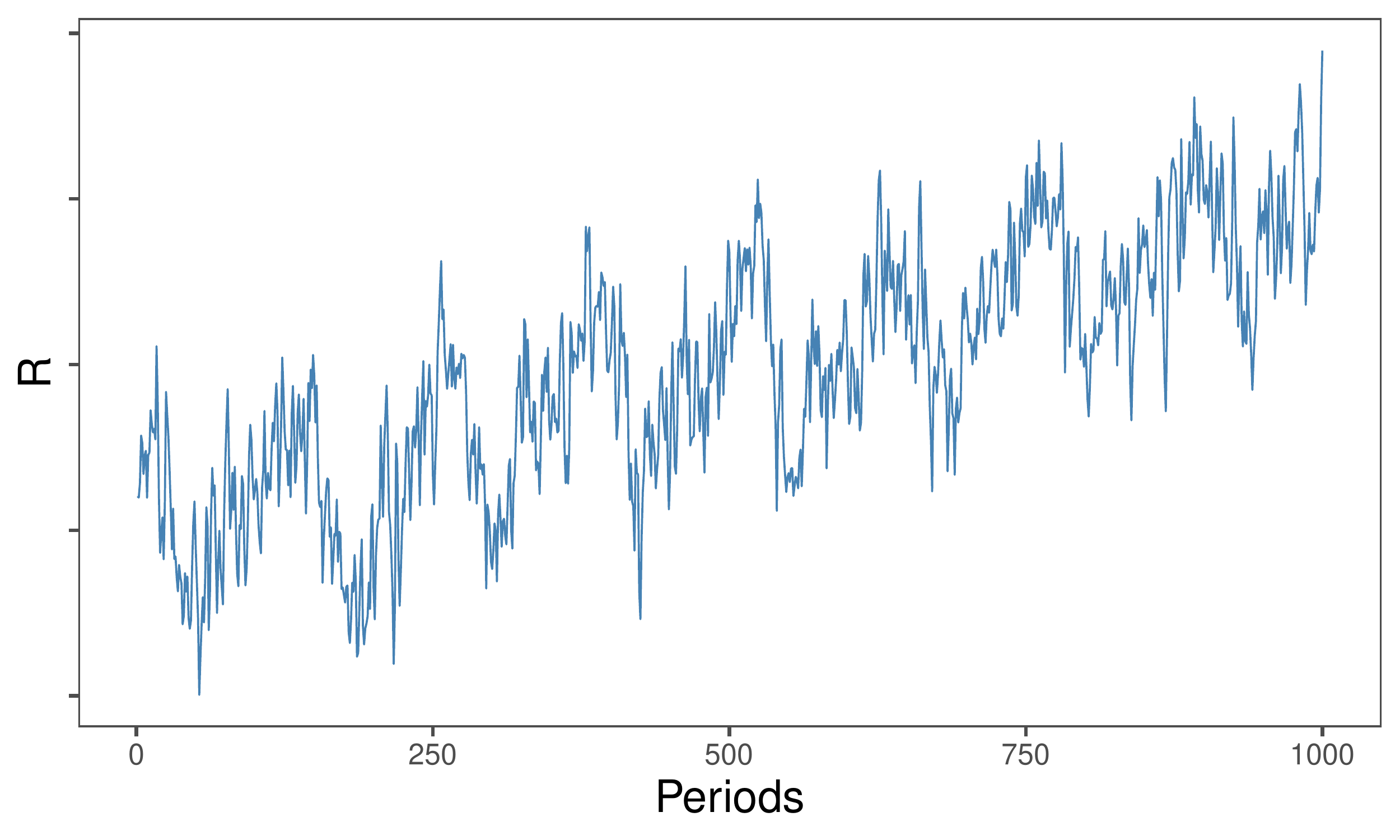}
\par\end{centering}
\caption{{\small{}Dynamic Entry and Exit - Exogenous Operating Profit\label{fig:DDCRt}}}
\end{figure}

For each experiment, we simulate $N=10,000$ firms for $\mathcal{T}=1000$
periods. The first $T=500$ periods are used for training and the
last $\mathcal{T}-T=500$ periods are used to assess the out-of-domain
performance of our estimators. The training data thus consist of $\text{\ensuremath{\mathcal{D}}}=\left\{ \left\{ a_{it}\right\} _{i=1}^{N},R_{t}\right\} _{t=1}^{T}$.
We simulate $R_{t}$ to follow an autoregressive process with a time
trend so that the environment is non-stationary. Figure \ref{fig:DDCRt}
shows a realized path of $R_{t}$. A different $R_{t}$ process is
chosen for each experiment so that the entry and exit dynamics over
the first $T$ periods are significantly different from the last $\mathcal{T}-T$
periods, allowing us to better distinguish the performance of the
estimators. Appendix \hyperlink{APP}{B.1} reports the parameter values
we use as well as other details of the simulation. 

\paragraph*{Statistical Model}

To predict the number of firms operating in the market each period,
$n_{t}$, based on observed exogenous operating profits, $R_{t}$,
we adopt the following ARX model:
\begin{equation}
n_{t}=\gamma_{0}+\gamma_{1}R_{t}+\rho_{1}n_{t-1}+\rho_{2}n_{t-2}+e_{t}\label{eq:DDCStatModel}
\end{equation}

\paragraph*{Structural Model}

We estimate the DDC model given by \eqref{eq:DDCUtility}--\eqref{eq:DDCCCP}
assuming rational expectations. Our estimation strategy builds on
\citet{arcidiacono_conditional_2011} and estimates an Euler-type
equation constructed out of CCPs. Here we sketch the strategy while
presenting its details in Appendix \hyperlink{APP}{B.1}\footnote{See \citet{arcidiacono_practical_2011} for a review of related CCP
estimators. For empirical implementations, see, e.g. \citet{artuc_trade_2010,scott_dynamic_2014}.}. A key to our strategy is the assumption that because agents have
rational expectations, their expected continuation values do not deviate
systematically from the realized values, i.e. $\overline{V}_{t+1}^{j}=\mathbb{E}_{t}\left[\overline{V}_{t+1}^{j}\right]+\xi_{t}^{j}$,
where $\xi_{t}^{j}$ is a time-$t$ expectational error with $\mathbb{E}\left(\xi_{t}^{j}\right)=0$.
Given this assumption, and since our model has the finite dependence
property of \citet{arcidiacono_conditional_2011}, solution to \eqref{eq:DDCMValueFcn}
can be written in the form of the following Euler equation: 
\begin{align}
\ln\frac{p_{t}\left(k|j\right)}{p_{t}\left(j|j\right)} & =\left(\pi_{t}^{j,k}-\pi_{t}^{j,j}+\beta\left(\pi_{t+1}^{k,k}-\pi_{t+1}^{j,k}\right)\right)-\beta\ln\frac{p_{t+1}\left(k|k\right)}{p_{t+1}\left(k|j\right)}+\epsilon_{t}^{j,k}\label{eq:DDCMEuler}
\end{align}
, where $\epsilon_{t}^{j,k}=\beta\left(\xi_{t}^{k}-\xi_{t}^{j}\right)$. 

Replacing the CCPs with their sample analogues, i.e. let $\widehat{p}_{t}\left(k|j\right)=$
observed percentage of firms that are in state $j$ in $t-1$ and
state $k$ in time $t$, we obtain the following estimating equations:
for all $j\ne k$,
\begin{equation}
\ln\frac{\widehat{p}_{t}\left(k|j\right)}{\widehat{p}_{t}\left(j|j\right)}+\beta\ln\frac{\widehat{p}_{t+1}\left(k|k\right)}{\widehat{p}_{t+1}\left(k|j\right)}=\begin{cases}
\mu+\alpha R_{t}-\left(1-\beta\right)c+e_{t}^{01} & \left(j,k\right)=\left(0,1\right)\\
-\mu-\alpha R_{t}+e_{t}^{10} & \left(j,k\right)=\left(1,0\right)
\end{cases}\label{eq:DDCEstimate}
\end{equation}
, where $e_{t}=\left(e_{t}^{01},e_{t}^{10}\right)$ is an error term
that captures both the expectational errors in $\epsilon_{t}^{j,k}$
and the approximation errors in $\widehat{p}_{t}\left(k|j\right)$. 

We assume that the value of the discount factor $\beta$ is known.
Estimating \eqref{eq:DDCEstimate} gives us an estimate of the model
parameters $\left(\mu,\alpha,c\right)$. These estimates are consistent
for a model that assumes rational expectations. Our structural model
is therefore correctly specified for experiment 1, but misspecified
in experiment 2 and 3. 

\paragraph*{Results}

\begin{figure}
\subfloat[\label{fig:ddc11}]{\includegraphics[width=1\columnwidth]{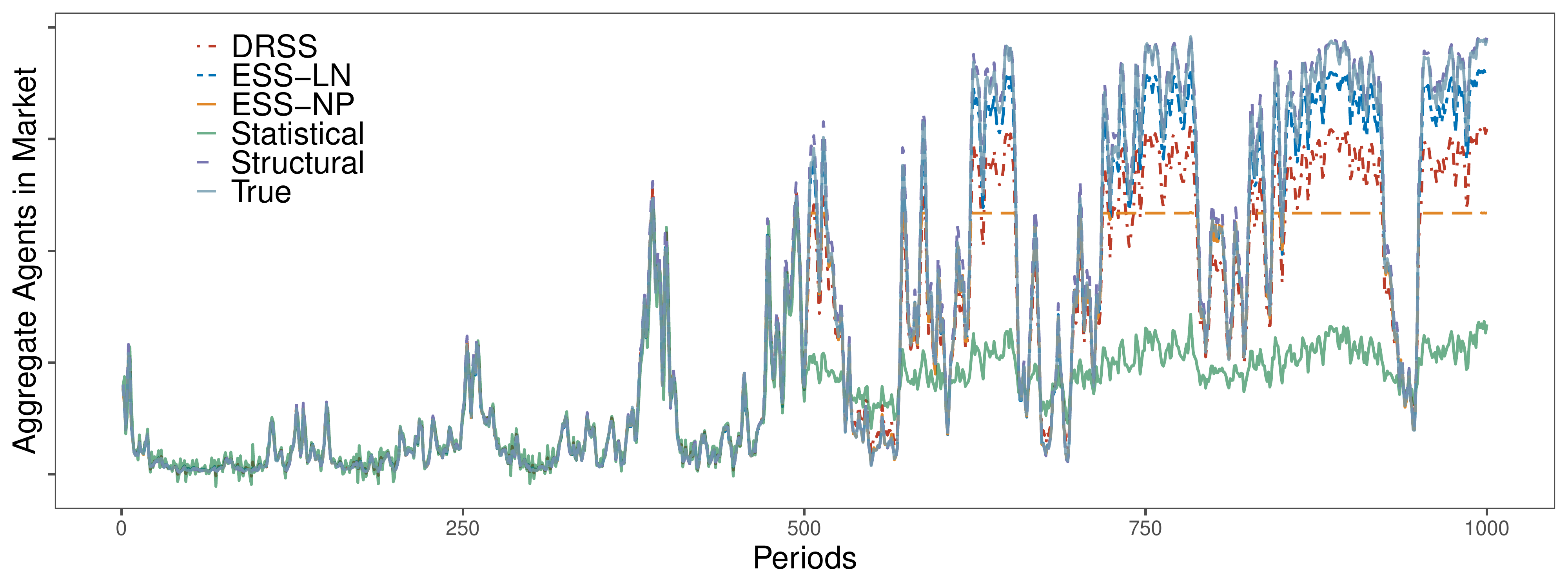}

}

\subfloat[\label{fig:ddc12}]{\includegraphics[width=0.5\columnwidth]{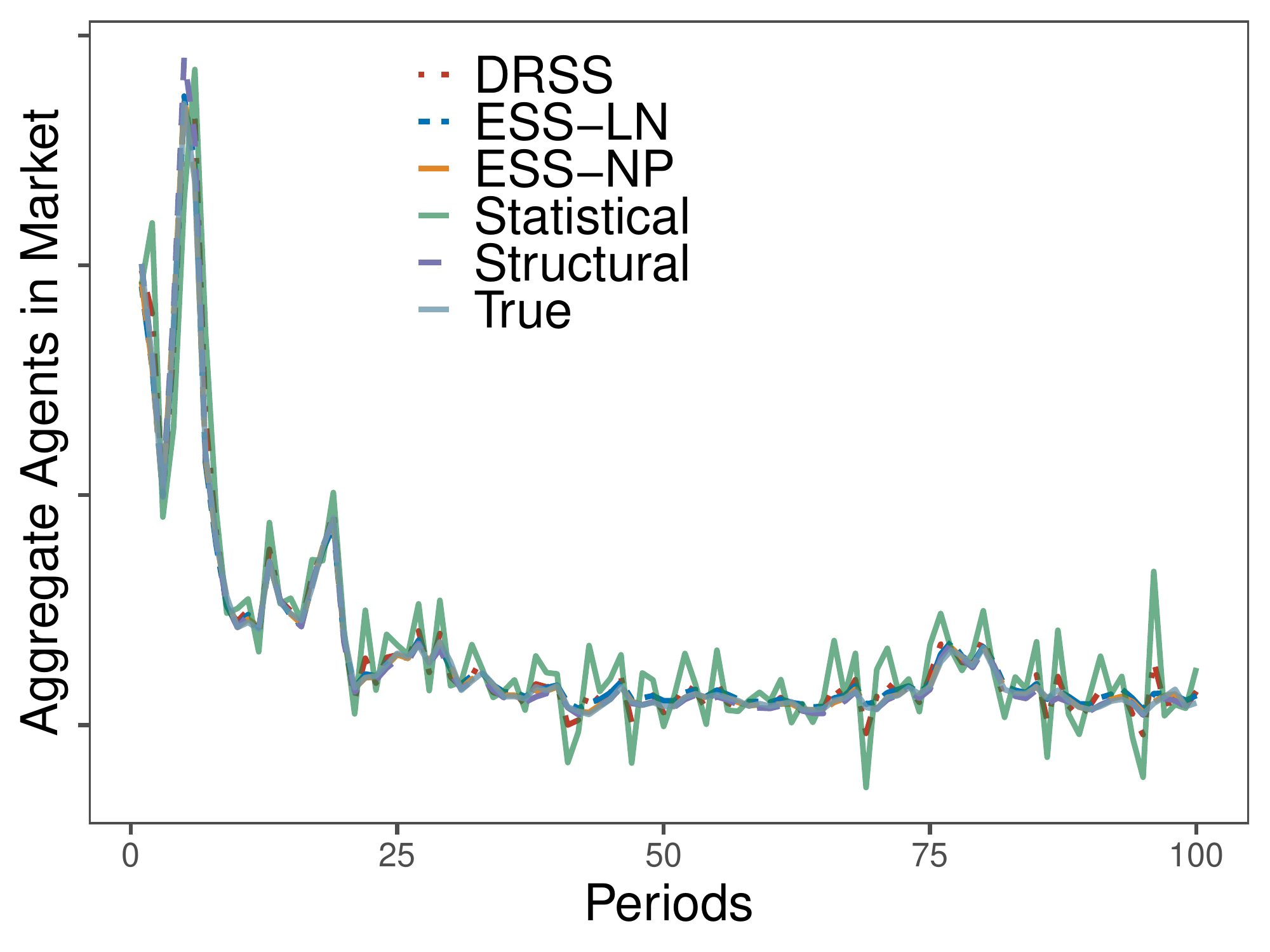}

}\subfloat[\label{fig:ddc13}]{\includegraphics[width=0.5\columnwidth]{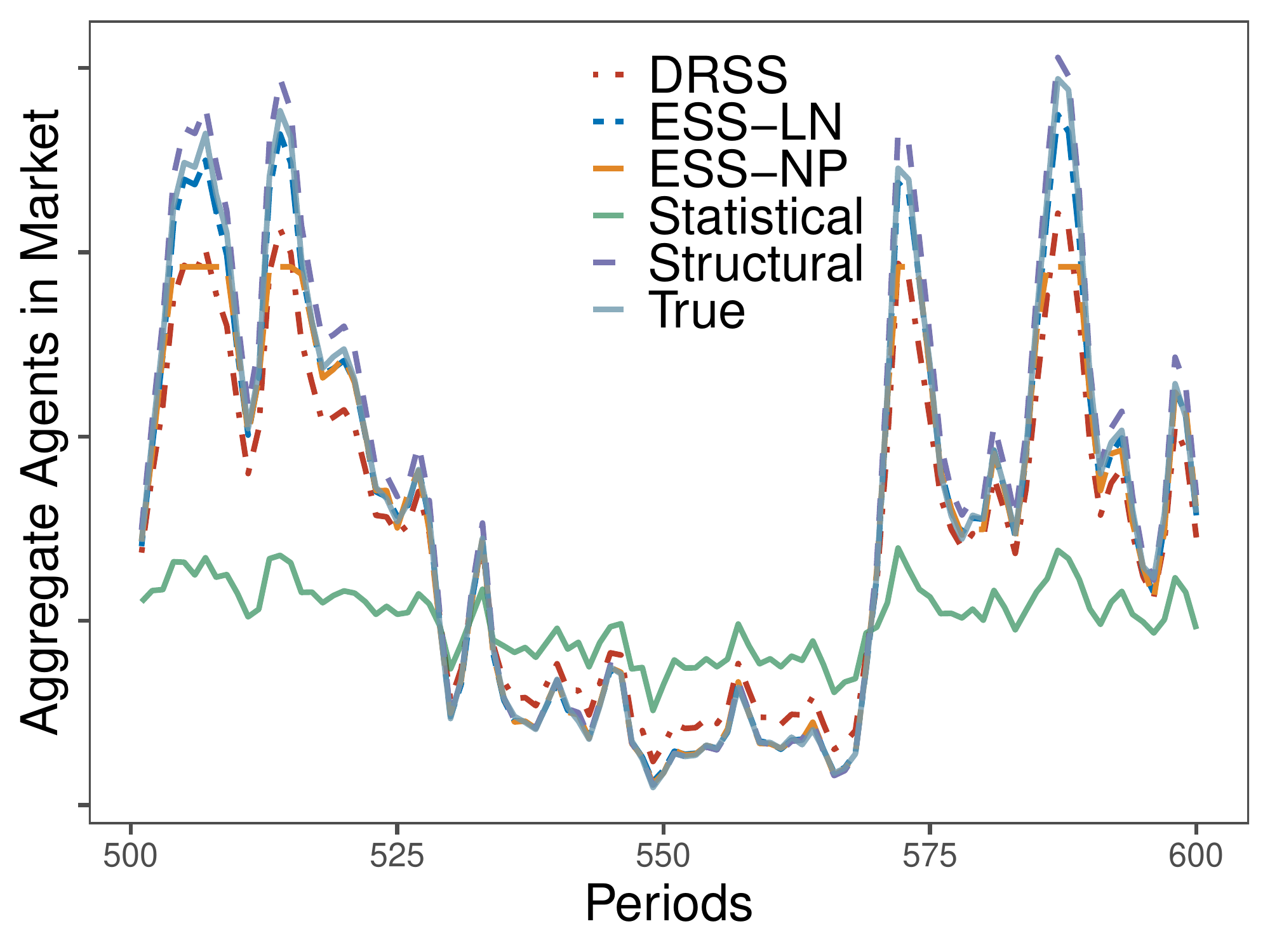}

}

\caption{{\small{}Dynamic Entry and Exit - Experiment 1. Plotted are the true
expected percentage of firms in the market along with model predictions.
Training data are not plotted for clarity. In (a), the entire periods
of $t=1-1000$ are plotted, which covers both the in-domain periods
of $t=1-500$ and the out-of-domain periods of $t=501-1000$. (b)
and (c) plot respectively the in-domain periods of $t=1-100$ and
the out-of-domain periods of $t=501-600$ in order to show a more
detailed picture. \label{fig:DDCI}}}
\end{figure}

\begin{figure}
\subfloat[\label{fig:ddc21}]{\includegraphics[width=1\columnwidth]{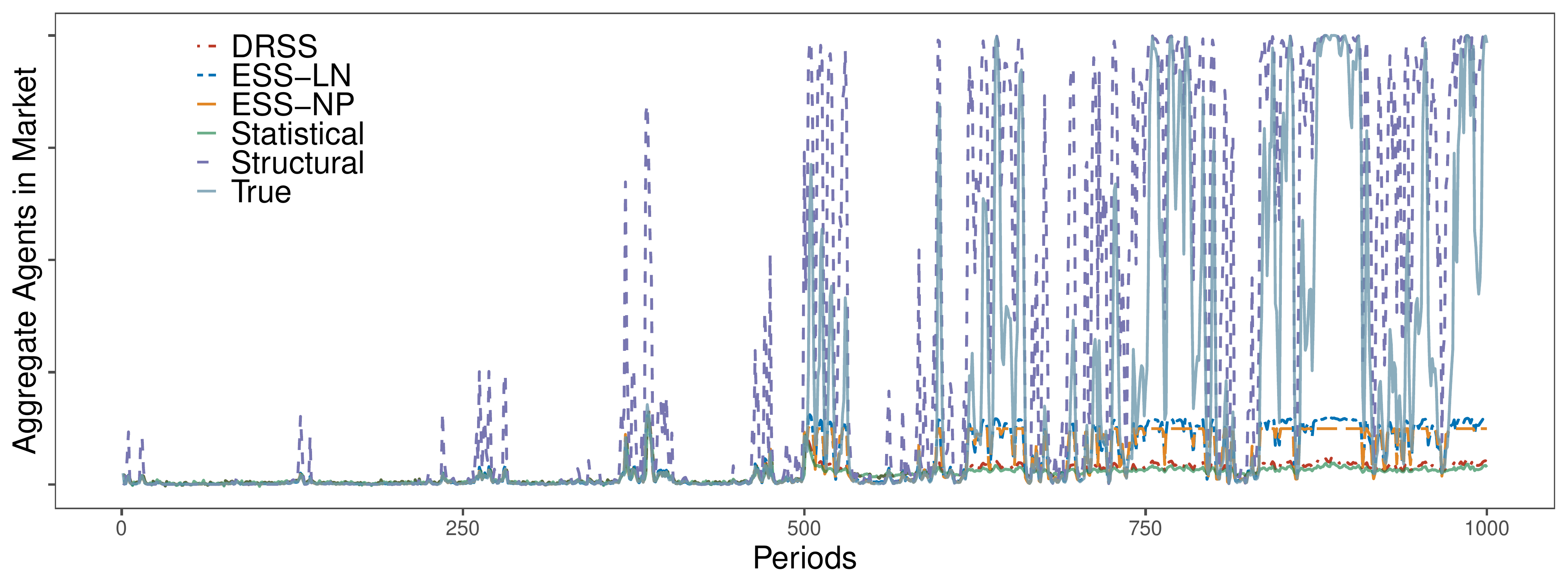}

}

\subfloat[\label{fig:ddc22}]{\includegraphics[width=0.5\columnwidth]{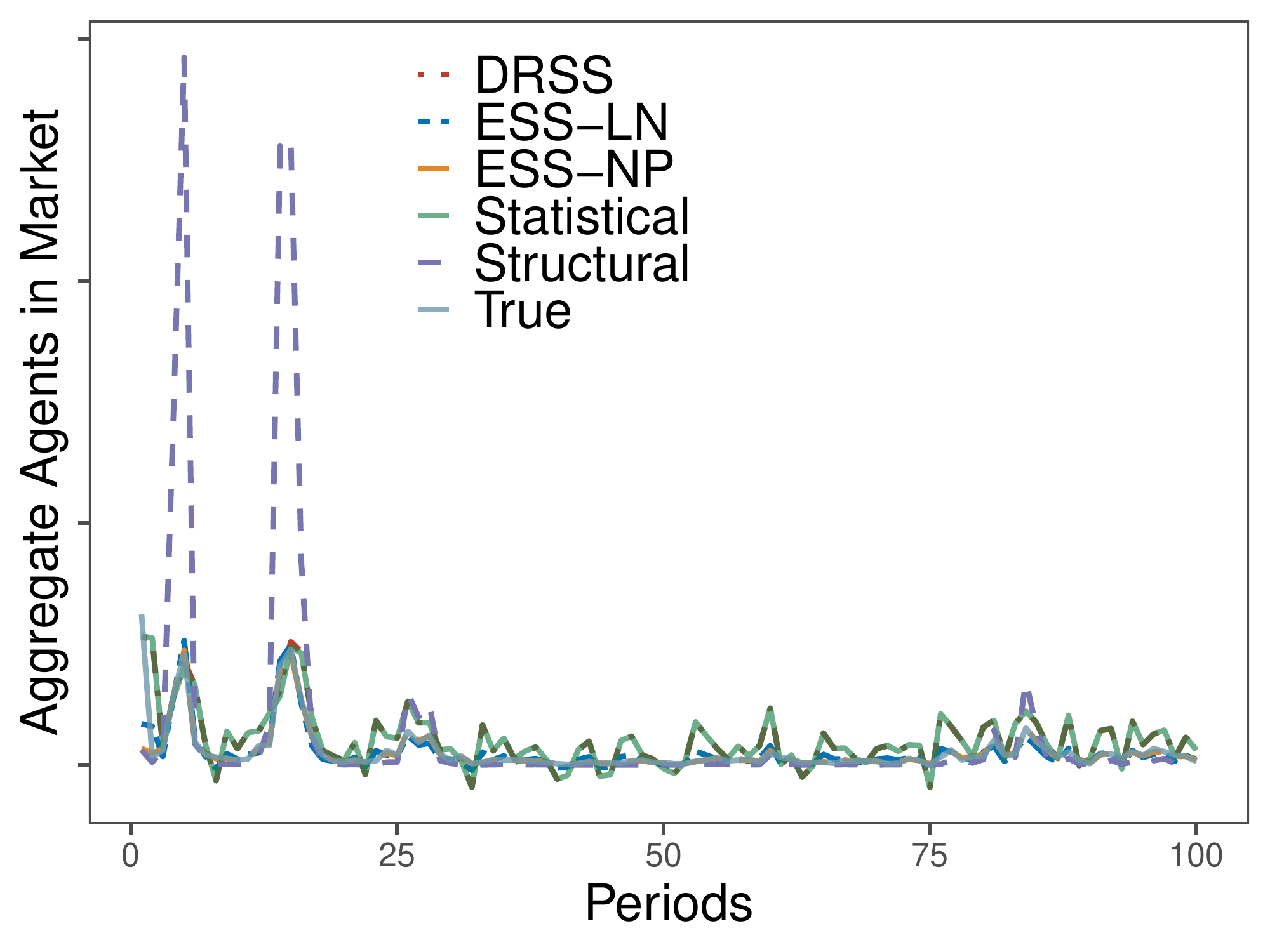}

}\subfloat[\label{fig:ddc23}]{\includegraphics[width=0.5\columnwidth]{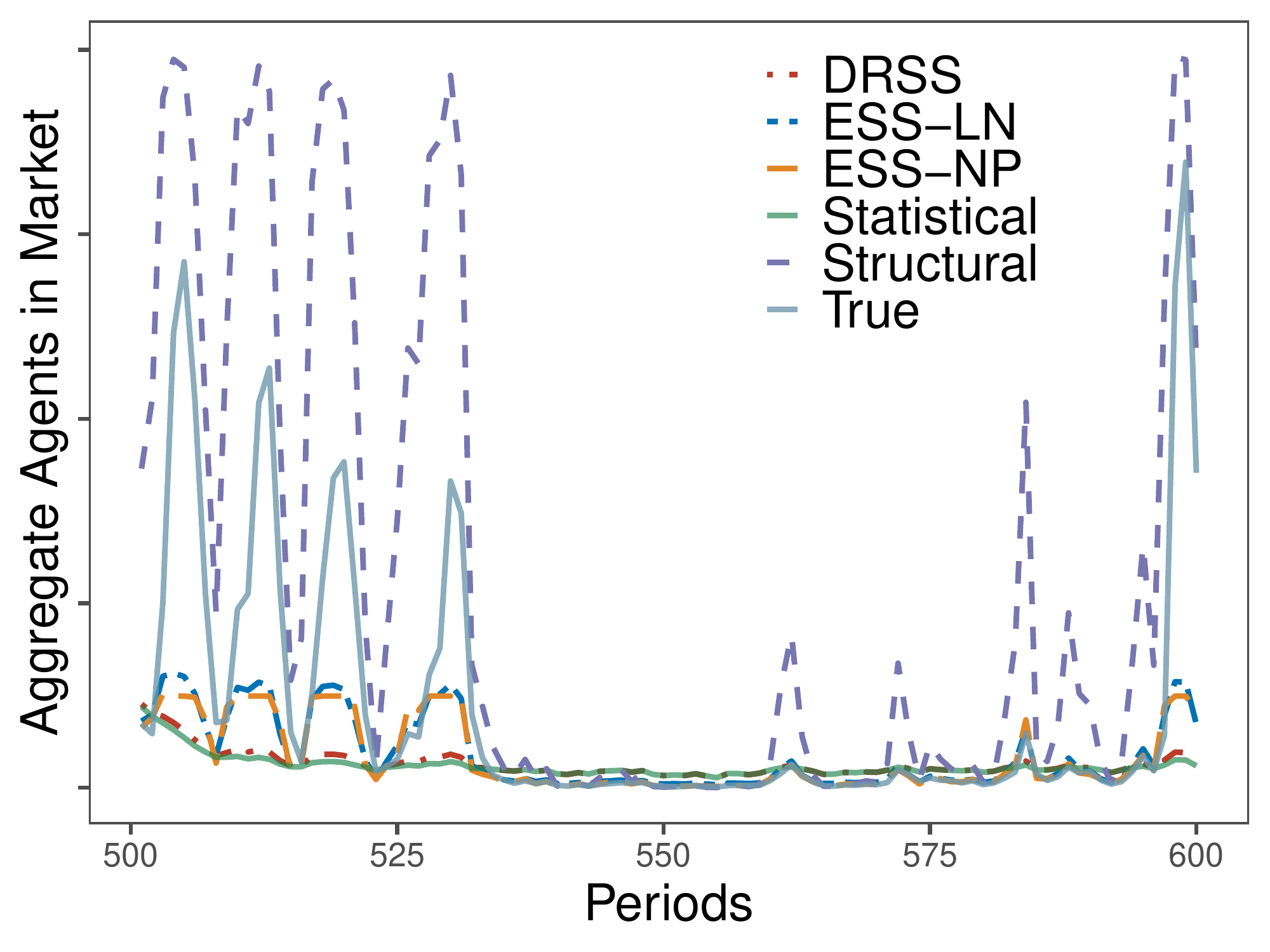}

}

\caption{{\small{}Dynamic Entry and Exit - Experiment 2. Plotted are the true
expected percentage of firms in the market along with model predictions.
Training data are not plotted for clarity. In (a), the entire periods
of $t=1-1000$ are plotted, which covers both the in-domain periods
of $t=1-500$ and the out-of-domain periods of $t=501-1000$. (b)
and (c) plot respectively the in-domain periods of $t=1-100$ and
the out-of-domain periods of $t=501-600$ in order to show a more
detailed picture. \label{fig:DDCII}}}
\end{figure}

\begin{figure}
\subfloat[\label{fig:ddc31}]{\includegraphics[width=1\columnwidth]{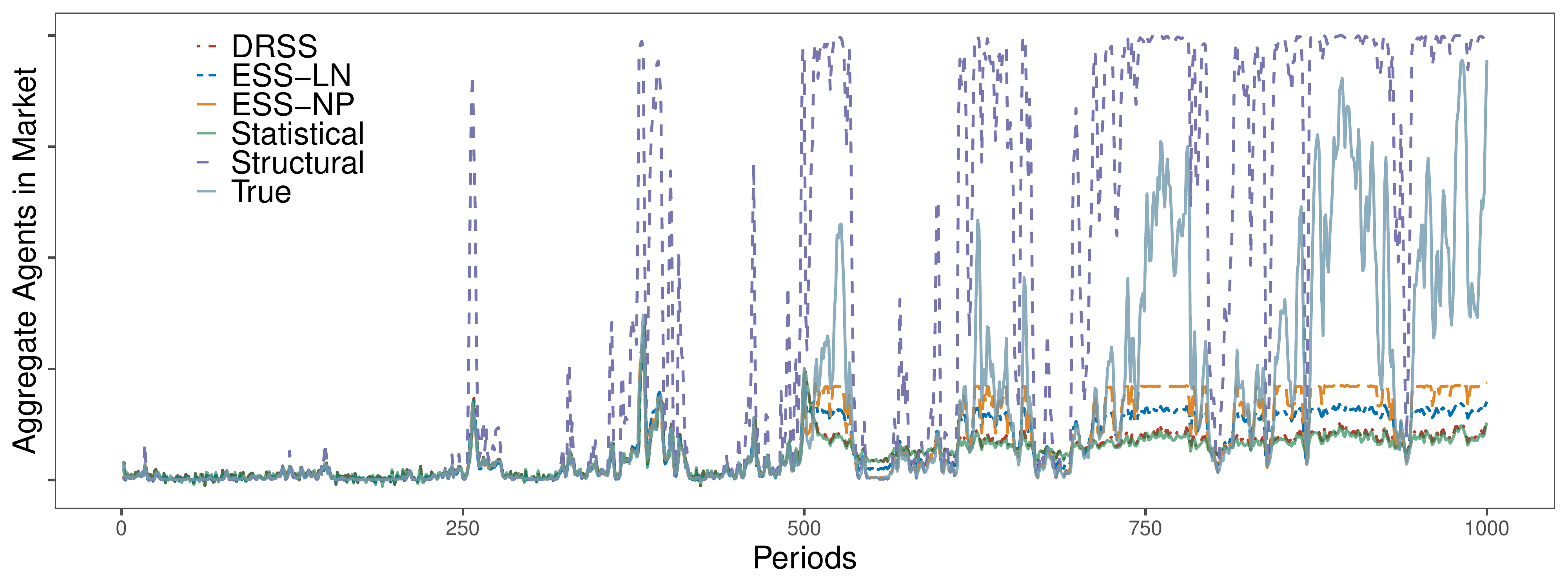}

}

\subfloat[\label{fig:ddc32}]{\includegraphics[width=0.5\columnwidth]{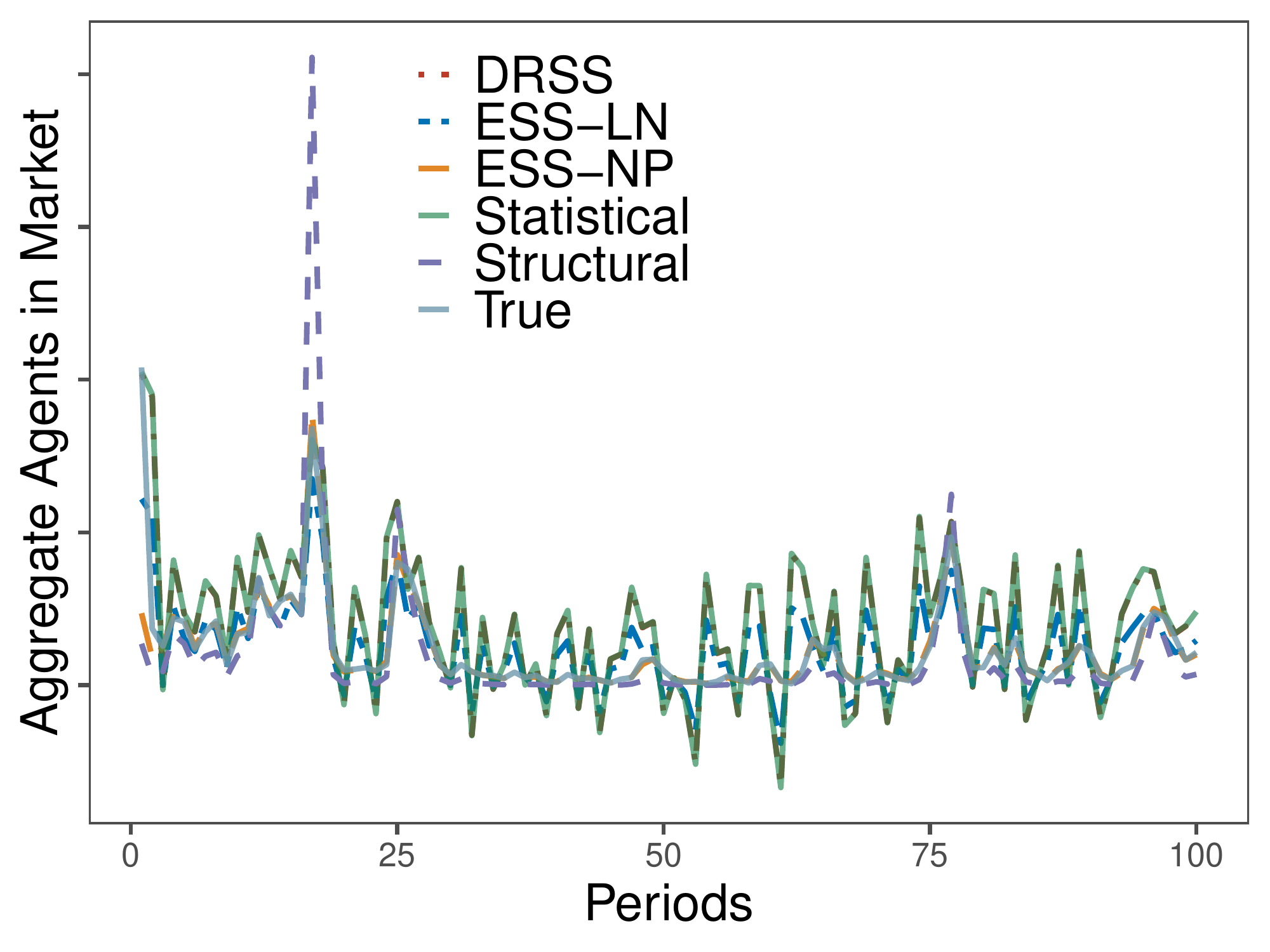}

}\subfloat[\label{fig:ddc33}]{\includegraphics[width=0.5\columnwidth]{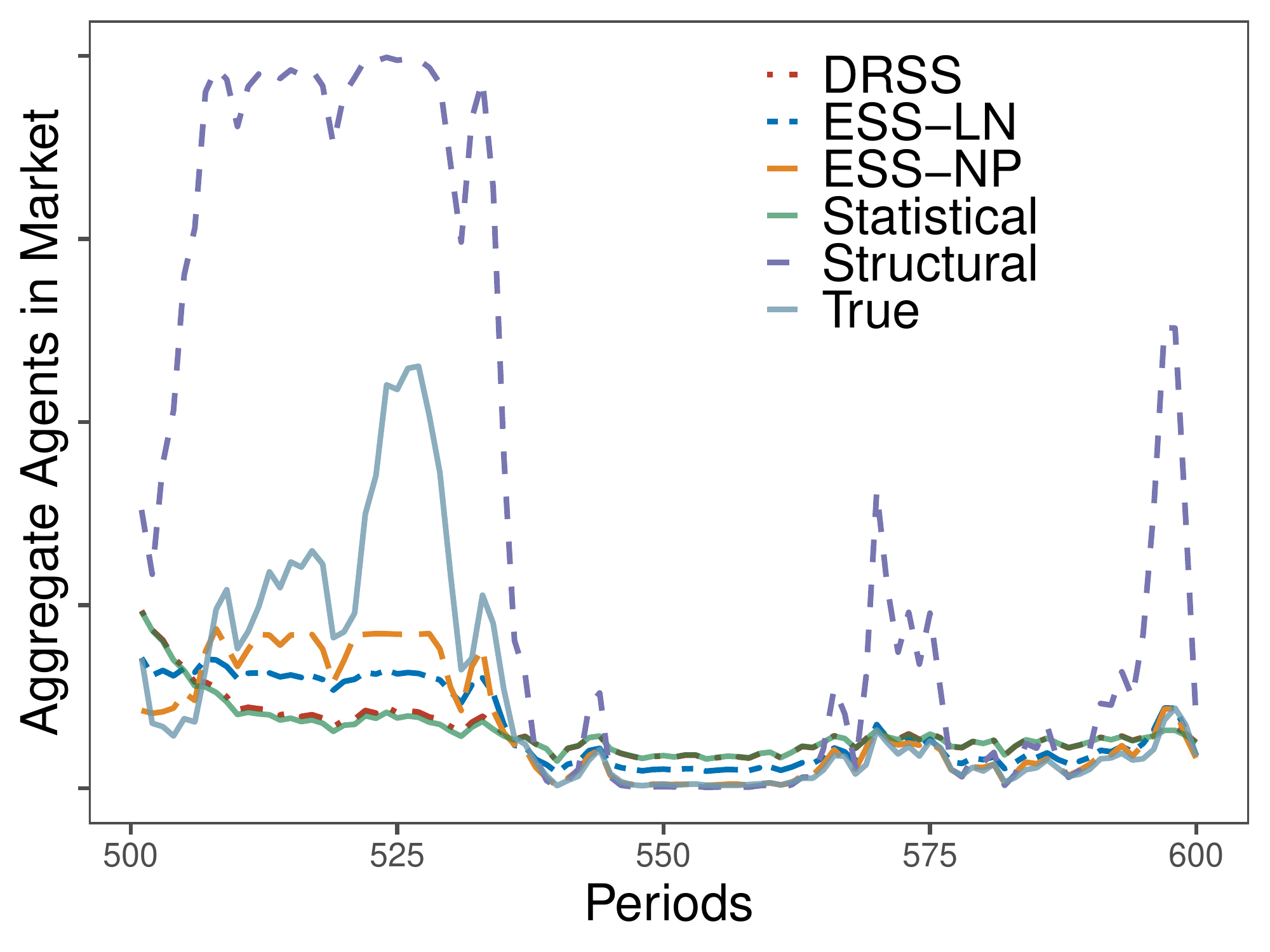}

}

\caption{{\small{}Dynamic Entry and Exit - Experiment 3. Plotted are the true
expected percentage of firms in the market along with model predictions.
Training data are not plotted for clarity. In (a), the entire periods
of $t=1-1000$ are plotted, which covers both the in-domain periods
of $t=1-500$ and the out-of-domain periods of $t=501-1000$. (b)
and (c) plot respectively the in-domain periods of $t=1-100$ and
the out-of-domain periods of $t=501-600$ in order to show a more
detailed picture. \label{fig:DDCIII}}}
\end{figure}

\begin{table}
\centering

\begin{threeparttable}

\caption{Dynamic Entry and Exit - Results\tnote{a}\label{tab:DDCTable}}
\medskip{}

\begin{tabular}{lrrrrrrr}
\toprule 
\toprule & \multicolumn{3}{c}{In-Domain} &  & \multicolumn{3}{c}{Out-of-Domain}\tabularnewline
\cmidrule{2-4} \cmidrule{3-4} \cmidrule{4-4} \cmidrule{6-8} \cmidrule{7-8} \cmidrule{8-8} 
 & MSE & Bias & Var & \  & MSE & Bias & Var\tabularnewline
\midrule 
 &  &  &  &  &  &  & \tabularnewline
\multicolumn{8}{l}{\textit{Experiment 1}}\tabularnewline
 &  &  &  &  &  &  & \tabularnewline
Structural & 10.13 & 133.01 & 80.50 &  & 55.86 & 562.82 & 276.50\tabularnewline
Statistical & 5.53 & 160.32 & 56.79 &  & 1620.59 & 3271.40 & 13.99\tabularnewline
DRSS & 4.12 & 135.05 & 57.87 &  & 1197.93 & 2707.97 & 105.57\tabularnewline
ESS-LN & 0.44 & 38.57 & 54.49 &  & 110.22 & 631.67 & 185.57\tabularnewline
ESS-NP & 0.12 & 14.09 & 53.95 &  & 379.99 & 1254.47 & 134.72\tabularnewline
 &  &  &  &  &  &  & \tabularnewline
\midrule
 &  &  &  &  &  &  & \tabularnewline
\multicolumn{8}{l}{\textit{Experiment 2}}\tabularnewline
 &  &  &  &  &  &  & \tabularnewline
Structural & 144.36 & 376.53 & 116.18 &  & 1199.94 & 2514.04 & 605.93\tabularnewline
Statistical & 3.22 & 67.78 & 7.26 &  & 1744.25 & 2569.35 & 2.97\tabularnewline
DRSS & 4.00 & 10.11 & 12.30 &  & 1502.54 & 2350.33 & 67.57\tabularnewline
ESS-LN & 1.45 & 35.32 & 7.50 &  & 1332.06 & 2126.37 & 16.39\tabularnewline
ESS-NP & 0.38 & 74.47 & 7.55 &  & 1146.09 & 1926.19 & 48.50\tabularnewline
 &  &  &  &  &  &  & \tabularnewline
\midrule
 &  &  &  &  &  &  & \tabularnewline
\multicolumn{8}{l}{\textit{Experiment 3}}\tabularnewline
 &  &  &  &  &  &  & \tabularnewline
Structural & 361.75 & 685.71 & 196.53 &  & 2670.64 & 4378.27 & 499.04\tabularnewline
Statistical & 1.89 & 78.72 & 7.36 &  & 890.14 & 1952.56 & 3.09\tabularnewline
DRSS & 1.88 & 78.35 & 8.14 &  & 849.67 & 1891.20 & 6.35\tabularnewline
ESS-LN & 0.99 & 49.24 & 6.78 &  & 762.69 & 1689.28 & 6.56\tabularnewline
ESS-NP & 0.24 & 14.24 & 6.85 &  & 628.23 & 1470.74 & 18.64\tabularnewline
 &  &  &  &  &  &  & \tabularnewline
\bottomrule
\end{tabular}

\medskip{}

\begin{tablenotes} 
\small
\item [a] Results are based on 100 simulation trials. All numbers are on the scale of $10^{-4}$.
\end{tablenotes}

\end{threeparttable}
\end{table}

Figure \ref{fig:DDCI} shows the results of the first experiment.
Figure \ref{fig:ddc11} plots the expected percentage of firms in
the market, $\mathbb{E}\left[n_{t}\right]$, for entire periods of
$t=1-1000$, including both the in-domain periods of $t=1-500$ and
the out-of-domain periods of $t=501-1000$, together with the predictions
of the five estimators. Predictions are made using one-step ahead
forecasting\footnote{Given an estimated model, in each period $t$, we predict $n_{t}$
based on $\left\{ \left(n_{t-1},n_{t-2},\ldots\right),\left(R_{t},R_{t-1},\ldots\right)\right\} $.
To generate predictions for the structural model, we also assume agents
have perfect foresight regarding $\left(R_{t+1},R_{t+2},\ldots\right)$
.}. A closer look at in-domain and out-of-domain results are presented
in Figure \ref{fig:ddc12} and \ref{fig:ddc13} for chosen periods.

All estimators fit relatively well in-domain. However, out-of-domain,
the time series model is unable to capture the rising number of firms
as $R_{t}$ increases. This is partly by design: as we have discussed,
we intentionally choose parameter values so that out-of-domain dynamics
differ markedly from those in-domain. A statistical model that fits
to the in-domain data is unable to extrapolate well in this case.
On the other hand, the structural model, which is correctly specified
in this experiment, extrapolates very well, as expected. Since one
of its candidate models is correctly specified, the DRSS is also expected
to perform well. Here, the DRSS model successfully allocates most
of its weight on the structural model. However, because some weight
is still put on the statistical model, it systematically underestimates
the number of firms in out-of-domain periods as well. This is also
expected as inability to distinguish between competing models based
on limited data is what motivates doubly robust and model averaging
approaches in the first place. Like the DRSS, the two ensemble estimators
are able to largely capture the rising number of firms in out-of-domain
periods, offering significantly better predictions than the statistical
model. Out of the two ensemble models, the ESS-LN performs particularly
well, matching the true model closely. 

In Table \ref{tab:DDCTable} Panel 1, we report the bias, variance,
and mean squared error of all the estimators with respect to the true
$\mathbb{E}\left[n_{t}\right]$ over $100$ trials. Somewhat surprisingly,
the structural model, albeit correctly specified, performs the \emph{worst}
in terms of MSE out of the five estimators in-domain. This is perhaps
due to a loss of efficiency associated with our Euler-equation approach
in estimating the model \citep{aguirregabiria_euler_2013}. Out of
domain, though, it predictably delivers the best performance. Out
of the remaining four estimators, the ESS-NP produces the best in-domain
fit, while the ESS-LN produces the best out-of-domain fit.

Figure \ref{fig:DDCII} shows the results of the second experiment.
In Experiment 2, agents have adaptive expectations in the sense that
they always assume $R_{t'}=R_{t}\ \forall t'>t$. Since in our simulations,
$R_{t}$ follows a rising trend, this means that agents systematically
underestimate future profits. The realized dynamics show that for
most of the in-domain periods, there are few firms in the market.
Number of firms increases significantly during the out-of-domain periods.
This marked difference between in-domain and out-of-domain dynamics
pose significant challenges. Looking at the model fits, the time series
model again fits relatively well in-domain but is completely unable
to extrapolate out-of-domain. The structural model, being misspecified,
is able to capture the rising entries, but tends to have larger fluctuations
than the true model. This can be explained by the fact that agents
in the structural model assumes that future profits will be the same
as current profits, thus reacting more dramatically to any changes
in $R_{t}$. As both the statistical and the structural model are
misspecified, the DRSS does not perform well. It puts most of the
weight on the statistical model, leading to a bad extrapolation performance.
The ensemble models, ESS-LN and ESS-NP, are both able to fit well
in-domain and capture some part of the rising trend out-of-domain.
Compared to the structural model, they tend to underfit rather than
overfit the true expected number of firms in out-of-domain periods. 

Looking at Panel 2 of Table \ref{tab:DDCTable}, we see that the ESS-NP
achieves the smallest MSE both in-domain and out-of-domain, making
it the winner in this experiment. The structural model is a close
second in out-of-domain performance but is by far the worst in-domain.
Indeed, the DRSS, the ESS-LN, and the ESS-NP all achieve significantly
smaller MSEs in-domain. This experiment serves to illustrate a scenario
in which the complementarity between the structural and the statistical
model is especially pronounced, with the former fitting relatively
badly in-domain and the latter completely unable to extrapolate. By
combining the two, our ensemble models mainly rely on the former to
guide out-of-domain prediction and on the latter to regulate in-domain
fit. 

Figure \ref{fig:DDCIII} shows the results of the third experiment.
In this experiment, agents are myopic in that they only care about
current period returns when making entry and exit decisions. The data-generating
model is therefore static in nature. Looking at estimator performances,
the story is broadly similar to that of experiment 2, with the difference
being that, in this experiment, the true model exhibits less dramatic
difference between its in-domain and out-of-domain dynamics and the
misspecified structural model tends to more significantly overestimate
the number of firms in the market. As a consequence, according to
Panel 3 of Table \ref{tab:DDCTable}, the structural model is the
worst performer both in-domain and out-of-domain in this experiment.
On the other hand, both ensemble estimators perform better than the
other estimators both in-domain and out-of-domain, with the ESS-NP
the clear winner. Thus, as in the auction experiments, our ensemble
methods are able to consistently outperform the other estimators when
both the structural and the statistical model are misspecified. 

\subsection{Demand Estimation}

\begin{table}
\caption{Demand Estimation - Setup\label{tab:DemandSetup}}

\medskip{}

\centering{}%
\begin{tabular}{clll}
\toprule 
\toprule Experiment & \multicolumn{1}{c}{True Mechanism} & \multicolumn{1}{c}{Reduced-Form} & \multicolumn{1}{c}{Structural Model}\tabularnewline
\midrule 
\multirow{2}{*}{1} & linear demand, optimal & \multirow{2}{*}{linear demand} & \tabularnewline
 & monopoly pricing &  & \tabularnewline
\multirow{2}{*}{2} & linear demand, non-optimal & \multirow{2}{*}{linear demand} & \tabularnewline
 & monopoly pricing &  & linear demand, optimal\tabularnewline
\multirow{2}{*}{3} & linear demand, optimal & \multirow{2}{*}{log-log demand} & monopoly pricing\tabularnewline
 & monopoly pricing &  & \tabularnewline
\multirow{2}{*}{4} & linear demand, non-optimal & \multirow{2}{*}{log-log demand} & \tabularnewline
 & monopoly pricing &  & \tabularnewline
\bottomrule
\end{tabular}
\end{table}

In our final application, we revisit the demand estimation problem
under a different setting. Suppose now that instead of observing consumer
demand under exogenously varying prices, the prices we observe are
set by a monopolist. In this case, changes in prices are endogenous
and the relationship between price and quantity sold is confounded.
We are interested in learning the true demand curve. To this end,
if we have access to a variable that shifts the cost of production
for the monopoly firm but does not affect demand directly, then it
can be used as an instrumental variable to help identify the demand
curve. This is the reduced-form approach. Alternatively, we can estimate
a structural model that fully specifies monopoly pricing behavior.
This is the structural approach. Finally, we can combine the two using
the DRSS and the ESS-LN\footnote{For instrumental variable estimation, we do not offer an ESS-NP estimator.}. 

In this exercise, we conduct four experiments. In all four experiments,
we assume that we have access to a valid instrument so that the demand
curve is identified. However, the functional form of the reduced-form
model may still be misspecified. On the other hand, using the structural
approach, we estimate a model that assumes the observed prices are
optimally set by a profit-maximizing monopoly firm. When this assumption
is violated, as when for example the firm's pricing is not optimal
or it does not have monopoly power, the structural model will also
be misspecified. The four experiments we conduct are thus arranged
as follows: in the first experiment, both the reduced-form and the
structural models are correctly specified. In experiment 2 and 3,
only one of the two is correctly specified. In experiment 4, both
are misspecified. Table \ref{tab:DemandSetup} summarizes this setup.
For each experiment, we also simulate both a slightly confounded data
set, in which the relationship between price and quantity does not
deviate too much from the demand curve, and a highly confounded data
set, in which they look nothing alike. 

In contrast to the previous two exercises, in this exercise, we focus
on comparisons of \emph{in-domain} performance. We show that when
either the reduced-form or the structural model is misspecified, the
DRSS and the ESS-LN will have better in-domain performance -- more
\emph{internal validity} -- than the misspecified model. When both
are misspecified, the ESS-LN outperforms them both. 

\paragraph*{Setup}

Consider $M$ geographical markets in which a product is sold. The
equilibrium price and quantity sold in market $m$ are $\left(p_{m},q_{m}\right)$.
Assume that all markets share the same aggregate demand function $Q^{d}\left(p\right)$:
\begin{equation}
q_{m}=Q^{d}\left(p_{m}\right)=\alpha-\beta\cdot p_{m}+\epsilon_{m}\label{eq:demandCurve}
\end{equation}

In experiment 1 and 3, we assume the product is sold by a monopoly
firm who sets the prices in each market to maximize its profit. The
firm has different marginal costs $c_{m}$ for operating in different
markets. Hence it sets 
\begin{align}
p_{m} & =\underset{p>0}{\arg\max}\left\{ \left(p-c_{m}\right)Q^{d}\left(p\right)\right\} \label{eq:demandP1}\\
 & =c_{m}+\frac{1}{\beta}q_{m}\label{eq:demandP2}
\end{align}

Assume that we also observe a cost-shifter $z_{m}$, e.g. transportation
costs, such that 
\begin{equation}
c_{m}=a+b\cdot z_{m}\label{eq:demandZ}
\end{equation}
, then $z_{m}$ can serve as an instrument for $p_{m}$ for identifying
the demand curve. 

In experiment 2 and 4, we assume the monopoly firm fails to set optimal
prices or does not have complete monopoly power. Its pricing decisions
are given by 
\begin{equation}
p_{m}=c_{m}+\frac{\lambda}{\beta}q_{m}\label{eq:demandNonOptP}
\end{equation}
, where $\lambda\in\left(0,1\right)$. The firm thus earns a lower
markup than a optimal price-setting monopoly.

\paragraph*{Simulation }

For each experiment, we simulate two data sets. Each data set consists
of prices, quantities, and cost shifters in $M=1000$ markets, i.e.
$\text{\ensuremath{\mathcal{D}}}=\left\{ \left(p_{m},q_{m},z_{m}\right)\right\} _{m=1}^{M}$.
One data set is only slightly confounded, so that $\mathbb{E}\left[\left.q_{m}\right|p_{m}\right]$
is close to the demand relation \eqref{eq:demandCurve}. The other
is highly confounded, so that they are completely different. See Appendix
\hyperlink{APP}{B.2} for the parameter values we use in simulation. 

\paragraph*{Reduced-Form Model}

Because $p_{m}$ is now endogenous -- $p_{m}$ and $\epsilon_{m}$
are correlated through \eqref{eq:demandP2} -- the statistical relation
between $p_{m}$ and $q_{m}$ is confounded and no longer represents
the demand function. To estimate the demand curve using the reduced-form
approach, we avail of the instrumental variable $z_{m}$ and estimate
$Q^{d}\left(p\right)$ by two-stage least squares (2SLS). In experiment
1 and 2, our reduced-form model is correctly specified, i.e. we fit
\eqref{eq:demandCurve} to the data by 2SLS. In experiment 3 an 4,
however, we assume the demand function takes on a log-log form:
\begin{equation}
\log q_{m}=\alpha-\beta\cdot\log p_{m}+\epsilon_{m}\label{eq:DemandRFWRONG}
\end{equation}
, and is therefore misspecified in these two experiments. 

\paragraph*{Structural Model}

We fit a structural model featuring linear demand function \eqref{eq:demandCurve}
and price-setting function \eqref{eq:demandP2}. This structural model
is correctly specified for experiment 1 and 3, but misspecified for
experiment 2 and 4. The structural parameters are $\left(\alpha,\beta,a,b\right)$
and can be estimated as follows: from \eqref{eq:demandCurve} and
\eqref{eq:demandP2}, we obtain 
\begin{equation}
p_{m}=a+b\cdot z_{m}+\frac{1}{\beta}q_{m}\label{eq:demandSol1}
\end{equation}

If our model is correct, \eqref{eq:demandSol1} is a deterministic
linear equation system from which we can solve directly for $\left(\widehat{a},\widehat{b},\widehat{\beta}\right)$.
Substituting $\widehat{\beta}$ into \eqref{eq:demandCurve}, we then
obtain $\widehat{\alpha}=\frac{1}{M}\sum_{m=1}^{M}\left(q_{m}+\widehat{\beta}p_{m}\right)$. 

\paragraph*{Results}

\begin{figure}
\subfloat[Experiment 1\label{fig:demand_a-slight}]{\includegraphics[width=0.5\columnwidth]{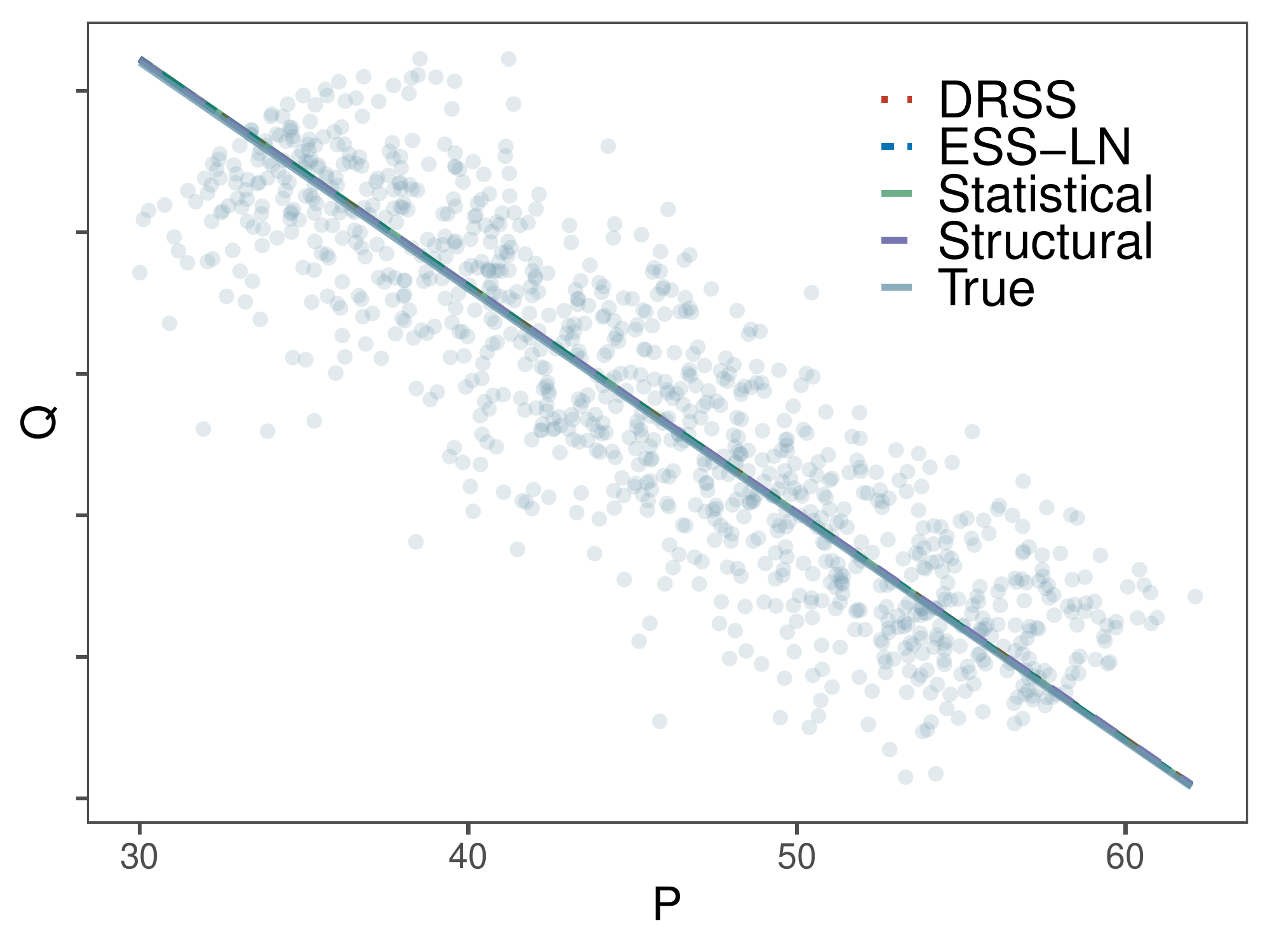}

}\subfloat[Experiment 2\label{fig:demand_b-slight}]{\includegraphics[width=0.5\columnwidth]{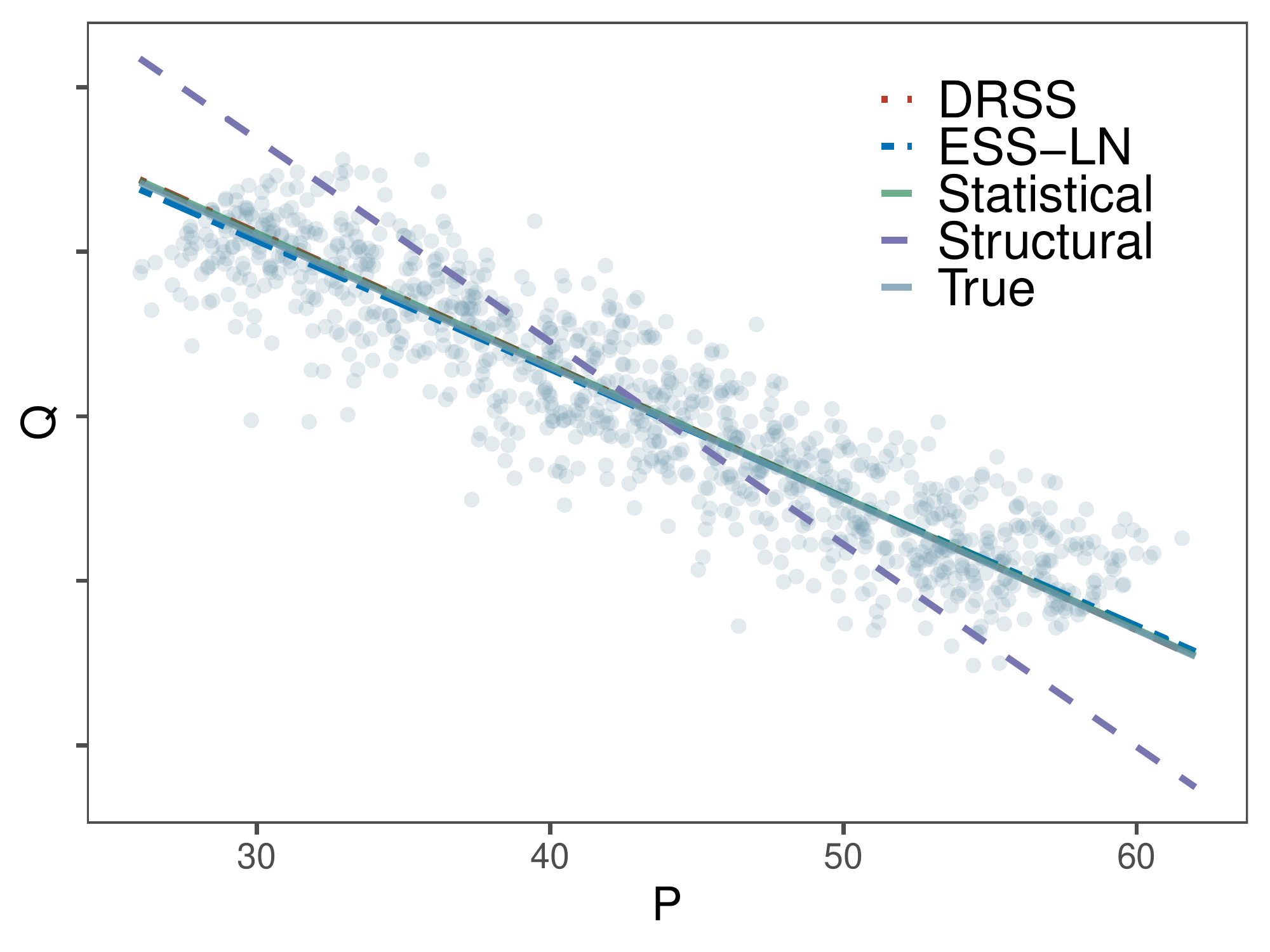}

}

\subfloat[Experiment 3\label{fig:demand_c-slight}]{\includegraphics[width=0.5\columnwidth]{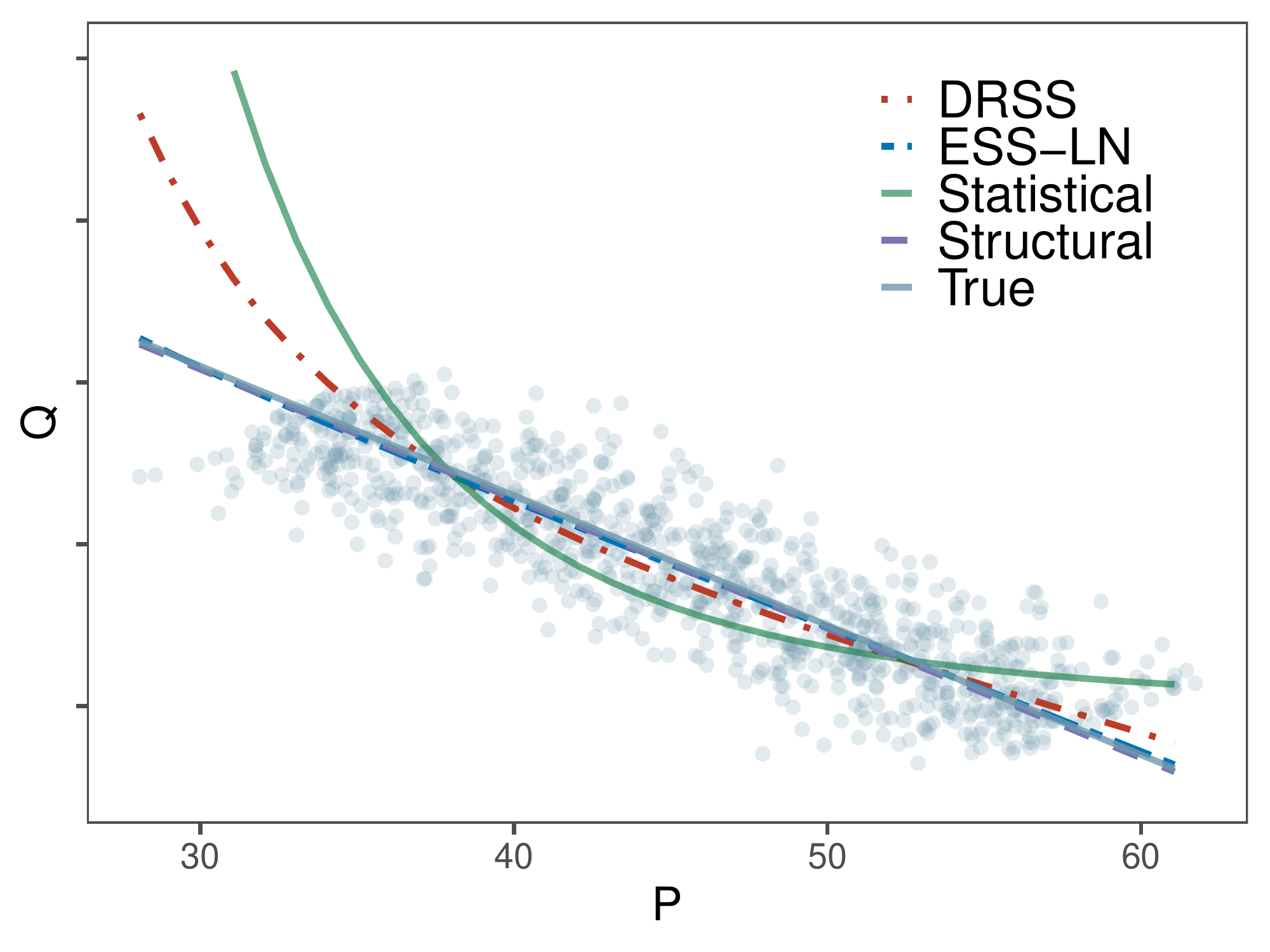}

}\subfloat[Experiment 4\label{fig:demand_d-slight}]{\includegraphics[width=0.5\columnwidth]{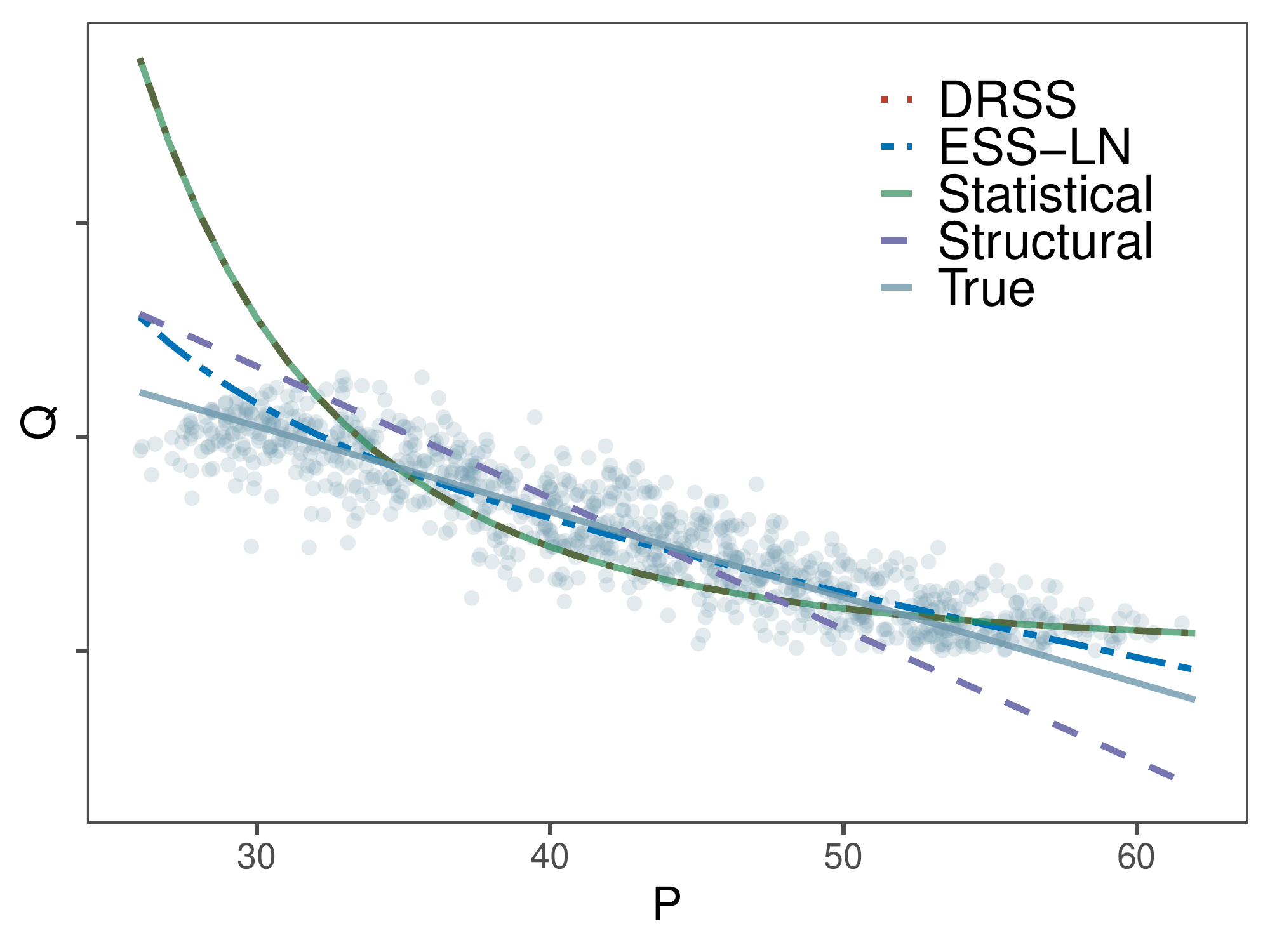}

}

\caption{{\small{}Demand Estimation -- Slightly Confounded Data\label{fig:Demand-Estimation-Slight}}}
\end{figure}

\begin{figure}
\subfloat[Experiment 1\label{fig:demand_a-high}]{\includegraphics[width=0.5\columnwidth]{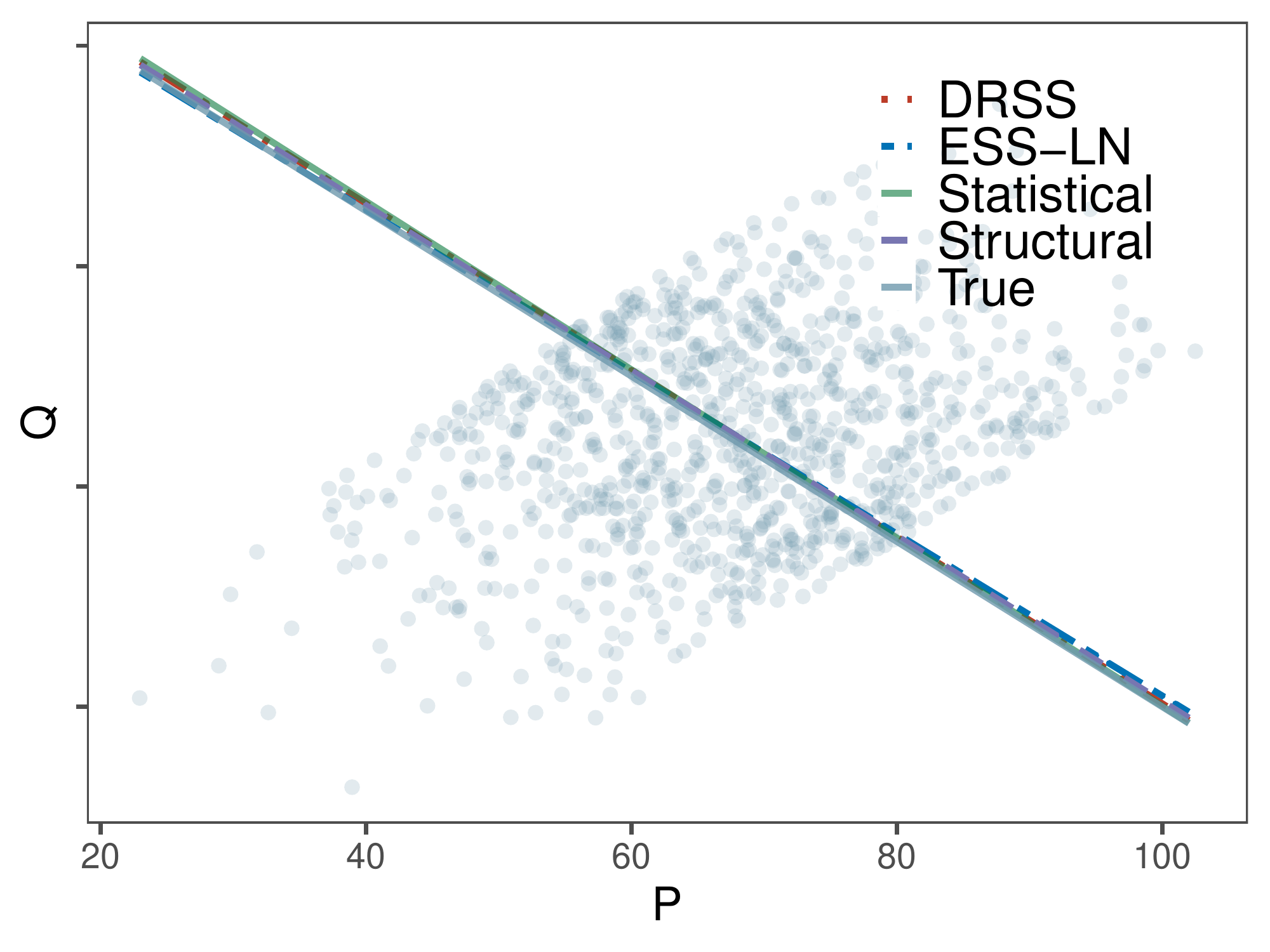}

}\subfloat[Experiment 2\label{fig:demand_b-high}]{\includegraphics[width=0.5\columnwidth]{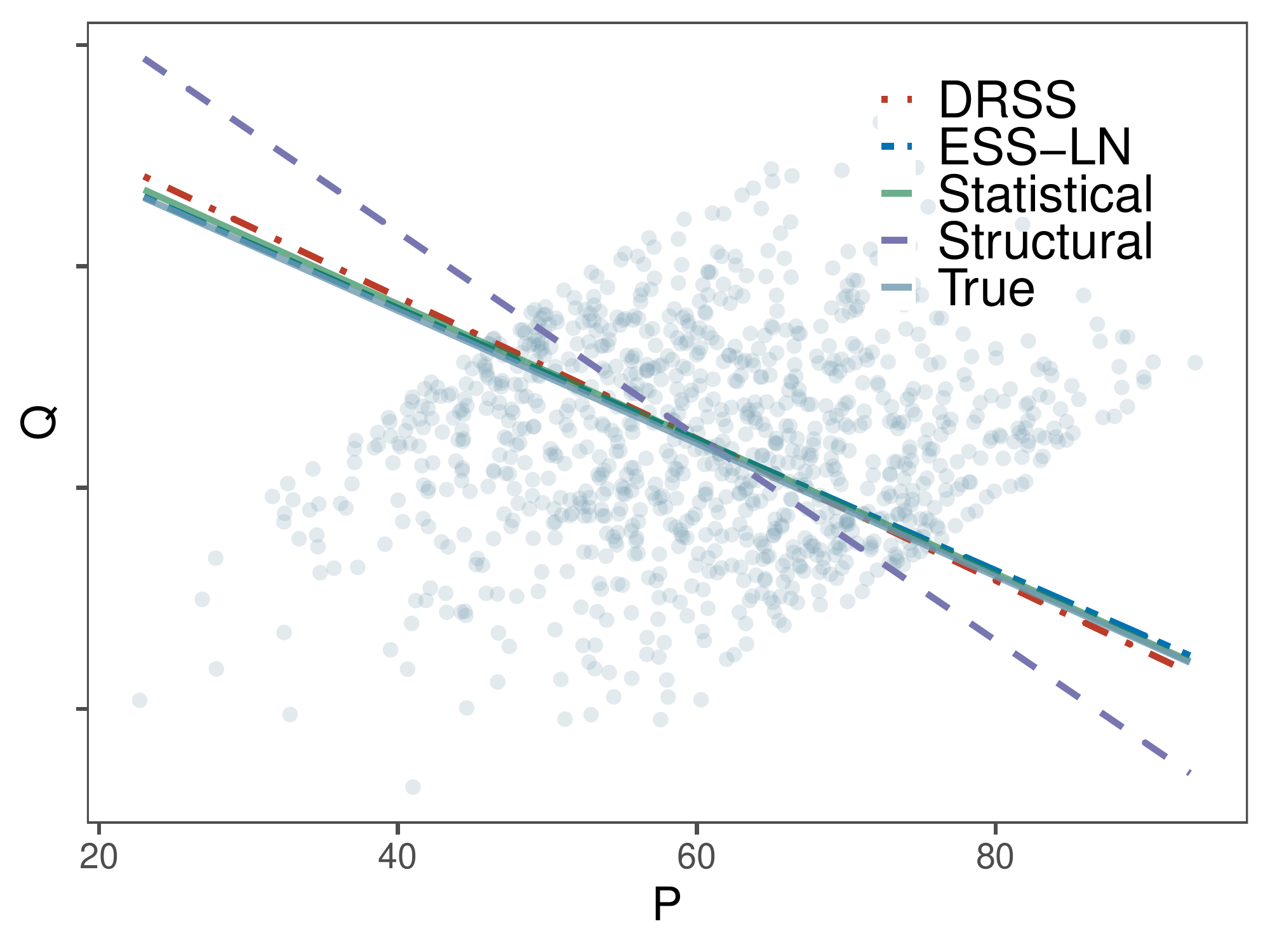}

}

\subfloat[Experiment 3\label{fig:demand_c-high}]{\includegraphics[width=0.5\columnwidth]{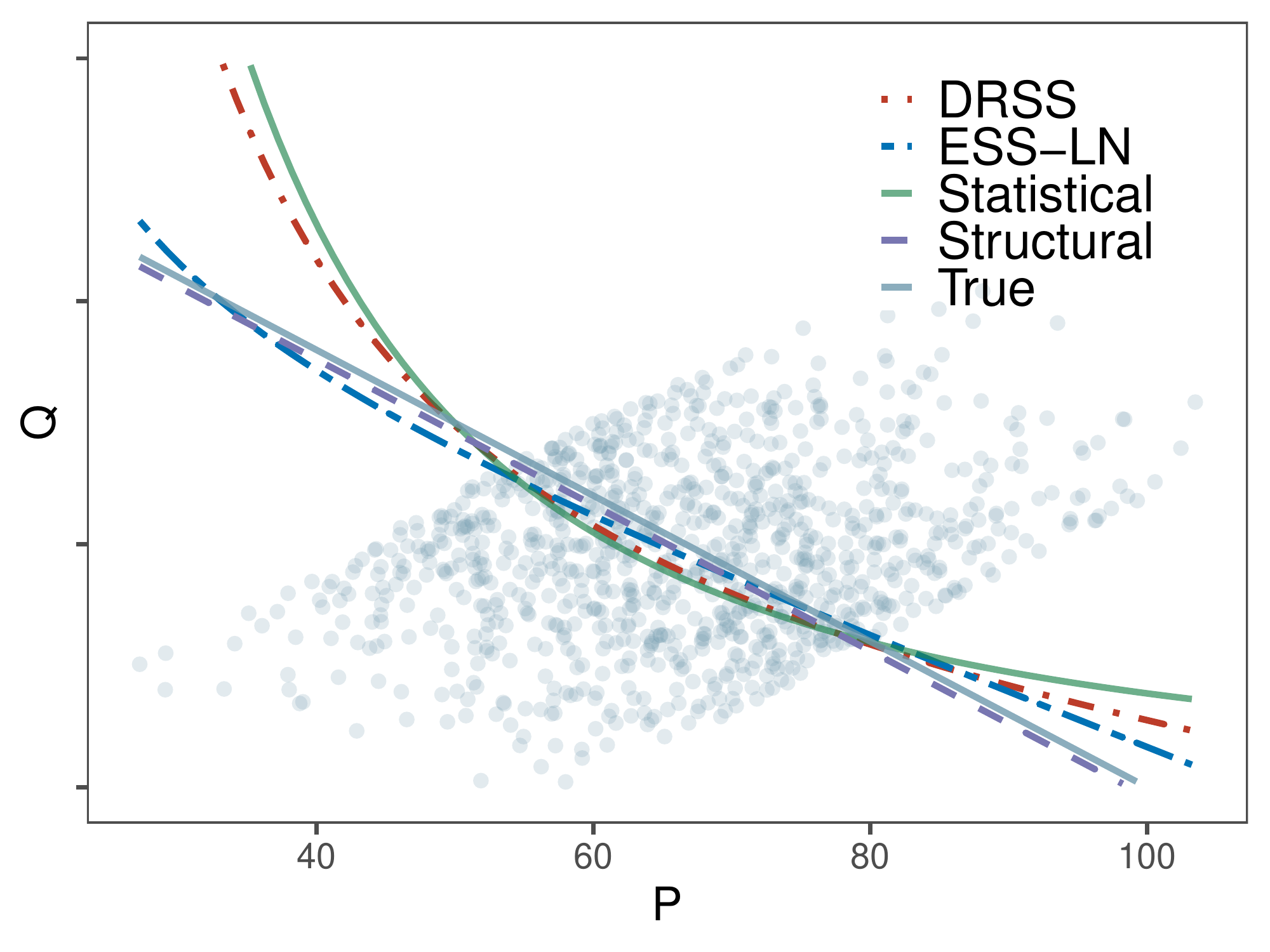}

}\subfloat[Experiment 4\label{fig:demand_d-high}]{\includegraphics[width=0.5\columnwidth]{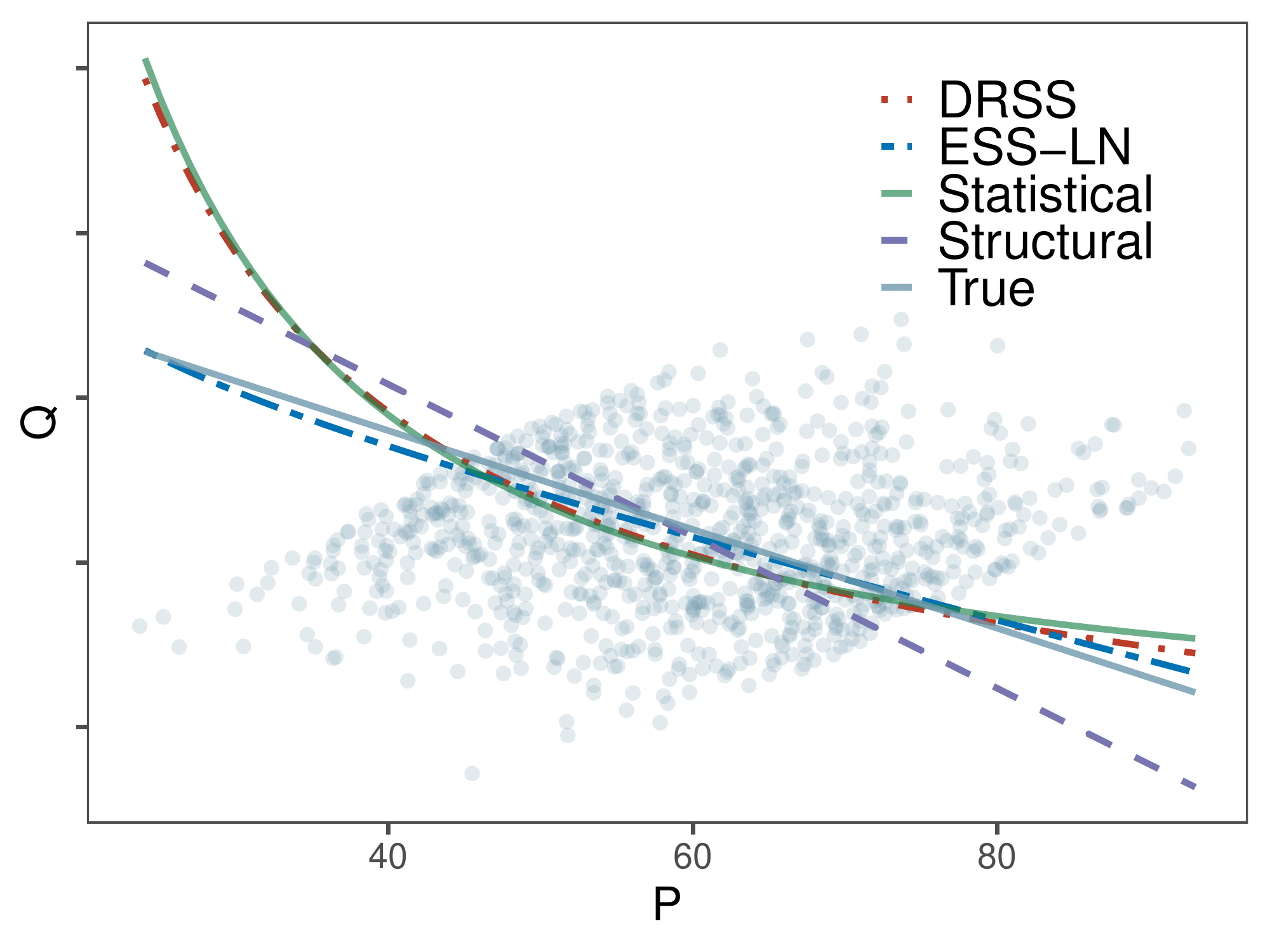}

}

\caption{{\small{}Demand Estimation -- Highly Confounded Data\label{fig:Demand-Estimation-High}}}
\end{figure}

In Figure \ref{fig:Demand-Estimation-Slight} and \ref{fig:Demand-Estimation-High},
we plot the results of the four experiments respectively for the slightly
and highly confounded scenarios. In the latter case, the observed
data $\left(p_{m},q_{m}\right)$ are significantly confounded such
that fitting a least squares model to the data would produce an upward-sloping
curve. Regardless of the level of confounding, however, the two groups
of plots tell a similar story. When correctly specified, both reduced-form
and structural estimation are able to identify the true demand curve
(Figure \ref{fig:demand_a-slight}, \ref{fig:demand_a-high})\footnote{This is because both use correctly specified models and $z$ is a
valid instrument.}. When only one of them is correctly specified, the misspecified model
produces fits that, while still managing to capture the downward-sloping
nature of the demand curve, can deviate significantly from the true
relationship (Figure \ref{fig:demand_b-slight}, \ref{fig:demand_b-high},
\ref{fig:demand_c-slight}, \ref{fig:demand_c-high}). In this case,
the ESS-LN generally still performs well, while the DRSS is able to
fit the demand curve well in Figure \ref{fig:demand_b-slight} and
\ref{fig:demand_b-high} but not in \ref{fig:demand_c-slight} and
\ref{fig:demand_c-high}. Finally, when both the reduced-form and
the structural models are misspecified, the ESS-LN becomes the only
method that is able to fit the true demand curve well (Figure \ref{fig:demand_d-slight},
\ref{fig:demand_d-high}).

\begin{table}
\centering

\begin{threeparttable}

\caption{Demand Estimation - Results\tnote{a}\label{tab:demandTable}}
\medskip{}

\begin{tabular}{lrrrrrrr}
\toprule 
\toprule & \multicolumn{3}{c}{Slightly Confounded} &  & \multicolumn{3}{c}{Highly Confounded}\tabularnewline
\cmidrule{2-4} \cmidrule{3-4} \cmidrule{4-4} \cmidrule{6-8} \cmidrule{7-8} \cmidrule{8-8} 
 & MSE & Bias & Var & \  & MSE & Bias & Var\tabularnewline
\midrule 
 &  &  &  &  &  &  & \tabularnewline
\multicolumn{8}{l}{\textit{Experiment 1}}\tabularnewline
 &  &  &  &  &  &  & \tabularnewline
Structural & 0.90 & 0.75 & 0.90 &  & 1.01 & 0.81 & 1.00\tabularnewline
Statistical & 2.34 & 1.16 & 2.36 &  & 14.37 & 2.70 & 14.43\tabularnewline
DRSS & 1.52 & 0.94 & 1.53 &  & 6.34 & 1.71 & 6.37\tabularnewline
ESS-LN & 2.06 & 1.07 & 2.05 &  & 13.99 & 2.53 & 13.93\tabularnewline
 &  &  &  &  &  &  & \tabularnewline
\midrule
 &  &  &  &  &  &  & \tabularnewline
\multicolumn{8}{l}{\textit{Experiment 2}}\tabularnewline
 &  &  &  &  &  &  & \tabularnewline
Structural & 2394.80 & 42.04 & 3.45 &  & 767.90 & 23.80 & 1.52\tabularnewline
Statistical & 1.377 & .898 & 1.38 &  & 17.83 & 2.93 & 17.99\tabularnewline
DRSS & 1.696 & .987 & 1.57 &  & 143.20 & 7.12 & 103.33\tabularnewline
ESS-LN & 1.701 & .990 & 1.72 &  & 18.11 & 2.97 & 18.29\tabularnewline
 &  &  &  &  &  &  & \tabularnewline
\midrule
 &  &  &  &  &  &  & \tabularnewline
\multicolumn{8}{l}{\textit{Experiment 3}}\tabularnewline
 &  &  &  &  &  &  & \tabularnewline
Structural & 0.85  & 0.76  & 0.86  &  & 1.01 & 0.81 & 1.00\tabularnewline
Statistical & 8062.98  & 50.29  & 99.73 &  & 329.33 & 13.60 & 2.43\tabularnewline
DRSS & 36.82  & 2.21 & 26.76 &  & 141.50 & 8.47 & 16.12\tabularnewline
ESS-LN & 11.25 & 1.97  & 10.96 &  & 137.87 & 7.07 & 139.20\tabularnewline
 &  &  &  &  &  &  & \tabularnewline
\midrule
 &  &  &  &  &  &  & \tabularnewline
\multicolumn{8}{l}{\textit{Experiment 4}}\tabularnewline
 &  &  &  &  &  &  & \tabularnewline
Structural & 2394.80 & 42.40 & 3.45 &  & 767.90 & 23.80 & 1.52\tabularnewline
Statistical & 1395.50 & 30.37 & 10.00 &  & 447.78 & 16.27 & 3.30\tabularnewline
DRSS & 1100.62 & 25.72 & 20.94 &  & 375.08 & 14.90 & 233.92\tabularnewline
ESS-LN & 3.55 & 1.40 & 3.53 &  & 168.19 & 8.41 & 169.70\tabularnewline
 &  &  &  &  &  &  & \tabularnewline
\bottomrule
\end{tabular}

\medskip{}

\begin{tablenotes} 
\small
\item [a] Results are based on 100 simulation trials. All numbers are on the scale of $10^{-4}$.
\end{tablenotes}

\end{threeparttable}
\end{table}

Table \ref{tab:demandTable} reports the bias, variance, and mean
squared error of the estimators with respect to the true demand curve
over 100 trials. In both the slightly and highly confounded scenarios,
when they are correctly specified, the reduced-form and the structural
models exhibit low biases. The structural model, by virtue of imposing
more structure on the data, attains a lower variance. When misspecified,
both types of models exhibit large biases and MSEs. The DRSS is able
to outperform the misspecified model in experiment 2 and 3, while
the ESS-LN consistently achieves the lowest MSE -- often significantly
lower than those of the other estimators, regardless of which model
-- the reduced-form or the structural or even both -- is misspecified.
Note, however, for all experiments, the DRSS and the ESS-LN perform
better on the slightly confounded data. This is not surprising. In
particular, as Figure \ref{fig:Demand-Estimation-High} reveals, when
the data are highly confounded, the structural and the reduced-form
models can behave similarly on the observed data, even when their
predicted demand curves are actually very different due to one or
both of them being misspecified, making it difficult for the DRSS
method to distinguish between them and for the ESS-LN to leverage
their differences in functional form. More confounding thus presents
more challenges for our methods to work well. 

\section{Conclusion\label{sec:Conclusion}}

In this paper, we propose a set of methods for combining statistical
and structural models for improved prediction and causal inference.
We demonstrate the effectiveness of our methods in a number of economic
applications including first-price auctions, dynamic models of entry
and exit, and demand estimation with instrumental variables. Our methods
offer a way to bridge the gap between the (reduced-form) statistical
approach and the structural approach in economic analysis and have
potentially wide applications in addressing problems for which significant
concerns about model misspecification exist.

\bibliographystyle{apa}
\bibliography{mybib}

\begin{thebibliography}{}

\bibitem[\protect\astroncite{Aguirregabiria and
  Magesan}{2013}]{aguirregabiria_euler_2013}
Aguirregabiria, V. and Magesan, A. (2013).
\newblock Euler equations for the estimation of dynamic discrete choice
  structural models.
\newblock {\em Advances in Econometrics}, 31:3--44.

\bibitem[\protect\astroncite{Aguirregabiria and
  Mira}{2010}]{aguirregabiria_dynamic_2010}
Aguirregabiria, V. and Mira, P. (2010).
\newblock Dynamic discrete choice structural models: {A} survey.
\newblock {\em Journal of Econometrics}, 156(1):38--67.
\newblock Publisher: Elsevier.

\bibitem[\protect\astroncite{Ando and Li}{2017}]{ando_weight-relaxed_2017}
Ando, T. and Li, K.-C. (2017).
\newblock A weight-relaxed model averaging approach for high-dimensional
  generalized linear models.
\newblock {\em The Annals of Statistics}, 45(6):2654--2679.
\newblock Publisher: Institute of Mathematical Statistics.

\bibitem[\protect\astroncite{Angrist and
  Krueger}{1995}]{angrist_split-sample_1995}
Angrist, J.~D. and Krueger, A.~B. (1995).
\newblock Split-{Sample} {Instrumental} {Variables} {Estimates} of the {Return}
  to {Schooling}.
\newblock {\em Journal of Business \& Economic Statistics}, 13(2):225--235.
\newblock Publisher: Taylor \& Francis.

\bibitem[\protect\astroncite{Angrist and
  Pischke}{2010}]{angrist_credibility_2010}
Angrist, J.~D. and Pischke, J.-S. (2010).
\newblock The credibility revolution in empirical economics: {How} better
  research design is taking the con out of econometrics.
\newblock {\em Journal of economic perspectives}, 24(2):3--30.

\bibitem[\protect\astroncite{Arcidiacono and
  Ellickson}{2011}]{arcidiacono_practical_2011}
Arcidiacono, P. and Ellickson, P.~B. (2011).
\newblock Practical {Methods} for {Estimation} of {Dynamic} {Discrete} {Choice}
  {Models}.
\newblock {\em Annual Review of Economics}, 3(1):363--394.
\newblock \_eprint: https://doi.org/10.1146/annurev-economics-111809-125038.

\bibitem[\protect\astroncite{Arcidiacono and
  Miller}{2011}]{arcidiacono_conditional_2011}
Arcidiacono, P. and Miller, R.~A. (2011).
\newblock Conditional choice probability estimation of dynamic discrete choice
  models with unobserved heterogeneity.
\newblock {\em Econometrica}, 79(6):1823--1867.
\newblock Publisher: Wiley Online Library.

\bibitem[\protect\astroncite{Arkhangelsky and
  Imbens}{2019}]{arkhangelsky_double-robust_2019}
Arkhangelsky, D. and Imbens, G.~W. (2019).
\newblock Double-robust identification for causal panel data models.
\newblock {\em arXiv preprint arXiv:1909.09412}.

\bibitem[\protect\astroncite{Artuc et~al.}{2010}]{artuc_trade_2010}
Artuc, E., Chaudhuri, S., and McLaren, J. (2010).
\newblock Trade {Shocks} and {Labor} {Adjustment}: {A} {Structural} {Empirical}
  {Approach}.
\newblock {\em American Economic Review}, 100(3):1008--1045.

\bibitem[\protect\astroncite{Athey}{2017}]{athey_beyond_2017}
Athey, S. (2017).
\newblock Beyond prediction: {Using} big data for policy problems.
\newblock {\em Science}, 355(6324):483--485.
\newblock Publisher: American Association for the Advancement of Science.

\bibitem[\protect\astroncite{Athey and Haile}{2007}]{athey_nonparametric_2007}
Athey, S. and Haile, P.~A. (2007).
\newblock Nonparametric approaches to auctions.
\newblock {\em Handbook of econometrics}, 6:3847--3965.
\newblock Publisher: Elsevier.

\bibitem[\protect\astroncite{Athey and Imbens}{2017}]{athey_state_2017}
Athey, S. and Imbens, G.~W. (2017).
\newblock The state of applied econometrics: {Causality} and policy evaluation.
\newblock {\em Journal of Economic Perspectives}, 31(2):3--32.

\bibitem[\protect\astroncite{Athey et~al.}{2019}]{athey_generalized_2019}
Athey, S., Tibshirani, J., and Wager, S. (2019).
\newblock Generalized random forests.
\newblock {\em Annals of Statistics}, 47(2):1148--1178.
\newblock Publisher: Institute of Mathematical Statistics.

\bibitem[\protect\astroncite{Bajari et~al.}{2013}]{bajari_game_2013}
Bajari, P., Hong, H., and Nekipelov, D. (2013).
\newblock Game theory and econometrics: {A} survey of some recent research.
\newblock In {\em Advances in economics and econometrics, 10th world congress},
  volume~3, pages 3--52.

\bibitem[\protect\astroncite{Bajari and Hortacsu}{2005}]{bajari_are_2005}
Bajari, P. and Hortacsu, A. (2005).
\newblock Are {Structural} {Estimates} of {Auction} {Models} {Reasonable}?
  {Evidence} from {Experimental} {Data}.
\newblock {\em Journal of Political Economy}, 113(4):703--741.
\newblock Publisher: The University of Chicago Press.

\bibitem[\protect\astroncite{Bang and Robins}{2005}]{bang_doubly_2005}
Bang, H. and Robins, J.~M. (2005).
\newblock Doubly robust estimation in missing data and causal inference models.
\newblock {\em Biometrics}, 61(4):962--973.
\newblock Publisher: Wiley Online Library.

\bibitem[\protect\astroncite{Ben-David et~al.}{2010}]{ben-david_theory_2010}
Ben-David, S., Blitzer, J., Crammer, K., Kulesza, A., Pereira, F., and Vaughan,
  J.~W. (2010).
\newblock A theory of learning from different domains.
\newblock {\em Machine Learning}, 79(1):151--175.

\bibitem[\protect\astroncite{Benkeser et~al.}{2017}]{benkeser_doubly_2017}
Benkeser, D., Carone, M., Laan, M. V.~D., and Gilbert, P.~B. (2017).
\newblock Doubly robust nonparametric inference on the average treatment
  effect.
\newblock {\em Biometrika}, 104(4):863--880.
\newblock Publisher: Oxford University Press.

\bibitem[\protect\astroncite{Bernardo and Smith}{2009}]{bernardo_bayesian_2009}
Bernardo, J.~M. and Smith, A.~F. (2009).
\newblock {\em Bayesian theory}, volume 405.
\newblock John Wiley \& Sons.

\bibitem[\protect\astroncite{Biau}{2012}]{biau_analysis_2012}
Biau, G. (2012).
\newblock Analysis of a {Random} {Forests} {Model}.
\newblock {\em Journal of Machine Learning Research}, 13(38):1063--1095.

\bibitem[\protect\astroncite{Biau and Scornet}{2016}]{biau_random_2016}
Biau, G. and Scornet, E. (2016).
\newblock A random forest guided tour.
\newblock {\em Test}, 25(2):197--227.
\newblock Publisher: Springer.

\bibitem[\protect\astroncite{Bishop and
  Lasserre}{2007}]{bishop_generative_2007}
Bishop, C.~M. and Lasserre, J. (2007).
\newblock Generative or discriminative? getting the best of both worlds.
\newblock {\em Bayesian statistics}, 8(3):3--24.

\bibitem[\protect\astroncite{Breiman}{1996a}]{breiman_bagging_1996}
Breiman, L. (1996a).
\newblock Bagging predictors.
\newblock {\em Machine learning}, 24(2):123--140.
\newblock Publisher: Springer.

\bibitem[\protect\astroncite{Breiman}{1996b}]{breiman_stacked_1996}
Breiman, L. (1996b).
\newblock Stacked regressions.
\newblock {\em Machine learning}, 24(1):49--64.
\newblock Publisher: Springer.

\bibitem[\protect\astroncite{Breiman}{2001}]{breiman_random_2001}
Breiman, L. (2001).
\newblock Random forests.
\newblock {\em Machine learning}, 45(1):5--32.
\newblock Publisher: Springer.

\bibitem[\protect\astroncite{Chernozhukov
  et~al.}{2017}]{chernozhukov_doubledebiasedneyman_2017}
Chernozhukov, V., Chetverikov, D., Demirer, M., Duflo, E., Hansen, C., and
  Newey, W. (2017).
\newblock Double/debiased/neyman machine learning of treatment effects.
\newblock {\em American Economic Review}, 107(5):261--65.

\bibitem[\protect\astroncite{Chernozhukov
  et~al.}{2016}]{chernozhukov_double_2016}
Chernozhukov, V., Chetverikov, D., Demirer, M., Duflo, E., Hansen, C., and
  Newey, W.~K. (2016).
\newblock Double machine learning for treatment and causal parameters.
\newblock Technical report, cemmap working paper.

\bibitem[\protect\astroncite{Chetty}{2009}]{chetty_sufficient_2009}
Chetty, R. (2009).
\newblock Sufficient {Statistics} for {Welfare} {Analysis}: {A} {Bridge}
  {Between} {Structural} and {Reduced}-{Form} {Methods}.
\newblock {\em Annual Review of Economics}, 1(1):451--488.

\bibitem[\protect\astroncite{Chopra et~al.}{2013}]{chopra_dlid_2013}
Chopra, S., Balakrishnan, S., and Gopalan, R. (2013).
\newblock Dlid: {Deep} learning for domain adaptation by interpolating between
  domains.
\newblock In {\em {ICML} workshop on challenges in representation learning},
  volume~2.

\bibitem[\protect\astroncite{Claeskens and
  Hjort}{2003}]{claeskens_focused_2003}
Claeskens, G. and Hjort, N.~L. (2003).
\newblock The focused information criterion.
\newblock {\em Journal of the American Statistical Association},
  98(464):900--916.
\newblock Publisher: Taylor \& Francis.

\bibitem[\protect\astroncite{Clyde and Iversen}{2013}]{clyde_bayesian_2013}
Clyde, M. and Iversen, E.~S. (2013).
\newblock Bayesian model averaging in the {M}-open framework.
\newblock {\em Bayesian theory and applications}, pages 483--498.
\newblock Publisher: Oxford University Press Oxford, UK.

\bibitem[\protect\astroncite{Deaton}{2010}]{deaton_instruments_2010}
Deaton, A. (2010).
\newblock Instruments, randomization, and learning about development.
\newblock {\em Journal of economic literature}, 48(2):424--55.

\bibitem[\protect\astroncite{Dietterich}{2000}]{dietterich_ensemble_2000}
Dietterich, T.~G. (2000).
\newblock Ensemble methods in machine learning.
\newblock In {\em International workshop on multiple classifier systems}, pages
  1--15. Springer.

\bibitem[\protect\astroncite{Farrell}{2015}]{farrell_robust_2015}
Farrell, M.~H. (2015).
\newblock Robust inference on average treatment effects with possibly more
  covariates than observations.
\newblock {\em Journal of Econometrics}, 189(1):1--23.
\newblock Publisher: Elsevier.

\bibitem[\protect\astroncite{Fessler and Kasy}{2019}]{fessler_how_2019}
Fessler, P. and Kasy, M. (2019).
\newblock How to {Use} {Economic} {Theory} to {Improve} {Estimators}:
  {Shrinking} {Toward} {Theoretical} {Restrictions}.
\newblock {\em The Review of Economics and Statistics}, 101(4):681--698.
\newblock Publisher: MIT Press.

\bibitem[\protect\astroncite{Freund and
  Schapire}{1996}]{freund_experiments_1996}
Freund, Y. and Schapire, R.~E. (1996).
\newblock Experiments with a new boosting algorithm.
\newblock In {\em icml}, volume~96, pages 148--156. Citeseer.

\bibitem[\protect\astroncite{Ganin and
  Lempitsky}{2014}]{ganin_unsupervised_2014}
Ganin, Y. and Lempitsky, V. (2014).
\newblock Unsupervised domain adaptation by backpropagation.
\newblock {\em arXiv preprint arXiv:1409.7495}.

\bibitem[\protect\astroncite{Glorot et~al.}{2011}]{glorot_domain_2011}
Glorot, X., Bordes, A., and Bengio, Y. (2011).
\newblock Domain adaptation for large-scale sentiment classification: {A} deep
  learning approach.

\bibitem[\protect\astroncite{Gopalan et~al.}{2011}]{gopalan_domain_2011}
Gopalan, R., Li, R., and Chellappa, R. (2011).
\newblock Domain adaptation for object recognition: {An} unsupervised approach.
\newblock In {\em 2011 international conference on computer vision}, pages
  999--1006. IEEE.

\bibitem[\protect\astroncite{Guerre et~al.}{2000}]{guerre_optimal_2000}
Guerre, E., Perrigne, I., and Vuong, Q. (2000).
\newblock Optimal nonparametric estimation of first-price auctions.
\newblock {\em Econometrica}, 68(3):525--574.
\newblock Publisher: Wiley Online Library.

\bibitem[\protect\astroncite{Hansen}{2007}]{hansen_least_2007}
Hansen, B.~E. (2007).
\newblock Least squares model averaging.
\newblock {\em Econometrica}, 75(4):1175--1189.
\newblock Publisher: Wiley Online Library.

\bibitem[\protect\astroncite{Hansen and Racine}{2012}]{hansen_jackknife_2012}
Hansen, B.~E. and Racine, J.~S. (2012).
\newblock Jackknife model averaging.
\newblock {\em Journal of Econometrics}, 167(1):38--46.
\newblock Publisher: Elsevier.

\bibitem[\protect\astroncite{Hansen}{1982}]{hansen_large_1982}
Hansen, L.~P. (1982).
\newblock Large sample properties of generalized method of moments estimators.
\newblock {\em Econometrica: Journal of the Econometric Society}, pages
  1029--1054.
\newblock Publisher: JSTOR.

\bibitem[\protect\astroncite{Hansen}{2015}]{hansen_method_2015}
Hansen, L.~P. (2015).
\newblock Method of {Moments} and {Generalized} {Method} of {Moments}.
\newblock In Wright, J.~D., editor, {\em International {Encyclopedia} of the
  {Social} \& {Behavioral} {Sciences} ({Second} {Edition})}, pages 294--301.
  Elsevier, Oxford.

\bibitem[\protect\astroncite{Hastie et~al.}{2009}]{hastie_elements_2009}
Hastie, T., Tibshirani, R., and Friedman, J. (2009).
\newblock {\em The elements of statistical learning: data mining, inference,
  and prediction}.
\newblock Springer Science \& Business Media.

\bibitem[\protect\astroncite{Heckman}{2000}]{heckman_causal_2000}
Heckman, J.~J. (2000).
\newblock Causal parameters and policy analysis in economics: {A} twentieth
  century retrospective.
\newblock {\em The Quarterly Journal of Economics}, 115(1):45--97.

\bibitem[\protect\astroncite{Heckman}{2010}]{heckman_building_2010}
Heckman, J.~J. (2010).
\newblock Building bridges between structural and program evaluation approaches
  to evaluating policy.
\newblock {\em Journal of Economic literature}, 48(2):356--98.

\bibitem[\protect\astroncite{Heckman and
  Vytlacil}{2007}]{heckman_econometric_2007}
Heckman, J.~J. and Vytlacil, E.~J. (2007).
\newblock Econometric {Evaluation} of {Social} {Programs}, {Part} {I}: {Causal}
  {Models}, {Structural} {Models} and {Econometric} {Policy} {Evaluation}.
\newblock In Heckman, J.~J. and Leamer, E.~E., editors, {\em Handbook of
  {Econometrics}}, volume~6, pages 4779--4874. Elsevier.

\bibitem[\protect\astroncite{Hickman et~al.}{2012}]{hickman_structural_2012}
Hickman, B.~R., Hubbard, T.~P., and Saglam, Y. (2012).
\newblock Structural econometric methods in auctions: {A} guide to the
  literature.
\newblock {\em Journal of Econometric Methods}, 1(1):67--106.
\newblock Publisher: De Gruyter.

\bibitem[\protect\astroncite{Hjort and
  Claeskens}{2003}]{hjort_frequentist_2003}
Hjort, N.~L. and Claeskens, G. (2003).
\newblock Frequentist model average estimators.
\newblock {\em Journal of the American Statistical Association},
  98(464):879--899.
\newblock Publisher: Taylor \& Francis.

\bibitem[\protect\astroncite{Hoeting et~al.}{1999}]{hoeting_bayesian_1999}
Hoeting, J.~A., Madigan, D., Raftery, A.~E., and Volinsky, C.~T. (1999).
\newblock Bayesian {Model} {Averaging}: {A} {Tutorial}.
\newblock {\em Statistical Science}, 14(4):382--401.
\newblock Publisher: Institute of Mathematical Statistics.

\bibitem[\protect\astroncite{Huang et~al.}{2007}]{huang_correcting_2007}
Huang, J., Gretton, A., Borgwardt, K., Scholkopf, B., and Smola, A.~J. (2007).
\newblock Correcting {Sample} {Selection} {Bias} by {Unlabeled} {Data}.
\newblock In Scholkopf, B., Platt, J.~C., and Hoffman, T., editors, {\em
  Advances in {Neural} {Information} {Processing} {Systems} 19}, pages
  601--608. MIT Press.

\bibitem[\protect\astroncite{Imbens and Rubin}{2015}]{imbens_causal_2015}
Imbens, G.~W. and Rubin, D.~B. (2015).
\newblock {\em Causal inference in statistics, social, and biomedical
  sciences}.
\newblock Cambridge University Press.

\bibitem[\protect\astroncite{Jebara}{2012}]{jebara_machine_2012}
Jebara, T. (2012).
\newblock {\em Machine learning: discriminative and generative}, volume 755.
\newblock Springer Science \& Business Media.

\bibitem[\protect\astroncite{Jiang and Zhai}{2007}]{jiang_instance_2007}
Jiang, J. and Zhai, C. (2007).
\newblock Instance weighting for domain adaptation in {NLP}.
\newblock In {\em Proceedings of the 45th annual meeting of the association of
  computational linguistics}, pages 264--271.

\bibitem[\protect\astroncite{Kang and Schafer}{2007}]{kang_demystifying_2007}
Kang, J.~D. and Schafer, J.~L. (2007).
\newblock Demystifying double robustness: {A} comparison of alternative
  strategies for estimating a population mean from incomplete data.
\newblock {\em Statistical science}, 22(4):523--539.
\newblock Publisher: Institute of Mathematical Statistics.

\bibitem[\protect\astroncite{Keane}{2010a}]{keane_structural_2010}
Keane, M.~P. (2010a).
\newblock A structural perspective on the experimentalist school.
\newblock {\em Journal of Economic Perspectives}, 24(2):47--58.

\bibitem[\protect\astroncite{Keane}{2010b}]{keane_structural_2010-1}
Keane, M.~P. (2010b).
\newblock Structural vs. atheoretic approaches to econometrics.
\newblock {\em Journal of Econometrics}, 156(1):3--20.
\newblock Publisher: Elsevier.

\bibitem[\protect\astroncite{Kellogg et~al.}{2020}]{kellogg_combining_2020}
Kellogg, M., Mogstad, M., Pouliot, G., and Torgovitsky, A. (2020).
\newblock Combining {Matching} and {Synthetic} {Controls} to {Trade} off
  {Biases} from {Extrapolation} and {Interpolation}.
\newblock Technical report, National Bureau of Economic Research.

\bibitem[\protect\astroncite{Kitagawa and Muris}{2016}]{kitagawa_model_2016}
Kitagawa, T. and Muris, C. (2016).
\newblock Model averaging in semiparametric estimation of treatment effects.
\newblock {\em Journal of Econometrics}, 193(1):271--289.
\newblock Publisher: Elsevier.

\bibitem[\protect\astroncite{Kuang et~al.}{2020}]{kuang_stable_2020}
Kuang, K., Xiong, R., Cui, P., Athey, S., and Li, B. (2020).
\newblock Stable {Prediction} with {Model} {Misspecification} and {Agnostic}
  {Distribution} {Shift}.
\newblock {\em arXiv:2001.11713 [cs, stat]}.
\newblock arXiv: 2001.11713.

\bibitem[\protect\astroncite{Lewbel et~al.}{2019}]{lewbel_general_2019}
Lewbel, A., Choi, J.-Y., and Zhou, Z. (2019).
\newblock General {Doubly} {Robust} {Identification} and {Estimation}.
\newblock {\em Working Paper}.

\bibitem[\protect\astroncite{Li}{1987}]{li_asymptotic_1987}
Li, K.-C. (1987).
\newblock Asymptotic optimality for {Cp}, {CL}, cross-validation and
  generalized cross-validation: discrete index set.
\newblock {\em The Annals of Statistics}, pages 958--975.
\newblock Publisher: JSTOR.

\bibitem[\protect\astroncite{Loh}{2014}]{loh_fifty_2014}
Loh, W.-Y. (2014).
\newblock Fifty years of classification and regression trees.
\newblock {\em International Statistical Review}, 82(3):329--348.
\newblock Publisher: Wiley Online Library.

\bibitem[\protect\astroncite{Long et~al.}{2015}]{long_learning_2015}
Long, M., Cao, Y., Wang, J., and Jordan, M.~I. (2015).
\newblock Learning transferable features with deep adaptation networks.
\newblock {\em arXiv preprint arXiv:1502.02791}.

\bibitem[\protect\astroncite{Low and Meghir}{2017}]{low_use_2017}
Low, H. and Meghir, C. (2017).
\newblock The use of structural models in econometrics.
\newblock {\em Journal of Economic Perspectives}, 31(2):33--58.

\bibitem[\protect\astroncite{Mao and Zheng}{2020}]{mao_structural_2020}
Mao, J. and Zheng, Z. (2020).
\newblock Structural {Regularization}.
\newblock {\em arXiv:2004.12601 [econ]}.
\newblock arXiv: 2004.12601.

\bibitem[\protect\astroncite{Minka}{2000}]{minka_bayesian_2000}
Minka, T.~P. (2000).
\newblock Bayesian model averaging is not model combination.
\newblock {\em Available electronically at http://www. stat. cmu.
  edu/minka/papers/bma. html}, pages 1--2.

\bibitem[\protect\astroncite{Moral-Benito}{2015}]{moral-benito_model_2015}
Moral-Benito, E. (2015).
\newblock Model {Averaging} in {Economics}: {An} {Overview}.
\newblock {\em Journal of Economic Surveys}, 29(1):46--75.
\newblock \_eprint: https://onlinelibrary.wiley.com/doi/pdf/10.1111/joes.12044.

\bibitem[\protect\astroncite{Muandet et~al.}{2013}]{muandet_domain_2013}
Muandet, K., Balduzzi, D., and Scholkopf, B. (2013).
\newblock Domain generalization via invariant feature representation.
\newblock In {\em Proceedings of the 30th {International} {Conference} on
  {International} {Conference} on {Machine} {Learning} - {Volume} 28},
  {ICML}'13, pages I--10--I--18, Atlanta, GA, USA. JMLR.org.

\bibitem[\protect\astroncite{Muth}{1961}]{muth_rational_1961}
Muth, J.~F. (1961).
\newblock Rational expectations and the theory of price movements.
\newblock {\em Econometrica: Journal of the Econometric Society}, pages
  315--335.
\newblock Publisher: JSTOR.

\bibitem[\protect\astroncite{Nevo and Whinston}{2010}]{nevo_taking_2010}
Nevo, A. and Whinston, M.~D. (2010).
\newblock Taking the dogma out of econometrics: {Structural} modeling and
  credible inference.
\newblock {\em Journal of Economic Perspectives}, 24(2):69--82.

\bibitem[\protect\astroncite{Newey}{2013}]{newey_nonparametric_2013}
Newey, W.~K. (2013).
\newblock Nonparametric {Instrumental} {Variables} {Estimation}.
\newblock {\em American Economic Review}, 103(3):550--556.

\bibitem[\protect\astroncite{Ng and Jordan}{2002}]{ng_discriminative_2002}
Ng, A.~Y. and Jordan, M.~I. (2002).
\newblock On discriminative vs. generative classifiers: {A} comparison of
  logistic regression and naive bayes.
\newblock In {\em Advances in neural information processing systems}, pages
  841--848.

\bibitem[\protect\astroncite{Okui et~al.}{2012}]{okui_doubly_2012}
Okui, R., Small, D.~S., Tan, Z., and Robins, J.~M. (2012).
\newblock Doubly robust instrumental variable regression.
\newblock {\em Statistica Sinica}, pages 173--205.
\newblock Publisher: JSTOR.

\bibitem[\protect\astroncite{Paarsch and
  Hong}{2006}]{paarsch_introduction_2006}
Paarsch, H.~J. and Hong, H. (2006).
\newblock An introduction to the structural econometrics of auction data.
\newblock {\em MIT Press Books}, 1.
\newblock Publisher: The MIT Press.

\bibitem[\protect\astroncite{Pan et~al.}{2010}]{pan_domain_2010}
Pan, S.~J., Tsang, I.~W., Kwok, J.~T., and Yang, Q. (2010).
\newblock Domain adaptation via transfer component analysis.
\newblock {\em IEEE Transactions on Neural Networks}, 22(2):199--210.

\bibitem[\protect\astroncite{Pan and Yang}{2010}]{pan_survey_2010}
Pan, S.~J. and Yang, Q. (2010).
\newblock A {Survey} on {Transfer} {Learning}.
\newblock {\em IEEE Transactions on Knowledge and Data Engineering},
  22(10):1345--1359.

\bibitem[\protect\astroncite{Pearl}{2009}]{pearl_causality_2009}
Pearl, J. (2009).
\newblock {\em Causality}.
\newblock Cambridge university press.

\bibitem[\protect\astroncite{Perrigne and
  Vuong}{2019}]{perrigne_econometrics_2019}
Perrigne, I. and Vuong, Q. (2019).
\newblock Econometrics of {Auctions} and {Nonlinear} {Pricing}.
\newblock {\em Annual Review of Economics}, 11(1):27--54.
\newblock \_eprint: https://doi.org/10.1146/annurev-economics-080218-025702.

\bibitem[\protect\astroncite{Reiss and Wolak}{2007}]{reiss_structural_2007}
Reiss, P.~C. and Wolak, F.~A. (2007).
\newblock Structural {Econometric} {Modeling}: {Rationales} and {Examples} from
  {Industrial} {Organization}.
\newblock In Heckman, J.~J. and Leamer, E.~E., editors, {\em Handbook of
  {Econometrics}}, volume~6, pages 4277--4415. Elsevier.

\bibitem[\protect\astroncite{Robins and
  Rotnitzky}{1995}]{robins_semiparametric_1995}
Robins, J.~M. and Rotnitzky, A. (1995).
\newblock Semiparametric efficiency in multivariate regression models with
  missing data.
\newblock {\em Journal of the American Statistical Association},
  90(429):122--129.
\newblock Publisher: Taylor \& Francis.

\bibitem[\protect\astroncite{Robins et~al.}{1994}]{robins_estimation_1994}
Robins, J.~M., Rotnitzky, A., and Zhao, L.~P. (1994).
\newblock Estimation of regression coefficients when some regressors are not
  always observed.
\newblock {\em Journal of the American statistical Association},
  89(427):846--866.
\newblock Publisher: Taylor \& Francis.

\bibitem[\protect\astroncite{Rojas-Carulla
  et~al.}{2018}]{rojas-carulla_invariant_2018}
Rojas-Carulla, M., Scholkopf, B., Turner, R., and Peters, J. (2018).
\newblock Invariant models for causal transfer learning.
\newblock {\em The Journal of Machine Learning Research}, 19(1):1309--1342.

\bibitem[\protect\astroncite{Rosenbaum and
  Rubin}{1983}]{rosenbaum_central_1983}
Rosenbaum, P.~R. and Rubin, D.~B. (1983).
\newblock The central role of the propensity score in observational studies for
  causal effects.
\newblock {\em Biometrika}, 70(1):41--55.
\newblock Publisher: Oxford University Press.

\bibitem[\protect\astroncite{Rubin}{1974}]{rubin_estimating_1974}
Rubin, D.~B. (1974).
\newblock Estimating causal effects of treatments in randomized and
  nonrandomized studies.
\newblock {\em Journal of educational Psychology}, 66(5):688.
\newblock Publisher: American Psychological Association.

\bibitem[\protect\astroncite{Rust}{2014}]{rust_limits_2014}
Rust, J. (2014).
\newblock The {Limits} of {Inference} with {Theory}: {A} {Review} of {Wolpin}
  (2013).
\newblock {\em Journal of Economic Literature}, 52(3):820--850.

\bibitem[\protect\astroncite{Scharfstein
  et~al.}{1999}]{scharfstein_adjusting_1999}
Scharfstein, D.~O., Rotnitzky, A., and Robins, J.~M. (1999).
\newblock Adjusting for nonignorable drop-out using semiparametric nonresponse
  models.
\newblock {\em Journal of the American Statistical Association},
  94(448):1096--1120.
\newblock Publisher: Taylor \& Francis Group.

\bibitem[\protect\astroncite{Scornet}{2016}]{scornet_asymptotics_2016}
Scornet, E. (2016).
\newblock On the asymptotics of random forests.
\newblock {\em Journal of Multivariate Analysis}, 146:72--83.

\bibitem[\protect\astroncite{Scornet et~al.}{2015}]{scornet_consistency_2015}
Scornet, E., Biau, G., and Vert, J.-P. (2015).
\newblock Consistency of random forests.
\newblock {\em Annals of Statistics}, 43(4):1716--1741.
\newblock Publisher: Institute of Mathematical Statistics.

\bibitem[\protect\astroncite{Scott}{2014}]{scott_dynamic_2014}
Scott, P. (2014).
\newblock Dynamic discrete choice estimation of agricultural land use.
\newblock Publisher: TSE Working Paper.

\bibitem[\protect\astroncite{Shalizi}{2013}]{shalizi_advanced_2013}
Shalizi, C. (2013).
\newblock {\em Advanced data analysis from an elementary point of view}.
\newblock Cambridge University Press Cambridge.

\bibitem[\protect\astroncite{Steel}{2019}]{steel_model_2019}
Steel, M.~F. (2019).
\newblock Model averaging and its use in economics.
\newblock {\em arXiv preprint arXiv:1709.08221}.

\bibitem[\protect\astroncite{Sugiyama et~al.}{2008}]{sugiyama_direct_2008}
Sugiyama, M., Nakajima, S., Kashima, H., Buenau, P.~V., and Kawanabe, M.
  (2008).
\newblock Direct {Importance} {Estimation} with {Model} {Selection} and {Its}
  {Application} to {Covariate} {Shift} {Adaptation}.
\newblock In Platt, J.~C., Koller, D., Singer, Y., and Roweis, S.~T., editors,
  {\em Advances in {Neural} {Information} {Processing} {Systems} 20}, pages
  1433--1440. Curran Associates, Inc.

\bibitem[\protect\astroncite{Tan}{2010}]{tan_bounded_2010}
Tan, Z. (2010).
\newblock Bounded, efficient and doubly robust estimation with inverse
  weighting.
\newblock {\em Biometrika}, 97(3):661--682.
\newblock Publisher: Oxford University Press.

\bibitem[\protect\astroncite{Tzeng et~al.}{2014}]{tzeng_deep_2014}
Tzeng, E., Hoffman, J., Zhang, N., Saenko, K., and Darrell, T. (2014).
\newblock Deep domain confusion: {Maximizing} for domain invariance.
\newblock {\em arXiv preprint arXiv:1412.3474}.

\bibitem[\protect\astroncite{Van~der Laan
  et~al.}{2007}]{van_der_laan_super_2007}
Van~der Laan, M.~J., Polley, E.~C., and Hubbard, A.~E. (2007).
\newblock Super learner.
\newblock {\em Statistical applications in genetics and molecular biology},
  6(1).
\newblock Publisher: De Gruyter.

\bibitem[\protect\astroncite{Vermeulen and
  Vansteelandt}{2015}]{vermeulen_bias-reduced_2015}
Vermeulen, K. and Vansteelandt, S. (2015).
\newblock Bias-reduced doubly robust estimation.
\newblock {\em Journal of the American Statistical Association},
  110(511):1024--1036.
\newblock Publisher: Taylor \& Francis.

\bibitem[\protect\astroncite{Wang and Deng}{2018}]{wang_deep_2018}
Wang, M. and Deng, W. (2018).
\newblock Deep visual domain adaptation: {A} survey.
\newblock {\em Neurocomputing}, 312:135--153.

\bibitem[\protect\astroncite{Wolpert}{1992}]{wolpert_stacked_1992}
Wolpert, D.~H. (1992).
\newblock Stacked generalization.
\newblock {\em Neural networks}, 5(2):241--259.
\newblock Publisher: Elsevier.

\bibitem[\protect\astroncite{Wolpin}{2013}]{wolpin_limits_2013}
Wolpin, K.~I. (2013).
\newblock {\em The {Limits} of {Inference} without {Theory}}.
\newblock MIT Press.
\newblock Google-Books-ID: ueXxCwAAQBAJ.

\bibitem[\protect\astroncite{Yao et~al.}{2018}]{yao_using_2018}
Yao, Y., Vehtari, A., Simpson, D., and Gelman, A. (2018).
\newblock Using {Stacking} to {Average} {Bayesian} {Predictive} {Distributions}
  (with {Discussion}).
\newblock {\em Bayesian Analysis}, 13(3):917--1007.
\newblock Publisher: International Society for Bayesian Analysis.

\bibitem[\protect\astroncite{Zadrozny}{2004}]{zadrozny_learning_2004}
Zadrozny, B. (2004).
\newblock Learning and evaluating classifiers under sample selection bias.
\newblock In {\em Proceedings of the twenty-first international conference on
  {Machine} learning}, {ICML} '04, page 114, Banff, Alberta, Canada.
  Association for Computing Machinery.

\bibitem[\protect\astroncite{Zhang et~al.}{2016}]{zhang_optimal_2016}
Zhang, X., Yu, D., Zou, G., and Liang, H. (2016).
\newblock Optimal model averaging estimation for generalized linear models and
  generalized linear mixed-effects models.
\newblock {\em Journal of the American Statistical Association},
  111(516):1775--1790.
\newblock Publisher: Taylor \& Francis.

\end{thebibliography}

\end{document}